\documentclass[pdflatex,sn-mathphys-num,iicol]{sn-jnl}

\usepackage{stix} 
\usepackage{helvet} 
\usepackage{multirow}
\usepackage{natbib}
\usepackage{array}
\setcitestyle{numbers, sort&compress,open={[},close={]},comma}
\usepackage[greek,english]{babel}
\usepackage{footnote}
\usepackage{placeins}
\usepackage{nicefrac}

\usepackage{graphicx} 
\usepackage[font={sf, footnotesize}]{caption}
\usepackage[dvipsnames]{xcolor} 
\usepackage{hyperref}
\usepackage{enumitem}
\usepackage{threeparttable}
\usepackage{amsmath,amssymb,amsfonts}%
\usepackage{manyfoot}%
\usepackage{algorithm}%
\usepackage{algorithmicx}%
\usepackage{algpseudocode}%
\usepackage{listings}%
\usepackage{arcs}
\usepackage{centernot}
\usepackage{upgreek}
\usepackage{siunitx}
\usepackage{alphabeta}
\usepackage{todonotes}
\usepackage{booktabs}
\usepackage{bm}
\usepackage{cancel}
\usepackage[title]{appendix}%

\usepackage{empheq}

\theoremstyle{thmstyleone}%
\theoremstyle{thmstyletwo}%

\theoremstyle{thmstylethree}%

\newcommand{\p}[2]{p\big(#1   \boldsymbol{\mid}  #2 \big)}
\newcommand{\pr}[1]{p \big( #1 \big) }
\newcommand{\vv}[1]{{\boldsymbol #1}}
\newcommand{\x}{\vv x}

\newcommand{\vtheta}{\vv \theta}

\newcommand{\var}{\mathrm{var}}

\newcommand{\Xset}{\underline{\mathrm{\bm{x}}}}

\begin{document}
\title[Uncertainty quantification in mechanics: A unified Bayesian perspective]{%
  \protect\centering
  Uncertainty quantification in mechanics:\\
  A unified Bayesian perspective}

%

%
\author*[1,2]{\fnm{Sascha} \sur{Ranftl}}\email{sranftl@purdue.edu}

\author[3,4]{\fnm{Malte} \sur{Rolf}}

\author[4,5]{\fnm{Gerhard A.} \sur{Holzapfel}}

\author[3,6]{\fnm{Ellen} \sur{Kuhl}}

\affil[1]{\orgdiv{Courant Institute of Mathematical Sciences}, \orgname{New York University}, \orgaddress{\city{New York City}, \state{NY}, \country{USA}}}

\affil[2]{\orgdiv{Division of Applied Mathematics}, \orgname{Brown University}, \orgaddress{\city{Providence}, \state{RI}, \country{USA}}}

\affil[3]{\orgdiv{Institute of Applied Mechanics}, \orgname{Friedrich-Alexander-Universität Erlangen-Nürnberg}, \orgaddress{\city{Erlangen}, \country{Germany}}}

\affil[4]{\orgdiv{Institute of Biomechanics}, \orgname{Graz University of Technology}, \orgaddress{\city{Graz}, \country{Austria}}}

\affil[5]{\orgdiv{Department of Structural Engineering}, \orgname{Norwegian University of Science and Technology (NTNU)}, \orgaddress{\city{Trondheim}, \country{Norway}}}

\affil[6]{\orgdiv{Department of Mechanical Engineering}, \orgname{Stanford University}, \orgaddress{\city{Stanford}, \state{CA}, \country{USA}}}

\abstract{
Uncertainty quantification (UQ) is essential to experimental mechanics, but has become particularly relevant in computational mechanics, manifesting in two fundamental problem types: forward and inverse problems. The former addresses how input uncertainties propagate to the quantities of interest, whereas the latter aims to infer unknown parameters from experimental observations or simulations. Since efficient propagation typically requires a prohibitive number of evaluations to compute marginal output distributions, the development of fast, data-driven surrogate models becomes necessary. Thus, we can distinguish between two inverse tasks: (i) the identification and calibration of input uncertainties, and (ii) the construction of surrogates, a methodology collectively referred to as surrogate-based UQ.
Building on probabilistic reasoning and the concept of partial belief, we demonstrate that Bayesian probability theory provides a unified theoretical framework for addressing both problem types.
We further show that Bayesian inference allows for the seamless incorporation of essential subproblems, including model selection for identifying the most probable model specifications and experimental design for optimizing data collection by identifying experiments or simulations that maximize expected information gain about parameters, among others such as connections to sensitivity analysis or the use of special priors like random fields.
While this theoretical framework is presented for general mechanical problems, particular emphasis is placed on biomechanics, where variability and uncertainty is especially pronounced due to inherent biological heterogeneity, patient-specific variability, and noisy data.}


\maketitle

\vspace*{3cm}

\newpage
\tableofcontents

\newpage
\section{Introduction}\label{sec:introduction}
The integration of uncertainty quantification (UQ) has become indispensable to the fields of experimental and computational mechanics alike. As computational models increasingly inform engineering design and safety-critical decisions, deterministic output predictions are no longer sufficient. Instead, ensuring the reliability and robustness of simulation results requires a rigorous assessment of how input uncertainties, such as material properties, geometry, and boundary conditions, propagate through mechanical systems to affect the quantities of interest.

Biomechanical systems represent a prime example of where these uncertainties become most apparent. While physics-based mathematical models have substantially advanced our understanding of complex biomechanical systems, such as cardiovascular, respiratory, or cerebral physiology, their typically deterministic nature struggles to account for patient-specific variability, inherent biological heterogeneity, and noisy data. Collectively, these limitations remain a significant bottleneck for clinical translation.

In particular, two key challenges in the clinical application of biomechanical models persist~\cite{Eck2016}. First, quantifying uncertainty when adapting model inputs to patient-specific conditions is inherently difficult, as available data are often uncertain, incomplete, or noisy. Second, a persistent trade-off between model complexity and input uncertainty requires balancing errors introduced by model simplifications against those arising from uncertain parameters. Although many computational models aim to support clinical decision-making and accelerate medical device development, this potential remains only partially realized despite substantial progress over the past decade, largely due to the lack of a consistent framework.

\begin{figure*}[t!]
    \centering
    \includegraphics[scale=1.0]{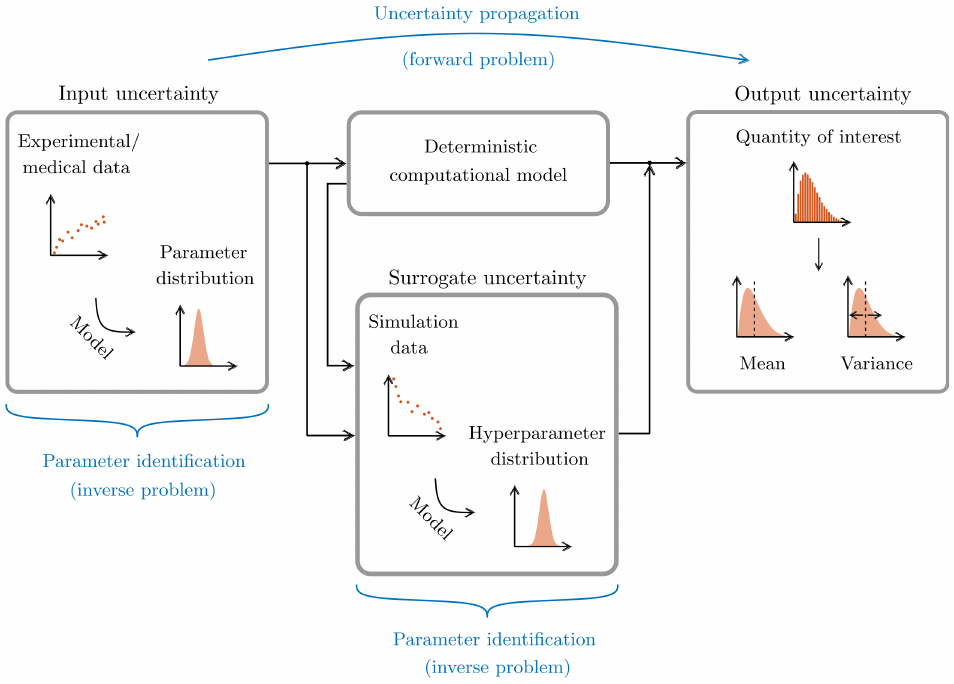}
    \caption{Overview of surrogate-based uncertainty quantification by propagating uncertainty from input to output and addressing the two inverse problems involved.}
    \label{fig-overview}
\end{figure*}

Together, these challenges motivate a unified theoretical framework for UQ. Indeed, Bayesian probability theory has been mathematically proven as the unique, consistent approach for inference under uncertainty~\cite{Cox1946, sivia2006}. By treating probability as a measure of partial belief, this theory provides a rigorous foundation to explicitly quantify all uncertainty sources, including noisy data, missing information, and uncertain models, while naturally incorporating prior knowledge. Building on these probabilistic principles, we demonstrate that Bayesian theory effectively addresses the two primary problem types: (i) \emph{forward problems} and (ii) \emph{inverse problems}, with many practical cases involving a combination of both, as shown in Fig.~\ref{fig-overview}.

Forward problems in UQ refer to quantifying how uncertainties in the model inputs affect the uncertainty in a model output or quantity of interest. For instance, one may ask: \emph{What is the uncertainty in a particular Cauchy stress component at a certain point in the aorta wall due to uncertainties in the parameters of our constitutive model?} 
Thus, we seek to \emph{propagate} uncertainty from the model inputs to the output. Indeed, it will soon become apparent that uncertainty propagation often involves solving inverse problems for two distinct practical purposes: (i) quantification of the input uncertainty and (ii) surrogate modeling.

The term \emph{quantification of the input uncertainty}, whether based on experimental data or on medical data in biomechanics, may require a parameter estimation problem for the constitutive model before uncertainty propagation can commence. It is therefore often referred to as \emph{model calibration} and can also be reformulated as an \emph{optimization problem}. Similar to before, one may ask: \emph{What are the most likely value and associated uncertainties for the stiffness parameter of a constitutive model, given a set of, say, uniaxial tension tests?} The estimated parameters, which link the measured experimental quantities to the parameters of interest, can then serve as input for computational simulations to propagate these uncertainties through the model.

The term \emph{surrogate modeling} requires further explanation. In order to propagate uncertainty, we need a sufficient number of samples to compute the \emph{marginal probability} of the model output. We obtain this marginal probability by integrating the joint probability with respect to the model input parameters.
In biomechanics, however, obtaining a single model output for one specific parameter set can demand hours, days, or even weeks of computational time. Unfortunately, reliable statistical estimates of model output uncertainty typically require thousands to millions of samples. This issue arises directly from the convergence behavior of numerical integration techniques used to evaluate the marginal probability (cf.~Wollner et al.~\cite{Wollner2025}).

Nevertheless, not all hope is lost. A common remedy is the use of a surrogate model, also known as an \emph{emulator} or \emph{meta-model}, which provides a fast, data-driven approximation of the original model.

Constructing such a surrogate follows a standard three-step procedure. This involves (i) generating a manageable dataset of model outputs by solving the original model for selected input parameter samples, (ii) fitting a parameterized function to these simulations to form the surrogate, and (iii) using this surrogate in place of the original model to efficiently approximate the marginal probability or output uncertainty.
In essence, we have demonstrated that the forward UQ task frequently incorporates the inverse problem of identifying surrogate parameters. The surrogate should (i) approximate the original model as accurately and faithfully as possible so that the resulting uncertainty estimates remain meaningful and statistically significant, even with limited data, and (ii) be inexpensive and fast to evaluate, so that the computation becomes practically feasible.

In this review article, we show that Bayesian probability theory provides the unifying theoretical framework, that incorporates all these problems and concepts into a single coherent whole. 
For this purpose, we begin by introducing the concepts of logic of belief and partial truth in Section~\ref{sec:logic}. We then propose a workflow for UQ in Section~\ref{sec:UQ}, ranging from inverse problems, namely model calibration and parameter estimation as well as surrogate modeling, to uncertainty propagation. Building on this foundation, we review popular surrogate models and selected alternatives and discuss model selection within the Bayesian framework in Section~\ref{sec:types-of-surrogates}. Next, we introduce Bayesian experimental design and provide a brief excursion into Bayesian optimization for inverse design in Section~\ref{sec:experimental-design}. This is followed by a brief exposition of the connection to sensitivity analysis in Section~\ref{sec:sensitivity-analysis}. It shows that sensitivity analysis can be understood as a special case and interpretation of the foregoing Bayesian theory, including a Bayesian generalization of Sobol' indices. We conclude with the introduction of random fields for heterogeneous materials modeling as a special statistical prior within the Bayesian theory in Section~\ref{sec:random-fields}. While many topics are accompanied by illustrative mechanical examples, detailed computational aspects and sampling methods are only briefly touched upon, as they remain beyond the scope of this review article.

\section{The logic of belief and partial truth}\label{sec:logic}
Here, we introduce probability theory as an extension of Aristotelian \emph{logic}. Aristotle's logic is formalized through logical statements called \emph{propositions} ($A, B, C, \ldots$). 

For example, let $A$ denote the proposition $A \equiv $ \lq Julius Caesar was left-handed.' Its negation, $\neg A$, would then mean $\neg {A} \equiv $ \lq Julius Caesar was not left-handed.' Other propositions might include: $B \equiv $ \lq Julius Caesar was right-handed' and $C \equiv $ \lq Julius Caesar was ambidextrous.' New propositions can be formed by combining existing ones with logical operations. The logical AND operation ($\wedge$) yields statements such as $A \wedge B$, meaning $A\wedge B \equiv$ \lq Julius Caesar was left-handed and right-handed.' which is logically equivalent to $A\wedge B \equiv$ \lq Julius Caesar was ambidextrous $\equiv C$.' Likewise, the logical OR operation ($\vee$) allows combinations like $A \vee B$, meaning $A\vee B \equiv$ \lq Caesar was left- or right-handed.' In fact, the logical OR is already one concept more than strictly necessary, since De Morgan’s law allows it to be expressed in terms of negation and conjunction: $\neg(A \vee B) = \neg A \wedge \neg B$.

Propositions can take one of two logical values: \emph{true} or \emph{false}. For instance, consider the statement $A \equiv $ \lq Julius Caesar was left-handed.' If this proposition is true, its negation must necessarily be false. Thus, $A = \text{True} \implies \neg A = \text{False}$ and \textit{vice versa} $A = \text{False} \implies \neg A = \text{True}$. Importantly, both $A$ and its negation $\neg A$ cannot be true at the same time. That is, the compound statement $A\wedge \neg A \equiv$ \lq Julius Caesar was left-handed and not left-handed' is false regardless of which is option is actually true. In Boolean algebra, the logical value \emph{true} is conventionally represented by the number $1$, and \emph{false} by the number $0$. Hence, if $A = 1$, then its negation satisfies $\neg A = 0$; equivalently, $A = 1 \implies \neg A = 0$. The conjunction of two propositions, $ C \equiv A \wedge B$, is true only when both $A$ and $B$ are true. In other words, the statement \lq Caesar was left- and right-handed' can only be true if both statements --- \lq Caesar was left-handed' and \lq Caesar was right-handed' --- are true. By contrast, the proposition, or disjunction (logical OR), \lq Caesar was left- or right-handed' is true if either (or both) of the statements is true. Formally, $A =1 \implies A\vee B =1$, and, independently, $B =1 \implies A\vee B =1$. The possible combinations of truth values for propositions and their logical operations can be conveniently summarized using truth tables, see Table~\ref{tab: truth}.
\begin{table}[t!]
\centering
\caption{Truth table according to Boolean (Aristotelian) \emph{logic}.}
\begin{tabular}{ccccc}
\toprule
$A$ & $B$ & & $A\wedge B $ & $A\vee B$\\
\midrule
1 & 0 && 0 & 1  \\
0 & 1 && 0 & 1  \\
1 & 1 && 1 & 1  \\
0 & 0 && 0 & 0 \\
\bottomrule
\end{tabular}
\label{tab: truth}
\end{table}

Propositions can also be conditioned on other propositions, denoted by the vertical bar ($\mid$). For example, $A\mid B =1$ is read as \lq $A$ is true given $B$ is true,' \lq $A$ is true on the condition that $B$ is true,' or simply \lq $A$ is true \emph{if} $B$ is true.' In our running example, this corresponds to the statement \lq Caesar was left-handed, assuming that he was right-handed.' In essence, a \emph{condition} specifies a prerequisite that is taken to be true before evaluating the truth of another proposition. Within the above truth table, conditioning on $B=1$ means restricting our attention to those rows where $B$ is fixed at $1$. In other words, we consider only the second and third rows, where $B$ is true, and assess the truth value of $A$ under that condition.

We will shortly see how Bayesian probability theory extends Boolean logic, in which a proposition is \emph{either} true ($1$) or false ($0$), to allow any real value in between. Such values can be interpreted as \emph{partial truths} or as a degree of \emph{belief} in the truthfulness of a given proposition. In fact, it was mathematically proven by Cox~\cite{Cox1946}, and later presented in a more modern formulation in~\cite{garrett1998whence,sivia2006}, that under certain natural assumptions\footnote{Specifically, Cox assumed that (i) degrees of belief are represented by real numbers, (ii) the system reduces to classical logic in the deterministic limit, and (iii) reasoning is internally consistent, i.e., equivalent inference paths yield identical results. Under these desiderata, the probability calculus is the unique solution.}, the probability calculus is, up to a constant exponent, the unique and only consistent extension of logic to partial truths. In this sense, the fundamental laws of probability arise directly from the requirements of logical consistency. They emerge naturally from Boolean algebra~\cite{Jaynes2003,VonToussaint2011,vonderLinden2014}. 

Let $P(A)$ denote the probability, or degree of belief, that the proposition $A$ is true. Let a belief of $1$ correspond to certainty, and a belief of $0$ to impossibility. In principle, any other numerical representation could have been chosen for \emph{certainty} or \emph{truth}; the present choice is simply the most convenient. Logical consistency requires that either $A$ or its negation $\neg A$ must be true, and conversely, that both cannot be true at the same time, i.e.,
\begin{align} \label{eq:negation}
    P(A \vee \neg A) = 1\;,\quad  P(A \wedge \neg A) = 0\;,
\end{align}
where these relations capture the most fundamental constraints of Boolean logic.

Cox~\cite{Cox1946}, building on Boole's work, identified six basic properties of logic: (i) double negation: $\neg (\neg A) = A$; (ii) commutativity: 
\newcommand{\wedgevee}{  \mathbin{{\wedge}\mkern-2mu{\vee}}}%
\newcommand{\veewedge}{  \mathbin{{\vee}\mkern-2mu{\wedge}}}%
$A\wedgevee B = B \wedgevee A$; (iii) idempotence: stating a proposition twice is equivalent to stating it once, $A\wedgevee A = A$; (iv) associativity: $A \wedgevee(B\wedgevee C) = (A\wedgevee B) \wedgevee C$; (v) De Morgan's law: $\neg (A\wedgevee B) = \neg A \veewedge \neg B$; and (vi) absorption: $A\wedgevee (A\veewedge B) = A$, where $\wedgevee$ and $\veewedge$ denotes two possible alternatives of each axiom using, consistently, either the first or second symbol each. In other words, each of the axioms: (ii) to (vi), has two equivalent variants.
Cox argued that these basic requirements of logical consistency must also hold for partial truths. For example, the law of double negation, $\neg (\neg A) = A$, implies that the corresponding degree of belief satisfies $P(\neg (\neg A)) = P(A)$, or formally, $\neg (\neg A) = A \implies P(\neg (\neg A)) = P(A)$. He further introduced the \emph{principle of transitivity}: If we believe more strongly in $A$ than in $B$, and more in $B$ than in $C$, then we must also believe more strongly in $A$ than in $C$; formally, $\big(P(A) > P(B)\big) \wedge \big(P(B)>P(C)\big) \implies P(A) > P(C)$.

Based on these logical axioms, Cox derived the so-called \emph{sum rule} and \emph{product rule} without assuming any particular functional form. This derivation relies on two key assumptions: (i) different paths of logical reasoning given the same context $\mathcal{K}$ must lead to the same conclusions, and (ii) the probability of a statement being true is directly related to the probability of its negation. The resulting expressions are the sum rule,
\begin{align} \label{eq:sum-rule}
   P(A \wedge B \mid \mathcal{K}) &=  P(A\mid \mathcal{K}) + P(B\mid \mathcal{K}) \nonumber\\
   &\hphantom{=}\,\, - P(A \vee B\mid \mathcal{K})\;,
\end{align}
and the product rule,
\begin{align} \label{eq:product-rule}
    P(A \wedge B\mid \mathcal{K}) = P(A\mid B, \mathcal{K})P(B\mid \mathcal{K})\;.
\end{align}
Here, $P(A \mid B, \mathcal{K})$ denotes the belief that $A$ is true, given that statement $B$ is assumed to be true and given context $\mathcal{K}$. For complete proofs, the interested reader is referred to related literature~\cite{Cox1946,garrett1998whence,knuth2012foundations}.

These two rules (Eqs.~\eqref{eq:sum-rule} and~\eqref{eq:product-rule}) form the mathematical foundation of Bayesian probability theory and provide a consistent framework for reasoning under uncertainty. To apply them meaningfully, however, one must always specify the context or background information $\mathcal{K}$ on which the probabilities are conditioned.
In any given problem, including those in the broad field of mechanics, there is always some background information available, at the very least in the form of the problem definition. This context is not a mere formality. It implies that (i) there is no such thing as an \emph{unconditional probability} and (ii) omitting relevant background knowledge often leads to inconsistent reasoning.
In our earlier discussion of Caesar’s dexterity, the background information implicitly included the assumptions that Caesar existed and that he was a human with two hands. To illustrate the importance of such context, let us now consider a different example. Suppose we analyze the preferred hand of the Hindu deity Vishnu, who is often depicted with four arms.
Counting clockwise and starting at an arbitrary arm, let us define the proposition $A_i \equiv$ \lq Vishnu's preferred arm was number $i$,' where $i=1,2,3,4$. Then, by basic deductive logic, we have $\neg A_1 = A_2 \vee A_3 \vee A_4$, or more generally $\neg A_i = \bigvee_{j\neq i} A_j$ for $j=1,2,3,4$. This leads to four equations $P(A_i \mid \mathcal{K}) + P(\neg A_i\mid \mathcal{K}) = 1$, and consequently, $\sum_i P(A_i\mid \mathcal{K}) = 1$. We see that the \emph{context} $\mathcal{K}$ of either Caesar or Vishnu substantially changes the analysis. In the probability-theoretic literature, this context $\mathcal K$ is often denoted in the form $p(A\mid \mathcal{K})$ with an explicit conditional complex of propositions. For the remainder of this exposition, we will omit the explicit denotation of this context, the reader however be advised that probabilities are nevertheless and  always conditional on context.

As introduced above, Bayesian probability theory is formulated in terms of propositions, to which it assigns a non-negative measure of truth called \emph{probability}. Unlike in Boolean algebra, propositions are not limited to being either true or false, but may take any value in between. In this way, Bayesian probability theory extends inductive logic to situations involving uncertainty, where probabilities represent degrees of belief rather than absolute truth values. 
In contrast to frequentist statistics, this concept of probability is not an intrinsic property of a system or object that arises from the relative frequency of occurrence of a proposition in an experiment repeated \emph{ad infinitum}. Instead, it expresses a subjective degree of belief in a proposition, conditioned on the available information. Different observers, having access to different information, may therefore reach different conclusions. However, consistency requires that observers who possess the same or equivalent information must arrive at the same conclusions. Together, this principle of consistency, along with the sum and product rules, forms the logical foundation of Bayesian inference and determines how beliefs should be updated as new information becomes available.

By combining the rules of logical consistency for partial truths, we arrive at the central relationships of Bayesian reasoning. Using the sum rule, we can immediately derive from Eq.~\eqref{eq:negation} the intuitive result $P(A \vee \neg A) = P(A) + P(\neg A) =  1$. Another basic intuition, and one of Cox's axioms mentioned before, suggests that our belief in the statement $A\wedge B$ should yield the exact same truth value as the logically equivalent reversed statement $B \wedge A$. Formally: $A\wedge B = B\wedge A \Rightarrow P(A\wedge B) = P(B\wedge A)$. From this point forward, we adopt notation that is more common in the contemporary literature and denote the logical AND ($\wedge$) with a comma (,), so that $P(A,B) \equiv P(A \wedge B)$. Applying the product rule Eq.~\eqref{eq:product-rule}, and, carelessly, omitting the context $\mathcal{K}$, we immediately arrive at the well-known \emph{Bayes' theorem},
\begin{align}\label{eq:Bayes-theorem}
    P(A,B) & = P(B,A) \nonumber\\ 
           & = P(A \mid B) P(B) \nonumber\\
           & = P(B \mid A) P(A)\;.
\end{align}
Bayes’ theorem expresses how our belief in a proposition $A$ should be updated when new information $B$ becomes available. The term $P(A)$ represents the \emph{prior probability}, quantifying our degree of belief in $A$ before observing $B$. The term $P(B \mid A)$ is the \emph{likelihood}, describing how compatible the new information $B$ is with the assumption that $A$ is true. The updated belief, or \emph{posterior probability}, is given by $P(A \mid B)$, which combines prior knowledge with new information through the normalization factor $P(B)$, known as the \emph{evidence}. In this way, Bayes’ theorem provides a consistent and logically grounded method for learning from data.

Evaluating the normalization term $P(B)$ typically requires accounting for all mutually exclusive possibilities under which $B$ could occur. This is achieved through the \emph{marginalization rule}, which ensures that probabilities remain consistent when integrating over unknown or unobserved variables. From the sum and product rules above, and using the law of negation (Eq.~\eqref{eq:negation}), we obtain the marginalization rule in its simplest form,
\begin{align}
    P(B) & = P(B, A\vee\neg A) =  P(B, A)  + P(B, \neg A) \nonumber\\
    & = P(B \mid A) P(A) + P(B \mid \neg A) P(\neg A)\;.
\end{align}
The marginalization rule generalizes naturally to a set of mutually exclusive and exhaustive propositions $B_i$, where $B_i \wedge B_j =0$ for all $i\neq j$, with $i,j = 1,\ldots,N$ and $\bigvee_{i=1}^N B_i = 1$; that is, no two propositions $B_j$ can be true simultaneously, and the disjunction of all propositions is surely true,
\begin{align} \label{eq:marginalization-rule-discreet}
    P(A) = \sum_{i=1}^N P(A, B_i) = \sum_{i=1}^N P(A\mid B_i) P(B_i)\;.
\end{align}
Moreover, the marginalization rule can also be extended to continuous propositions $a,b,c$, where the proposition $a$ expresses \lq $a$ has value in $[a_{\rm min},a_{\rm max}]$,' or $a \in [a_{\rm min},a_{\rm max}]$. For continuous variables, the probability of a specific value becomes infinitesimally small~\cite{ChapterWollner}. In this case, the notion of probability is generalized to \emph{probability density functions} (PDF), denoted $p(a)$. The continuous form of the marginalization rule is then
\begin{align}\label{eq:marginalization-rule}
    p(a) & = \int p(a\mid b)p(b)\,\mathrm{d}b\;.
\end{align}

Together, the marginalization rule and Bayes' theorem provide all necessary tools for reasoning and computation with probabilities\footnote{Note the distinction between a random variable, an abstract quantity ranging over possible values, and its realization, the concrete value observed in a particular instance~\cite{ChapterWollner}.}, or what we may call the \emph{calculus of uncertainties}. Indeed, either the pair of the sum and product rules, or the pair of Bayes theorem (Eq.~\eqref{eq:Bayes-theorem}) and the marginalization rule (Eq.~\eqref{eq:marginalization-rule-discreet}), are the only two rules we need to remember to do Bayesian inference.

\section{Uncertainty quantification}\label{sec:UQ} 
Bayesian probability theory provides a robust framework for quantifying information and characterizing uncertainty across diverse contexts. It enables us to perform statistical inference, make controlled approximations, and switch seamlessly between forward and inverse problems. Furthermore, it allows for the analysis of how uncertainties interact within coupled physical models, even when informed by heterogeneous and noisy datasets.
Remarkably, models for data analysis and UQ are built upon just two fundamental rules: (i) Bayes’ theorem and (ii) the marginalization rule.

\begin{figure*}[t!]
    \centering
    \includegraphics[scale=1.0]{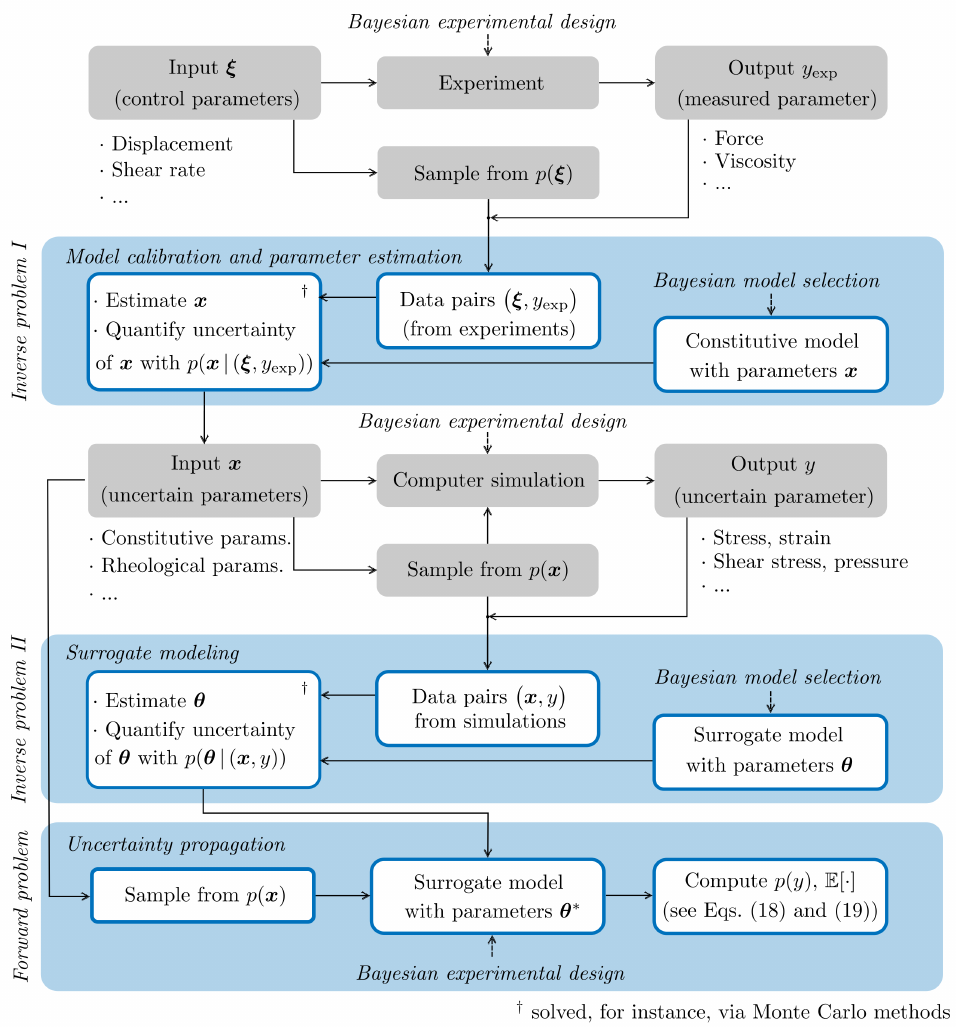}
    \caption{
    Typical workflow for uncertainty quantification in mechanics.
    An experiment provides input-output data pairs $(\bm{\xi},y_{\rm exp})$, which are used for model calibration and parameter estimation (inverse problem I), yielding uncertain parameters of a constitutive model $\bm{x}$. These uncertain parameters then serve as uncertain inputs to a computer simulation to predict uncertain outputs. Because such simulations are often computationally expensive, their sparse input–output data pairs $(\bm{x},y)$ are often used to train the parameters of a surrogate model $\bm{\theta}$ (inverse problem II) that replaces the computer simulation. In the final step, the uncertainties in the constitutive parameters --- and, if applicable, in the surrogate model itself --- are propagated from inputs to outputs using this fast surrogate approximation.
    The workflow is flexible: a surrogate may be unnecessary in some cases, and it can equally accommodate related experiments or coupled simulations without changing the formalism; see, e.g.,~\cite{avril2008overview}, where constitutive parameters are calibrated from finite element simulations of image-based extension tests, reversing the roles of experiment and simulation.
    }
    \label{fig-workflow}
\end{figure*}

To implement this framework effectively, we follow a structured workflow for UQ, as illustrated in Fig.~\ref{fig-workflow}. The following sections adhere strictly to the notation and logic established in this process.

\subsection{Uncertainty propagation} \label{ssec:propagation}
In the workflow of UQ, the core objective of uncertainty propagation can be formulated as:

\begin{quotation}
    \textit{Quantification of how input uncertainties propagate through a computational model to yield a predictive PDF for the quantity of interest, or a summary proxy such as the variance thereof.}
\end{quotation}
\subsubsection{General concept}
We first introduce (surrogate-based) uncertainty propagation in a Bayesian framework~\cite{Haylock1997,OHagan1999,Kennedy2000,oakley2002bayesian,OHagan2006,ranftl2021bayesian}.
Let $D\in\mathbb{N}$ denote the number of uncertain parameters. Moreover, we define $\bm{x} = (x_1, \ldots, x_{D})^{\rm T} \in \mathcal{X}$ as the vector of uncertain parameters or random variables within the parameter space~$\mathcal{X}$ and let $y = y(\bm{x})$\footnote{Note that $y$ here simultaneously plays the role of a random variable, a function of $\bm{x}$, and a realization thereof. We keep this simplified notation to avoid further overloading an already extensive formalism, and expect the intended meaning to remain clear from context.}, whose image $\mathcal{Y}$ represents the set of all possible values of $y$, represent the quantity of interest.
Following the notation in \cite{Jaynes2003,vonderLinden2014}, the fundamental object that we are interested in, and that fully describes the uncertainty, is the probability (density) for $y$, denoted as
\begin{align} \label{eq:uq-posterior}
    p(y)\;,
\end{align}
and the probability for $\bm{x}$, denoted as
\begin{align}  \label{eq:uq-prior}
    p(\bm{x})\;.
\end{align}
Per the marginalization rule (Eq.~\eqref{eq:marginalization-rule}), these two quantities are related as follows,
\begin{align} \label{eq:marginalization2}
 p(y) = \int_{\bm{x} \in \mathcal{X}} p(y \mid \vv x)p(\vv x) \, \mathrm{d}\bm{x}\;,   
\end{align}
where $\mathrm{d}\vv x$ denotes integration with respect to the volume measure (the \lq\lq infinitesimal volume element'') on $\mathcal{X}$.
In the forward problem, where the uncertainty is propagated from the parameters $\bm{x}$ to the quantity of interest $y$, the main task is marginalization, since it allows us to obtain the probability for $y$ by integrating over the uncertain inputs $\vv x $, i.e., averaging over all possible parameter values weighted by their probabilities. In the inverse problem, however, where we infer the probability for $\bm{x}$ from the probability for $y$, we rely on Bayes’ theorem.

Having established the most general framework, we now turn to how uncertainty is commonly represented and interpreted in practical application in the field of mechanics. While the complete description of uncertainty is provided by the probability $p(y)$, it is often more convenient to express uncertainty through simplified statistical summary measures or proxies derived from this probability. Alas, researchers often interpret \textit{uncertainty} in a narrower sense, typically associating it with the \emph{variance}. In fact, this narrow notion of uncertainty is merely derived from the underlying probability, as will become apparent after defining these quantities below.

Let us first introduce the expected value, which represents the average or mean value of $y$, weighted by its probability. The expected value is commonly denoted by $\big\langle \cdot \big\rangle$ or $\mathbb{E}[\cdot]$. Here, we follow the latter notation, and the expectation value of $y$ is defined as
\begin{align}
    \mathbb{E}[y] =  \int_{y \in \mathcal{Y}} y p(y)\: \mathrm{d}y\;. 
\end{align}
Next, we introduce the variance of $y$, commonly denoted as $\mathrm{var}[y]$, $\mathbb{E}[(y-\mathbb{E}[y])^2]$, or $\mathbb{V}[y]$, which measures the spread or dispersion of possible outcomes around the mean. Often referred to colloquially as the \emph{uncertainty} of $y$, it is defined as
\begin{align}
\mathrm{var}[y] = \int_{y \in \mathcal{Y}} \big(y - \mathbb{E}[y]\big)^2p(y)\:\mathrm{d}y\;. \label{eq:def-2ndmoment}
\end{align}
These two quantities, the expectation and the variance, provide statistical metrics for describing uncertainty in a compact form. This reduction is however one-directional, as infinitely many distributions can share the same mean and variance. In mechanics, UQ is therefore often understood as the computation of the integral in Eq.~\eqref{eq:def-2ndmoment}. While the probability $p(y)$ holds \textit{all} information about the uncertainty by capturing all possible outcomes and their likelihoods, the variance offers a simplified statistical measure that summarizes this information into a single scalar quantity. This simplification is often pragmatic, since computing the probability $p(y)$ using the marginalization rule (Eq.~\eqref{eq:marginalization2}) can be significantly more involved than evaluating the variance $\mathrm{var}[y]$ through Eq.~\eqref{eq:def-2ndmoment}, which is to be shown now.

Having defined these simple statistical measures, we now turn to the practical question of how to compute $\mathrm{var}[y]$ when prior knowledge about the input parameters $\bm{x}$ is available. This knowledge about $\bm{x}$ is described by the probability $p(\bm{x})$. However, the probability $p(\bm{x})$ does not appear explicitly in Eq.~\eqref{eq:def-2ndmoment}. Furthermore, the computational model output $y$ is typically defined only as a function of the inputs, $y = y(\bm{x})$. To incorporate the probability $p(\bm{x})$ into Eq.~\eqref{eq:def-2ndmoment}, we apply the marginalization rule as in Eq.~\eqref{eq:marginalization2} and rearrange the terms as follows
\begin{empheq}[box=\fbox]{align}
    \mathbb{E}[y] &= \int_{\bm{x} \in \mathcal{X}}  \int_{y \in \mathcal{Y}}  y \, \p{y}{\bm{x}} \pr{\bm{x}}\: \mathrm{d}{\bm{x}} \: \mathrm{d}y\;, \\
    \mathrm{var}[y] &= \int_{\bm{x} \in \mathcal{X}}  \int_{y \in \mathcal{Y}}  \Big(y  - \mathbb{E}[y]\Big)^2\p{y}{\bm{x}}\nonumber\\
    &\hphantom{=}\,\, \times \pr{\bm{x}} \: \mathrm{d}{\bm{x}} \: \mathrm{d}y\;. \label{eq:def-uncertainty-quantification}
\end{empheq}
In the following, we will discuss the aspects and components of Eq.~\eqref{eq:def-uncertainty-quantification} one by one, namely the likelihood and the prior, as already introduced in Eq.~\eqref{eq:Bayes-theorem}. Henceforth, for the sake of notational simplicity and clarity, integrals are understood to be taken over the appropriate parameter or state space, with domains omitted unless stated otherwise.

\subsubsection{The likelihood}\label{sssec:likelihood}
The \textit{likelihood}, or the conditional probability in Eq.~\eqref{eq:marginalization2}, $p(y \mid \bm{x})$, represents, in the context of experimental mechanics, the \emph{measurement noise of the instrument} when $y(\bm{x})$ is obtained from a mechanical experiment such as a uniaxial tensile test. Suppose this noise $\eta$ is additive and Gaussian\footnote{Noise can also be non-additive and non-Gaussian, alas this model is a default choice and particularly convenient.}. Then, we can write
\begin{equation}
    y = f(\bm{x}) + \eta\;, \quad \eta \sim \mathcal{N}(0, \sigma^2)\;,
\end{equation}
where $\mathcal{N}(\mu,\sigma^2)$ denotes a normal distribution with mean $\mu=0$ and variance $\sigma^2$. This implies
\begin{equation}
    y \mid \bm{x} \sim \mathcal{N}(f(\bm{x}) , \sigma^2)\;.
\end{equation}
In other words, the likelihood quantifies how closely the measured data cluster around the prediction of the forward model $f(\bm{x})$.
If $y(\bm{x})$ instead comes from solving a computational model, we distinguish two cases: (i) \textit{deterministic} models, where the same input always yields the same output --- here, the likelihood is effectively a Dirac delta centered at $f(\bm{x})$; and (ii) \textit{stochastic} models, where randomness is inherent to the simulation --- then $p(y\mid\bm{x})$ again becomes a proper probability distribution, just like in the experimental setting.

In the first case, when the computational model is deterministic, the output $y$ is uniquely defined by the input parameters $\bm{x}$. Indeed, most models in computational mechanics fall into this category: even if the inputs contain random parameters, once a specific realization of $\bm{x}$ is fixed, the resulting output $y$ is fully determined. 
In probabilistic terms, this means the likelihood $p(y \mid \bm{x})$ simplifies and assigns all its probability to the single value predicted by the model. That is, the likelihood is zero everywhere except at $y=y(\bm{x})$, where it takes its entire probability mass. Formally,
\begin{align}
p(y \mid \bm{x}) =
\begin{cases}
1, & \text{if } y = y(\bm{x})\;, \\[4pt]
0, & \text{otherwise}\;.
\end{cases}
\end{align}
%
This simplification allows us to carry out the integration over $y$ in Eq.~\eqref{eq:def-uncertainty-quantification} analytically. The UQ equations for deterministic computer simulations then reduce to
\begin{align}
    \mathbb{E}[y] & = \int  y(\bm{x})p(\bm{x}) \:\mathrm{d}\bm{x}\;, \\
    \mathrm{var}[y] & =  \int  \big(y(\bm{x}) - \mathbb{E}[y]\big)^2 p(\bm{x}) \: \mathrm{d}\bm{x}\;, \label{eq:def-uncertainty-quantification-deterministic}  
\end{align}
which describe how uncertainty in the input parameters $\bm{x}$ propagates through a deterministic model to produce uncertainty in the output $y$.

\bigbreak
\textit{Remark.} The previous assertion is a simplification. More precisely, for a deterministic model, the likelihood takes the form of a Dirac delta function, $p(y \mid \bm{x}) = \delta\big( y - y(\bm{x})\big)$. Carrying this delta function through the integrals is slightly more involved, but ultimately leads to the same results derived above. For simplicity, we therefore keep the more intuitive rationale. However, this simplification yields incorrect results when the mapping $\bm{x}\mapsto y(\bm{x})$ is deterministic but not unique --- for example, when a single input $\bm{x}$ can lead to more than one possible outcome. This situation can arise, for instance, in polymer materials or biological tissues exhibiting hysteresis or path-dependent behavior under cyclic uniaxial tensile loading, where the same stress level may correspond to different strain states during loading and unloading. In such cases, the likelihood must be expressed more carefully~\cite[Sec.~4.5]{ChapterWollner}.

\bigbreak
The task of UQ for a quantity of interest $y$ given uncertain input parameters $\bm{x}$ typically involves evaluating the integral in Eq.~\eqref{eq:def-uncertainty-quantification}, or its simplified form for deterministic models in Eq.~\eqref{eq:def-uncertainty-quantification-deterministic}. Hence, Eq.~\eqref{eq:def-uncertainty-quantification} can be considered as a general formulation of UQ, which is in most practical cases solved numerically. For the purposes of this introduction, we will not go further into stochastic models. Instead, we continue the exposition with the simplified UQ equation for deterministic models, see Eq.~\eqref{eq:def-uncertainty-quantification-deterministic}.

\subsubsection{The prior}\label{sssec:prior}
The \textit{prior} $p(\bm{x})$, the second term in Eq.~\eqref{eq:def-uncertainty-quantification-deterministic}, represents the probability distribution of the uncertain input parameters $\bm{x}$; see Eqs.~\eqref{eq:def-uncertainty-quantification} and \eqref{eq:def-uncertainty-quantification-deterministic}. This distribution models variability in the input parameters; in Bayesian inference, it encodes knowledge about $\bm{x}$ before observing new data. The specification of appropriate priors requires careful consideration of available information. In mechanics, priors are commonly constructed according to three complementary principles: data-driven, structured, and principle-based approaches.
\bigbreak
\textit{Data-driven priors.}
A natural way to construct a prior is from experimental data, or, in the case of biomechanical modeling, from animal or clinical data. For instance, material parameters may be inferred from mechanical tests that provide a stress–strain relationship, population statistics, or imaging data. 
Therefore, the prior in the UQ formulation (Eq.~\eqref{eq:def-uncertainty-quantification-deterministic}) can itself be the posterior of a previous inference task. Formally, one may write $p(\bm{x}) \equiv p(\bm{x}\mid \mathcal{D}_{\rm exp})$, read as the probability for $\bm{x}$ given the experimental data $\mathcal{D}_{\rm exp}$, where $\mathcal{D}_{\rm exp} = (\bm{\xi}^{(n)}, y_{\rm exp}^{(n)})_{n=1}^N$ denotes experimental input--output pairs. In data-driven settings, defining the parameter distribution thus amounts to solving an inverse problem, as discussed in the following Section~\ref{ssec:inverseI}; see also Fig.~\ref{fig-workflow}.

\bigbreak
\textit{Generative models as data-driven priors.}
A particularly recent and emerging perspective interprets such data-driven priors through so-called generative models, often based on NNs. These typically lack a closed-form expression for $p(\vv x)$, but instead offer the ability to \emph{generate} realizations of $\vv x$, i.e., draw samples from $p(\vv x)$. As we learned~\cite{ChapterWollner}, Bayesian inference does not necessarily require a closed-form expression, as long as samples can be drawn from the distribution. 
Examples include generative adversarial networks to generate material microstructures ~\cite{fokina2020microstructure, hsu2021microstructure, lambard2023generation}, variational auto-encoders to generate domain shapes, e.g., blood vessels~\cite{feldman2025recursive}, and statistical shape models~\cite{liang2017machine,ChapterHuberts}, which can likewise be interpreted as generative models sampling from a \lq distribution' of shapes.
In regards to UQ, all these neural network (NN) architectures are specific, heuristic models that allow one to generate samples; other architectures, such as diffusion models~\cite{lyu2024microstructure, Colmenarez2025a, Vahidullah2026a} or normalizing flows~\cite{yang2019pointflow}, may equally be used for generating microstructures or shapes, and these architectural specifics matter for UQ only insofar as they affect approximation quality. Inference based on neural generative models can be expensive, and the physical consistency of samples, though implicit in the data, is not easily guaranteed.

\bigbreak
\textit{Structured priors: Random fields.}
In many mechanical systems, uncertainty is distributed spatially rather than localized in a few parameters. In such cases, the prior defines a field of random variables over the spatial domain, referred to as a \emph{random field}. Each point in the domain is associated with its own random variable, leading to spatially heterogeneous yet statistically correlated material properties. Typical examples include spatially varying mechanical properties of biological tissue, for instance due to localized pathological alterations~\cite{ranftl2022stochastic}, as well as random fiber orientations in fibrous tissues.
Random fields are infinite-dimensional and therefore require finite-dimensional approximations in practice, for instance via truncated Karhunen--Lo\`eve expansions. The assumed covariance structure and correlation length directly influence the predicted variability. We return to random fields later in this review article (Section~\ref{sec:random-fields}).

\bigbreak
\textit{Principle-based priors: Maximum entropy.}
When only partial information about a parameter is available --- such as certain moments (e.g., mean, variance), bounds, or physical constraints --- the maximum entropy principle presents a systematic and compelling framework for constructing a prior. According to this principle, one selects the distribution that maximizes entropy subject to the known constraints; this yields the least biased representation consistent with the available information.
For real-valued variables defined on the unbounded domain $\mathbb{R}$, if only normalization, mean, and variance are prescribed, the entropy-maximizing distribution is Gaussian. 
If the variable is instead defined on a bounded and periodic domain, the corresponding maximum-entropy distribution changes\footnote{As a practical example from biomechanics, consider the directional distribution of collagen-fiber angles $\varphi \in [0,2\pi)$ in arterial tissue~\cite{Holzapfel2015}. Since $\varphi$ is a circular random variable, its first circular trigonometric moment --- given by $\mathbb{E}[\cos\varphi]$ and $\mathbb{E}[\sin\varphi]$ --- provides the appropriate constraint. Under these conditions, the entropy-maximizing distribution is the von Mises distribution, $p(\varphi \mid a,b) \propto \exp\!\big(a \cos(\varphi-b)\big),$ which may be interpreted as the circular analogue of the Gaussian distribution~\cite{jammalamadaka2001topics}.}.
For further details on the maximum entropy principle, see~\cite{jaynes1957information}.


\subsubsection{Bayesian inference}\label{sssec:inference}
Once the prior $p(\bm{x})$ has been specified and a computational model $y(\bm{x})$ is available, we can attempt to quantify the resulting uncertainty in the model output by solving Eq.~\eqref{eq:def-uncertainty-quantification-deterministic}. To make this concept more intuitive, we illustrate it using a simple mechanics example from linear elasticity theory, in which the model complexity is increased step by step.

\bigbreak
\textit{Example: Linear elasticity}.
We consider a linear elasticity problem with three progressively increasing levels of complexity. First, let the forward model that describes the experimental stress--strain ($y$--$\varepsilon$) relationship be $y = E \varepsilon$, and define the input as $\bm{x} \coloneqq E$, where $E$ is the Young's modulus. The whole setup is conditioned on a prescribed strain $\varepsilon$. We assume a deterministic model, corresponding to a delta-type likelihood $p(y \mid \varepsilon, E) = \delta(y - E\varepsilon)$. Based on data reported in the literature, we prescribe a Gaussian prior for the Young’s modulus, $E \sim \mathcal{N}(\mu_E, \sigma^2_E)$ with density $p(E) = (2\pi\sigma_E^2)^{-1/2}\exp\big(-(E-\mu_E)^2/(2\sigma_E^2)\big)$, where $\mu_E$ denotes the prior mean and $\sigma_E^2$ the prior variance, both treated as
prescribed hyperparameters rather than inferred. Under these assumptions, the uncertainty in the stress as a function of strain follows for $\varepsilon \neq 0$ by marginalization,
$p(y \mid \varepsilon) = \int \delta(y - E\varepsilon)p(E)\:\mathrm{d}E$, which yields $y\sim \mathcal{N}(\mu_E \varepsilon, \sigma_E^2 \varepsilon^2)$, so that $\var[y \mid \varepsilon] =
\sigma_E^2 \varepsilon^2$ grows quadratically with the applied strain.

In general, however, analytical solutions of the UQ equations are rarely available. It is important to note that the relative ease of computation in the present example arises from the linear dependence on the uncertain parameter $E$, rather than from the linearity with respect to the strain $\varepsilon$; nonlinear constitutive relationships would, in this context, lead to a comparable level of inference complexity. To illustrate this, in a second step, we consider a Gaussian likelihood with prescribed noise level $\sigma_y$, $y\mid\varepsilon,E \sim \mathcal{N}(E\varepsilon,\sigma^2_y)$, and assign a log-uniform prior to the Young’s modulus, $p(E) \propto 1/E$ on the truncated support $E\in [E_{\rm min, max}]$ with $E_{\min} > 0$.\footnote{A uniform prior on a quantity is \emph{inconsistent} with the information that this quantity is strictly positive~\cite{Jeffreys1939-JEFTOP-5}. Instead, one assigns a uniform prior to the logarithm of the quantity, which directly leads to the log-uniform prior used here.}. In this case, no closed-form expression for the prior predictive $p(y \mid \varepsilon)$ exists. Its first two moments remain analytically accessible, $\mathbb{E}[y \mid \varepsilon] = \varepsilon (E_{\max} - E_{\min}) / \ln(E_{\max}/E_{\min})$ and $\var[y \mid \varepsilon] = \sigma_y^2 + \varepsilon^2 \var[E]$, whereas the
density itself requires numerical integration techniques, such as Monte Carlo methods~\cite{Wollner2025}.

Finally, in a third step, we consider the same model as in the second case, but change the quantity of interest from the stress at a single strain to the maximum absolute stress attained over a prescribed strain history, $y_{\rm max} = \max_{\varepsilon\in[\varepsilon_{\rm min},\varepsilon_{\rm max}]} |E\varepsilon|$.
Although the constitutive model remains linear elastic, numerical methods are now typically required even to compute $\mathbb{E}[y_{\rm max}]$ and $\var[y_{\rm max}]$. The computational difficulty is further amplified when dealing with large-strain settings or nonlinear material behavior, such as viscoelasticity or material anisotropy.

\bigbreak
In addition to the above three levels, in computational mechanics, we encounter yet another, fourth practical challenge: even with advanced numerical integration methods, computing the distribution $p(y)$ or its proxies typically requires a very large number of simulations, i.e., evaluations of $y(\vv x)$, often on the order of $10^5$ to $10^8$ samples. For complex mechanical models, this is computationally infeasible. As an illustration, consider a simple finite element simulation that takes only one minute to run. Performing $10^6$ such simulations would require roughly two years of continuous computation. Although vanilla Monte Carlo sampling~\cite{Wollner2025} is trivially parallelizable, the implied computational resources are still enormous.

A widely used strategy to address this fourth issue is the implementation of \emph{surrogate models}. In this approach, a relatively small number of simulations is first performed, and the resulting data are used to \emph{learn} a fast approximation of the model output, with predictions then available in milliseconds. In other words, we fit a parameterized function, called a surrogate, to the simulation data and use it as a substitute for the original, computationally expensive model. This allows for rapid evaluation of $y(\bm{x})$ millions of times, and enables numerical integration of Eq.~\eqref{eq:def-uncertainty-quantification-deterministic}.
We discuss surrogate construction and training in Section~\ref{ssec:inverseII}, and the zoo of available surrogate models, that often share mathematical foundations with machine learning methods, in Section~\ref{sec:types-of-surrogates}.

\begin{figure}[t!]
    \centering
    \includegraphics[scale=1.0]{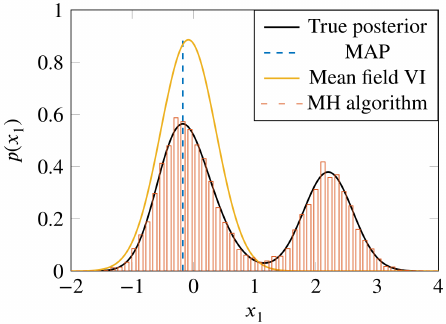}
    \caption{
    Comparison of standard methods for computing the posterior: (i) synthetic bimodal, asymmetric true posterior of a parameter $x_1$ (black); (ii) maximum a posteriori (MAP) estimate (blue), which provides only a point estimate without uncertainty quantification; (iii) mean-field Gaussian approximation via variational inference (VI) (yellow), which captures local structure around the dominant mode but fails to represent multimodality and asymmetry and underestimates uncertainty; and (iv) reference solution (asymptotically exact) obtained with the Metropolis–Hastings (MH) algorithm (red).
    }
    \label{fig-posterior}
\end{figure}

Several methods exist for computing the posterior, and each presents distinct trade-offs between computational cost and approximation accuracy, in specific contexts
. While point estimates --- such as maximum likelihood, or maximum a posteriori (MAP) and posterior mean from Bayesian inference that incorporate prior knowledge, but reduce the posterior to a single value --- provide single representative values, they fail to characterize the full parameter uncertainty inherent in mechanical models. 
To obtain a complete characterization, it is often necessary to sample from the full posterior using, for instance, Monte Carlo~\cite{ChapterWollner} or Markov chain Monte Carlo (MCMC) methods. Although MCMC methods are computationally expensive, they are asymptotically exact (under some conditions, as usual), and considered the most common approach. Finally, nested sampling is an evidence-focused distinct method that enables model comparison in high dimensions (cf. Section~\ref{ssec:model-selection}). Importantly, these methods can yield substantially different results, particularly for complex posteriors that are multi-modal, skewed, or heavy-tailed, as illustrated in Fig.~\ref{fig-posterior} for a representative synthetic bimodal posterior. Modelers should therefore be aware of these method-specific limitations; see related literature~\cite{Kroese2011a,ChapterPensalfini}.

\subsection{Inverse problems I: Model calibration and parameter estimation}\label{ssec:inverseI}
The core objective of model calibration and parameter estimation can be stated as:

\begin{quotation}
\textit{Determination of model parameters and uncertainty quantification thereof given experimental or simulation data.}
\end{quotation}

\subsubsection{General concept}
In mechanics, inverse problems typically arise in the following three situations: (i) experimental measurements are available, but the underlying constitutive or material parameters remain unknown; (ii) computational simulation data are available, and one aims to learn a surrogate model; or (iii) a combination of both (i) and (ii). In this section, we will mainly discuss case (i) and defer cases (ii) and (iii) to Section~\ref{ssec:inverseII}.

Let $\bm{x} \in \mathcal{X} \subseteq \mathbb{R}^{D_{\bm{x}}}$ denote the vector of constitutive parameters that characterize the material behavior, where $D_{\bm{x}}$ represents the number of constitutive parameters in the model. The deterministic forward model $f(\bm{x} ; \bm{\xi})$ describes the system's response for a given set of constitutive parameters $\bm{x}$ and experimental control parameters $\bm{\xi} \in {\Xi} \subseteq \mathbb{R}^{D_{\bm{\xi}}}$, as defined in Fig.~\ref{fig-workflow}. These control parameters include applied loads or displacements and $D_{\bm{\xi}}$ is the total number of control parameters. The model output is an observable quantity $y_{\rm exp} = f(\bm{x} ; \bm{\xi})$, such as a strain or stress component at a given location. This forward model typically represents the underlying physical laws of the system, for example through the numerical solution of partial differential equations (PDEs).
In parallel, experimental measurements provide corresponding observations of the same physical quantities under specified control conditions. The measurement can be modeled, e.g., by 
\begin{align} \label{eq:noise-model-example}
    y_{\rm exp} = f(\bm{x} ; \bm{\xi}) +\eta\;,
\end{align}
with additive noise $\eta$. 
The collected data are represented as
\begin{equation}
    \mathcal{D}_{\rm exp} = \big(\bm{\xi}^{(n)}, y_{\rm exp}^{(n)}\big)_{n=1}^N\;,
\end{equation}
where each pair $(\bm{\xi}^{(n)}, y_{\rm exp}^{(n)})$ corresponds to the $n$-th observation point in one experiment performed under the control parameters $\bm{\xi}^{(n)}$, and $N$ denotes the total number of observations. The goal is then to determine the parameters $\vv x$ given the data $\mathcal{D}_{\rm exp}$ and the physical model $f(\bm{x} ; \bm{\xi})$. 

Unfortunately, inverse problems are often ill-posed, meaning that multiple parameter sets may explain the data equally well, which is often exacerbated by the presence of noise.
To address these challenges, we can reformulate the inverse problem in a Bayesian inference framework and incorporate both measurement noise and prior knowledge about the parameters in a consistent manner. Instead of seeking a single \emph{optimal} estimate of the unknown parameters, we quantify our uncertainty about them using probability distributions. Consider, for example, the case of above assumptions with additive Gaussian noise modeled as $\eta \sim \mathcal{N}(0,\sigma^2)$ with measurements that are independently and identically distributed (i.i.d.). Then, the joint likelihood is
\begin{equation}
\begin{split}
    p(\mathcal{D}_{\rm exp} & \mid  \bm{x}, \sigma) \\
    & = \prod_{n=1}^N \frac{1}{\sigma\sqrt{2\pi}}
    \exp\left(-\frac{1}{2\sigma^2} \big(y_{\rm exp}^{(n)} - f(\bm{x}; \bm{\xi}^{(n)})\big)^2\right) \\
    &\propto \sigma^{-N} \exp\left(-\frac{1}{2\sigma^2}
    \sum_{n=1}^N\big(y_{\rm exp}^{(n)} - f(\bm{x}; \bm{\xi}^{(n)})\big)^2\right)\;.
\end{split}
\end{equation}
where $\sigma$ is the standard deviation and $\sigma^2$ is the variance of the measurement noise. Then, according to Bayes' theorem, with a prior $p(\bm{x}, \sigma)$, the posterior is
\begin{equation} \label{eq:inverser-problem-posterior}
p(\bm{x}, \sigma \mid \mathcal{D}_{\rm exp}) \propto p(\mathcal{D}_{\rm exp} \mid \bm{x}, \sigma) p(\bm{x}, \sigma)\;,
\end{equation}
which expresses our updated knowledge about the unknown parameters after observing the data. This distribution characterizes not only the expected parameter values but also their associated uncertainty or variance~\cite{ChapterWollner}. Note that methods for computing the posterior were outlined in Section~\ref{sssec:inference}.

Importantly, the posterior obtained from the inverse problem (cf. Eq.~\eqref{eq:inverser-problem-posterior}) often serves as the input distribution, i.e., the prior, for the subsequent uncertainty propagation analysis, as illustrated in Fig.~\ref{fig-workflow}. In other words, once the parameters have been inferred probabilistically, their posterior defines the space of plausible parameter values that should be propagated through the model to assess the resulting uncertainty in model predictions.

In summary, inverse problems play several multi-faceted roles in mechanics, with three prominent applications:
(i) calibration or, synonymously, \emph{parameter estimation}, such as fitting constitutive models to experimental data for predictive simulations; 
(ii) input characterization, i.e., determining distributions of input parameters $\bm{x}$ as a prerequisite for forward uncertainty propagation; and 
(iii) surrogate model calibration --- an additional inverse problem often arises, where the goal is to calibrate the hyperparameter of a surrogate model that replaces a more expensive computational model to quantify the uncertainty of the surrogate itself, as shown in Fig.~\ref{fig-workflow}.
Note that case (iii) is a variant of case (i) with different context. 
We first discuss case (iii) in the following Section~\ref{ssec:inverseII}, before presenting an illustrative mechanical example for cases (i) and (ii) in Section~\ref{sec:example1}.

For further context, the literature provides several (bio)mechanical examples of model calibration and parameter estimation using Bayesian inference.
In biomechanics, these studies encompass UQ in blood rheology models~\cite{Ranftl2022}, parameter estimation in soft biological tissues using both standard constitutive models~\cite{Madireddy2016a,Teferra2019a,Sundes2022a} and unsupervised discovery of constitutive models~\cite{Joshi2022a,Krijnen2025a}, as well as efficient inference of Windkessel parameter posteriors in three-dimensional aortic hemodynamics simulations through zero-dimensional surrogates~\cite{Richter2025a}, and amortized posterior estimation of boundary conditions in patient-specific cardiovascular models using conditional flow matching~\cite{Choi2026a}. Another example of amortized variational inference for real-time inference with potential applications in medical diagnosis or health monitoring of engineered systems is presented in~\cite{Karumuri2024a}, which learns a NN map from data to posterior, avoiding re-solving from scratch for each new dataset.

For general mechanical problems, recent work has addressed UQ of material parameters for different experimental setups under reparametrization~\cite{Wollner2025}, for constitutive parameters of fiber-reinforced polymer composites~\cite{Thomas2022a}, constitutive models including damage~\cite{Chakraborty2021a}, phase-field fracture models~\cite{Khodadadian2020a}, and coupled mechanics models based on interface deformations where only shape changes are observable~\cite{Willmann2022a}.
Furthermore, Bayesian inference has been employed to infer hyperparameters in physics-informed NNs~\cite{Yang2021a}, with applications to COVID-19 outbreak dynamics~\cite{Linka2022a} and constitutive artificial NNs for soft biological tissues~\cite{Linka2025a}. Here, the NNs with hyperparameters calibrated on experimental data can be seen as the forward model directly, in contrast to surrogate modeling where simplified models approximate expensive high-fidelity simulations; a subtle but important distinction.

\subsubsection{Data fusion and data assimilation}\label{ssec:fusion_assimilation}
In the context of model calibration and parameter estimation, we introduce the terms \textit{data fusion} and \textit{data assimilation}, noting that alternative definitions exist in the literature. We adopt these terms because they provide a useful distinction between inference problems involving a \emph{single} forward model (data fusion) and those that rely on \emph{multiple} forward models (data assimilation), which frequently arise in mechanics, multi-physics modeling, and multi-modal experimental setups.

\begin{figure*}[t!]
    \centering
    \includegraphics[scale=1.0]{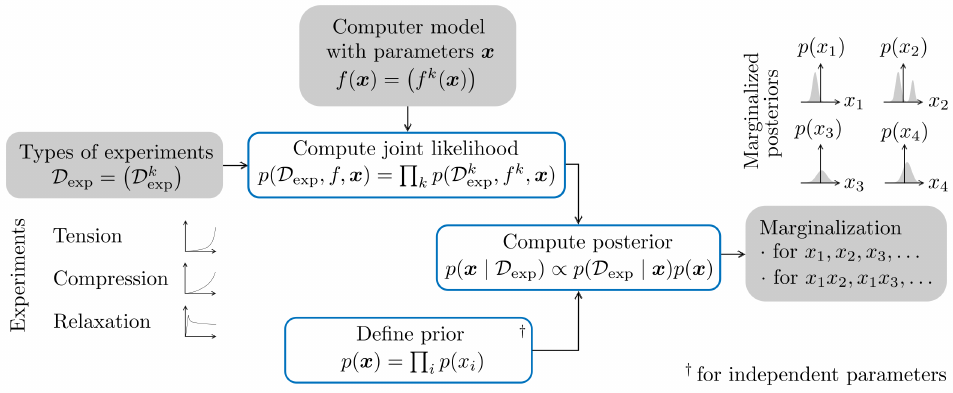}
    \caption{General workflow for model calibration and parameter estimation based on multiple experimental types under the simplifying assumption of independent parameters.
    The posterior is computed from the chosen prior and the likelihood, with the latter defined by the experimental data pairs $\mathcal{D}_{\rm exp}$ and the computational models associated with the respective experimental types $k$ and parameterized by $\bm{x}$. From the resulting posterior, marginalized posteriors for individual parameters or combined parameter combinations can be computed.}
    \label{fig-calibration}
\end{figure*}

\textit{Data fusion} refers to the consistent combination of information from multiple experiments, measurement modalities, or simulations within a single likelihood model, i.e., a single probabilistic model. This may include merging mechanical test data from uniaxial tensile, compression, or relaxation tests with medical imaging measurements, such as ultrasound, computed tomography, or magnetic resonance imaging. More generally, data fusion exploits complementary information and weights inconsistent evidence according to the uncertainty associated with each data source. Under the standard assumption of conditional independence, the joint likelihood factorizes and can be written as
\begin{align}
    p(y_{\rm exp}^{(1)}, y_{\rm exp}^{(2)}, \ldots \mid \bm{x}) = \prod_k p(y_{\rm exp}^{(k)} \mid \bm{x})\;,
\end{align}
where each $y_{\rm exp}^{(k)}$ denotes a distinct dataset from the $k$-th experiment. Factorization greatly simplifies inference, though it requires that measurement noise and experimental errors be independent across modalities or setups. When this assumption is reasonable, each dataset contributes multiplicatively to the posterior, and usually improves parameter identifiability, as exemplarily shown for parameter identification in constitutive modeling based on multiple experimental modes~\cite{Wollner2025}. 
We may generalize this notion also to multiple physical models, see also Fig.~\ref{fig-calibration}.

Building on this concept, we understand \textit{data assimilation} as the modeling of the likelihood for sequential state estimation in dynamical models. The system evolves according to $\vv x_t = F_t(\vv x_{t-1}) +\eta_t$, and observations are collected sequentially as $y_t = G_t(x_t) + \eta_t$, where $t$ denotes an ordered (typically time) index, and $F_t$ and $G_t$ describe the state transition and observation models.
In contrast to data fusion --- where all data are processed jointly without a particular order --- data assimilation explicitly accounts for sequential structure, as in time-series problems. Both the likelihood and the prior follow this ordering.
Given observations up to time $T$, the posterior over the full state trajectory $(\vv x_{0:T}, y_{0:T})$ typically factorizes as
\begin{align}
    p(\vv x_{0:T} \mid y_{1:T}) & \propto p(\vv x_0) \prod_{t=1}^T
    p(\vv x_t \mid \vv x_{t-1}) \nonumber \\
    & \hphantom{=}\,\, \times p(y_t \mid \vv x_t) \;,
    \label{eq:da_joint_posterior}
\end{align}
meaning that the likelihood at each new state depends only on the previous state (Markov property), and each observation depends only on the current state.

Examples from the literature include the dynamic estimation of hyperelastic material parameters across various mechanical tests using an extended Kalman filter~\cite{Xie2021a,SONG2023105553}, as well as dynamic soft tissue identification achieved by coupling a contact model with an iterated Kalman filter~\cite{Zhu2023a}.

\subsubsection{Accounting for model discrepancy}\label{sssec:discrepancy}
Even a carefully constructed computational model may not perfectly reproduce the true physical system. Simplifying assumptions, incomplete physics, and numerical approximations introduce systematic discrepancies between model predictions and experimental observations. These deviations are collectively referred to as \emph{model discrepancy}, or \emph{model inadequacy}; see, e.g.,~\cite{kennedy2001bayesian,Morrison2018} for Bayesian discussions.

To account for such effects in a most simple manner, the standard observation model is extended as
\begin{equation} \label{eq:model-discrepancy}
    y_{\rm exp} = f(\bm{x}, \bm{\xi}) + \delta(\bm{\xi}) + \eta\;,
\end{equation}
where $f(\bm{x}, \bm{\xi})$ is the forward model prediction under the control parameters $\bm{\xi}$, $\eta$ represents (Gaussian) measurement noise, and 
$\delta(\bm{\xi})$ captures the systematic model discrepancy between simulations and observations.
Within the Bayesian framework, both the model parameters $\bm{x}$ and the discrepancy function $\delta(\bm{\xi})$ are treated as uncertain quantities. A flexible and widely used approach is to model the discrepancy as a Gaussian process (GP). Alternatively, it could be represented using a low-dimensional basis expansion, which is formally introduced in the following Section~\ref{ssec:PCE}.

Including a model discrepancy term helps prevent biased parameter estimates, as it avoids attributing structural model errors to the model parameters $\bm{x}$. However, it also introduces additional uncertainty and potential identifiability issues, since both $f(\bm{x}, \bm{\xi})$ and $\delta(\bm{\xi})$ may explain similar residual patterns. Careful prior specification and model validation are therefore essential to ensure meaningful inference.

Introducing a model discrepancy is recommended when the residuals --- defined as the differences between observed data and model predictions --- after calibration exhibit a smooth or correlated pattern with respect to the control parameters $\bm{\xi}$. This approach is also advisable if the forward model $f(\bm{x}, \bm{\xi})$ involves simplifications or neglects important physical effects, such as anisotropy, nonlinear behavior, or three-dimensionality. Additionally, model discrepancy should be considered when predictive performance deteriorates under unseen control conditions or when the residuals reveal input-dependent bias, indicating a systematic error.
Residual diagnostics, posterior predictive checks, and the model evidence or its proxies (e.g., Akaike information criterion or leave-one-out cross-validation) can be used to evaluate whether the inclusion of $\delta(\bm{\xi})$ improves model adequacy.  

The need to address such discrepancies is a recurring theme in recent literature. For instance, Paun et al.~\cite{Paun2020a} demonstrated this type of model error in one-dimensional hemodynamic simulations of the pulmonary artery, where linear and nonlinear material models yielded notably different predictions. Similarly, Römer et al.~\cite{Romer2021a} used a discrepancy term to account for errors in a surrogate model, while Calvetti et al.~\cite{Calvetti2018a} proposed an iterative loop to simultaneously update both the unknown parameters and the model error, thereby refining the posterior distribution.

\subsubsection{Hierarchical modeling}\label{sssec:hierarchical}
Hierarchical Bayesian modeling provides a systematic way to represent complex uncertainty structures by introducing latent variables and hyperparameters that govern lower-level model components. In contrast to a single-level Bayesian model, hierarchical formulations explicitly distinguish between different sources of uncertainty, such as parameter variability, noise levels, and higher-level structural assumptions.

To illustrate the idea, consider a forward model
\begin{equation}
    y^{(n)}=f(\boldsymbol{x};\boldsymbol{\xi}^{(n)})+\eta^{(n)},\, \quad  \eta^{(n)}\sim\mathcal{N}(0,\sigma^2)\;.
\end{equation}
In a standard formulation, priors would be assigned directly to $\bm{x}$ and $\sigma$. In a hierarchical setting, however, these priors themselves depend on additional hyperparameters $\Psi$ and $\Sigma$, for example $p(\boldsymbol{x}\mid {\Psi})$ and $p(\sigma\mid \Sigma)$, together with a hyperprior $p(\Psi,\Sigma)= p(\Psi)p(\Sigma)$. This additional layer allows the model to learn not only the parameters $\bm{x}$ and $\sigma$, but also the statistical structure governing them.

The resulting joint posterior factorizes into a hierarchy
\begin{align}
    p(\boldsymbol{x},\sigma, {\Psi}, \Sigma\mid \mathcal{D}) & \propto
    p(\mathcal{D}\mid \boldsymbol{x},\sigma)
    p(\boldsymbol{x}\mid  {\Psi}) \nonumber \\
    & \hphantom{=}\,\, \times p(\sigma\mid \Sigma)
    p( {\Psi})p(\Sigma)\;.
\end{align}
To predict the outcome for a new control parameter $\boldsymbol{\xi}_\ast$, we marginalize the model likelihood over the full parameter space. By using the joint posterior as a weighting function, the predictive distribution accounts for uncertainty at every level of the hierarchy
\begin{align}
    p(y_\ast\mid \boldsymbol{\xi}_\ast,\mathcal{D}) &= 
    \iiiint 
    p(y_\ast\mid \boldsymbol{\xi}_\ast,\boldsymbol{x},\sigma, \cancel{\Psi, \Sigma}) \nonumber\\
    &\hphantom{=}\,\, \times p(\boldsymbol{x},\sigma, {\Psi}, \Sigma\mid \mathcal{D})\:
    \mathrm{d}\boldsymbol{x}\:\mathrm{d}\sigma\:\mathrm{d}{\Psi}\:\mathrm{d}\Sigma\;,
\end{align}
where we use the fact that the conditional of the first term contains redundant information, such that $p(y_\ast\mid \boldsymbol{\xi}_\ast,\boldsymbol{x},\sigma, {\Psi, \Sigma}) \equiv p(y_\ast\mid \boldsymbol{\xi}_\ast,\boldsymbol{x},\sigma)$. This simplification reflects that once the parameters $\bm{x}$ and $\sigma$ are known, the additional hyperparameters $\Psi$ and $\Sigma$ provide no additional information for the prediction $y_\ast$.
This nested marginalization highlights a key principle: uncertainty is propagated across all layers of the model, not only at the parameter level.

A practical advantage of this layered structure is flexibility. Different approximation strategies can be applied selectively at different levels. For instance, one may retain full inference for $\bm{x}$ and $\sigma$ while approximating the hyperparameter posterior via a Laplace or Dirac approximation. This reduces computational cost without collapsing the hierarchical structure.

Hierarchical formulations are particularly effective for modeling structured or heterogeneous systems. For instance, inter-patient variability can be represented by assuming that individual parameters follow a common distribution $p(\boldsymbol{x}_i\mid{\Psi})$, where $i$ here denotes the patient index and $\Psi$ represents the shared population-level hyperparameter. This framework allows the model to \textit{pool} information from all individuals while still respecting their unique differences. By sharing data across the entire group, the global population trends help guide the estimation for each specific case.\footnote{In practice, when characterizing biological tissue samples from multiple patient donors, experimental limitations often result in some specimens providing incomplete or noisy datasets. Instead of analyzing each specimen in isolation, \emph{pooling} allows the model to learn the characteristic mechanical response from the entire population. If an individual specimen's data is sparse, the model uses the group's collective behavior as an informative starting point (prior) to \emph{fill in the gaps}, ensuring that the identified parameters remain physically plausible and consistent with the broader patient group.}
This collective approach leads to much more reliable results than fitting each specimen separately, which is especially valuable when the data for a single individual are too sparse to stand on their own~\cite{Pensalfini2023a,ChapterPensalfini}.
Similarly, intra-patient variability or spatial heterogeneity can be addressed through random field priors~\cite{ranftl2022stochastic}, which will be introduced in Section~\ref{sec:random-fields}, where the statistical properties of a field are specified rather than prescribing a deterministic function.

Many other modeling strategies can be interpreted through this hierarchical lens. For instance, model discrepancy formulations (Section~\ref{sssec:discrepancy}) and multi-fidelity strategies (Section~\ref{sssec:MF}) introduce additional latent layers to account for structural model errors or varying levels of approximation. Rather than modifying the likelihood \emph{ad hoc}, these approaches embed such assumptions directly into the probabilistic hierarchy by treating the model's inadequacy as a structured uncertainty.\footnote{Notably, Koutsourelakis~\cite{koutsourelakis2009accurate} demonstrated that this hierarchical treatment allows for accurate uncertainty estimates even when the underlying computational models are themselves inaccurate, provided the discrepancy is modeled and propagated consistently.}
Even standard regression models admit such an interpretation: GP regression, for example, can be viewed as a hierarchical model where latent function values are marginalized under a conjugate prior, while hyperparameters are estimated by a higher-level distribution. This perspective will be discussed in more detail in Section~\ref{ssec:GP}.

In addition, several robust or sparse modeling strategies arise naturally from hierarchical constructions. The Student-$t$ distribution, for instance, can be derived by marginalizing a Gaussian likelihood with respect to an inverse-gamma prior on its variance parameter. This construction, which can be viewed as a regularized version of a non-informative Jeffreys prior results in a heavy-tailed formulation that acts as a scale mixture of Gaussians, providing inherent robustness against outliers~\cite{Ranftl2022,ranftl2022stochastic}, as discussed in the following example (cf. Section~\ref{sec:example1}).
Similarly, the horseshoe prior employs a hierarchy of local and global scale parameters. This structure enforces strong shrinkage on negligible model parameters or surrogate hyperparameters while leaving significant physical effects virtually unaffected --- a property particularly suited for high-dimensional problems where the underlying system is expected to be sparse.

From a broader perspective, UQ itself may be viewed as a hierarchical construction. Model calibration (inverse problem I) and surrogate modeling (inverse problem II) then constitute different levels within this probabilistic hierarchy, to explicitly separate distinct sources of uncertainty rather than conflating them.

\subsection{Inverse problems II: Surrogate modeling}\label{ssec:inverseII}
In the Bayesian context, the core objective of surrogate modeling is:

\begin{quotation}
    \textit{To construct computationally efficient approximations to high-fidelity forward models in order to enable uncertainty propagation.}
\end{quotation}

\subsubsection{General concept}
To evaluate the integral in Eq.~\eqref{eq:def-uncertainty-quantification-deterministic}, or equivalently Eq.~\eqref{eq:def-uncertainty-quantification}, we first need to construct a surrogate for the computational model. The surrogate is typically represented as a parameterized function,
\begin{align} \label{eq:def-surrogate}
     y(\bm{x}) \approx \hat{f}(\bm{x} ; \bm{\theta})\;,
\end{align}
where $\bm{\theta} \in \Theta$ is a vector of parameters parameterizing the surrogate. The functional form of the surrogate can vary widely; the only requirements are that it can be evaluated efficiently (fast) and that its parameters can be identified from a simulation dataset of input–output pairs $\mathcal{D} = (\bm{x}^{(n)}, y^{(n)})_{n=1}^{N}$. In the simplest case, the surrogate may be a constant, $\hat{f}(\bm{x} ; \bm \theta) = a$, or a linear function $\hat{f}(\bm{x} ; \bm \theta) = b\bm{x} + c$, fitted directly to the data $\mathcal{D}$. More generally, we can express the surrogate as an expansion,
\begin{align} \label{eq:def-expansion}
    \hat{f} (\bm{x} ; \bm{\theta}) = \sum_{p=1}^P c_p \phi_p(\bm{x} ; w_p)\;,
\end{align}
where $\phi_p$ denotes a set of basis functions parametrized by $w_p$ and $c_p$ are linear coefficients. The index $p$ enumerates the basis functions in the surrogate model, with a total of $P$ terms in the expansion.

We can thus distinguish between linear surrogate parameters $c_p$ and nonlinear surrogate parameters $w_p$, such that $\bm{\theta} = (c_p, w_p)^{\rm T}$. This distinction is important, since linear parameters are often easier to determine than non-linear parameters.\footnote{For a practical example, if we consider a Gaussian likelihood and a Gaussian prior on both $c_p$ and $w_p$, then we can obtain analytic expressions for $\mathbb{E}[c_p]$, conditioned on some $w_p$, but not for $\mathbb{E}[w_p]$, i.e., the non-linear parameters $w_p$ typically require numerical estimation.}
The basis functions~$\phi_p$ can be just about anything. Popular examples include polynomial and trigonometric functions, which lead to polynomial expansions or Fourier expansions, respectively. A particularly notable case is the polynomial chaos expansion (PCE), where the basis polynomials $\phi_p$ are constructed to be orthogonal with respect to the input distribution $p(\bm{x})$.
Furthermore, the surrogate model $\hat{f}(\bm{x} ; \bm{\theta})$ can also be interpreted as a machine learning model. If the surrogate were implemented as a NN, the functions $\phi_p$ would correspond to the outputs of the final layer. We introduce PCEs in Section~\ref{ssec:PCE}, while the special case of machine learning–based surrogate models is discussed in more detail later in Section~\ref{ssec:PINN-and-NO}.

After selecting a suitable parameterization for the surrogate model $\hat{f}(\bm{x} ; \theta)$, the next step is to determine the parameters $\bm{\theta}$ from the available data $\mathcal{D}$. To do so, we must define a \emph{fitting criterion}; that is, an optimization objective defining which surrogate parameters $\bm{\theta}$ are considered optimal given the data $\mathcal{D}$. In non-probabilistic settings and particularly in machine learning, this criterion is often referred to as a \emph{loss function}. 

The simplest example is the $L_2$-loss, which leads to the familiar least-squares criterion: minimizing the sum of squared errors between predictions and observations. However, least squares is merely a special case of a more general probabilistic framework. In this setting, we seek to maximize the posterior for the surrogate parameters $\bm{\theta}$, given the surrogate model $\hat{f}$ and the data $\mathcal{D}$,
\begin{align} \label{eq:def-machine-learning-equation}
    \bm{\theta}^{\ast} = \arg \max_{\bm{\theta}}\, p(\bm{\theta}\mid \mathcal{D}, \hat{f})\;,
\end{align}
where $\bm{\theta}^{\ast}$ represents the \emph{optimal} surrogate parameters given the ansatz $\hat{f}$.
Under the assumption of a Gaussian likelihood with fixed variance and uniform prior\footnote{Specifically, by assuming a Gaussian likelihood $p(\mathcal{D} \mid \bm{\theta}, \hat{f}) = \mathcal{N}(\hat{f}(\bm{x} ; \bm{\theta}), \sigma^2)$ with known, constant variance $\sigma^2$, and a non-informative uniform prior over $p(\bm{\theta}) = \text{const.}$, Bayes' theorem implies that the posterior is proportional to the likelihood, $p(\bm{\theta} \mid \mathcal{D}, \hat{f}) \propto \mathcal{N}(\hat{f}(\bm{x} ; \bm{\theta}), \sigma^2)$.}, maximizing this posterior is mathematically equivalent to minimizing the $L_2$-loss, or the sum of squared residuals,
\begin{align}
    \bm{\theta}^{\ast} = \arg \min_{\bm{\theta}} 
    \sum_{n=1}^{N} \big(y^{(n)} - \hat{f}(\bm{x}^{(n)} ; \bm{\theta})\big)^2\;.
\end{align}
The terms \emph{training} or \emph{learning} typically refer to the --- often iterative --- process of solving Eq.~\eqref{eq:def-machine-learning-equation}. This optimization naturally extends to the Bayesian framework discussed in the following Section~\ref{sssec:Bayesian-surrogate-modeling}, where the prior $p(\vv \theta)$ effectively serves as a \emph{regularizer} for the objective function.

Once the optimal parameters $\bm{\theta}^{\ast}$ are identified, the resulting surrogate model, 
\begin{equation}
    \hat{f}(\bm{x} ; \bm{\theta}^\ast) \approx y(\bm{x})\;, 
\end{equation}
serves as a computationally efficient proxy for the original model. This substitution allows for the efficient approximation of the UQ solution in Eq.~\eqref{eq:def-uncertainty-quantification-deterministic} (or equivalently Eq.~\eqref{eq:marginalization2}).

\subsubsection{Bayesian surrogate modeling} \label{sssec:Bayesian-surrogate-modeling}
The probabilistic approach of surrogate \emph{learning} is more general than the use of simple loss functions. The approach discussed so far does not account for the uncertainty associated with the estimated optimal parameters $\bm{\theta}^{\ast}$. This simplification is valid as long as the data are sufficiently informative such that the posterior $p(\bm{\theta} \mid \mathcal{D}, \hat{f})$ is sharply peaked; we then have a high degree of belief about our choice of $\bm{\theta}^{\ast}$. This is typically the case when the following conditions are satisfied: (i) the dataset is large, of high quality, and informative with respect to the parameters; and (ii) the simulation and surrogate is not insensitive to certain parameters, the surrogate is not overparametrized, and no parameter-identifiability issues arise. 

For truly Bayesian inference, we would need to define a suitable prior and evaluate the full parameter posterior $p(\bm{\theta} \mid \mathcal{D}, \hat{f})$. To quantify the uncertainty in the parameters~$\bm{\theta}$, one would compute the first and second posterior moments as a proxy for the full distribution, analogous to Eq.~\eqref{eq:def-2ndmoment}, as
\begin{align} \label{eq:def-machine-learning-equation-Bayesian}
    \mathbb{E}[\vv \theta ] &= \int \bm{\theta} p(\bm{\theta}\mid \mathcal{D}, \hat{f}) \: \mathrm{d}\bm{\theta} \;, \\
    \var[\bm{\theta}] &= \int \big(\bm{\theta} - \mathbb{E} [\bm{\theta}] \big)^2 p(\bm{\theta}\mid \mathcal{D}, \hat{f}) \: \mathrm{d}\bm{\theta} \;.
\end{align}
In this light, the optimization following Eq.~\eqref{eq:def-machine-learning-equation} can be viewed as a point-estimate approximation of Eq.~\eqref{eq:def-machine-learning-equation-Bayesian}. This simplification effectively represents the posterior as a Dirac delta distribution centered at the mode, i.e.,~$ p(\bm{\theta}\mid \mathcal{D}, \hat{f}) \approx \delta(\vv \theta - \vv \theta^\ast)$. Such an approach is often preferred for its computational efficiency, particularly when epistemic uncertainty is expected to be negligible or is simply not of interest.

Estimating the uncertainty of the surrogate prediction $\hat{f}(\bm{x} ; \bm{\theta})$ itself is more involved: one must introduce the posterior $p(\bm{\theta} \mid \mathcal{D}, \hat{f})$ into the UQ equation (Eqs.~\eqref{eq:def-uncertainty-quantification-deterministic},~\eqref{eq:marginalization2}, or~\eqref{eq:def-uncertainty-quantification}) via the marginalization rule. This results in an additional integral of the form
\begin{align}
    p(\hat{y} \mid \vv x, \mathcal{D}, \hat{f})&= \int p(\hat{y}\mid \vv x, \vv \theta) \, p(\bm{\theta}\mid \mathcal{D}, \hat{f}) \:\mathrm{d}\vtheta\;, \\
    \mathbb{E}\big[\hat{f}(\bm{x})  \big] &= \int \hat{f}(\bm{x} ; \bm{\theta}) p(\bm{\theta}\mid \mathcal{D}, \hat{f}) \:\mathrm{d}\vtheta\;, \\
    \var\big[ \hat{f}(\bm{x})\big] & = \int \Big(\hat{f}(\bm{x} ; \bm{\theta}) - \mathbb{E}\big[\hat{f}(\bm{x}) \big]\Big)^2 \nonumber\\ 
    &\hphantom{=}\,\, \times p(\bm{\theta} \mid \mathcal{D}, \hat{f})\, \mathrm{d}\vtheta\;,
\end{align}
for the posterior uncertainty as a function of $\vv x$, or marginalized with respect to the inputs,
%
\begin{align} \label{eq:bayesian-surrogates-deterministic}
    p(\hat{y} \mid \mathcal{D}) & = \iint p(\hat{y}\mid \vv x, \vv \theta) \nonumber\\
    &\hphantom{=}\,\, \times p( \vv x, \bm{\theta} \mid \mathcal{D}, \hat{f}) \:  \mathrm{d}{\vv x}\:\mathrm{d}\vtheta\;,\\
    \mathbb{E}\big[y\big] \approx \mathbb{E}\big[\hat{f} \big] &= \iint \hat{f}(\bm{x} ; \bm{\theta}) \nonumber\\
    &\hphantom{=}\,\, \times p( \vv x, \bm{\theta}\mid \mathcal{D}, \hat{f}) \:\mathrm{d}{\vv x}\: \mathrm{d}\vtheta\;,\\
    \var\big[ y\big] \approx \var\big[  \hat{f} \big] & = \iint \big(\hat{f}(\bm{x} ; \bm{\theta}) - \mathbb{E}\big[\hat{f} \big]\big)^2 \nonumber \\ 
    &\hphantom{=}\,\, \times p( \vv x, \bm{\theta}\mid \mathcal{D}, \hat{f}) \:\mathrm{d}{\vv x}\, \mathrm{d}\vtheta\;, 
\end{align}
where, in surrogate modeling, Bayes’ theorem implies that the joint distribution of a new input $\bm{x}$ and the surrogate parameters $\bm{\theta}$, conditioned on the observed dataset $\mathcal{D}$ and the surrogate model $\hat{f}$, factorizes as $p(\vv x, \bm{\theta}\mid \mathcal{D}, \hat{f}) = p(\vv x) \, p(\bm{\theta}\mid \mathcal{D}, \cancel{\vv x}, \hat{f})$. This factorization shows that the posterior of the surrogate parameters $\boldsymbol{\theta}$ is independent of a new, unseen input $\vv x$. It depends solely on the calibration data in $\mathcal{D}$, i.e., the observed input--output data pairs from simulations $(\bm{x}^{(n)}, y^{(n)})$. A new input does not alter the uncertainty in $\boldsymbol{\theta}$; it only affects the model prediction at that input. The posterior changes only if additional measured data are added into $\mathcal{D}$.

Analogously, the posterior of the surrogate parameters can be approximated by a point mass at their \emph{optimal} value $\boldsymbol{\theta}^\ast$, i.e.,
\begin{equation}
  p(\boldsymbol{\theta} \mid \mathcal{D}, \hat{f})
  \approx \delta(\boldsymbol{\theta} - \boldsymbol{\theta}^\ast)\;.
\end{equation}
Inserting this into the marginalization over $\boldsymbol{\theta}$ and using the
sifting property of the Dirac delta, the predictive distribution collapses to a
single evaluation of the surrogate,
\begin{align}
  p(y \mid \bm{x}, \mathcal{D}, \hat{f})
  & = \int p(y \mid \bm{x}, \boldsymbol{\theta}, \hat{f})\,
         p(\boldsymbol{\theta} \mid \mathcal{D}, \hat{f})\:\mathrm{d}\boldsymbol{\theta} \nonumber\\
  & \approx p(y \mid \bm{x}, \boldsymbol{\theta}^\ast, \hat{f})\;.
\end{align}
This approximation neglects the uncertainty in the surrogate parameter estimates and treats them as known exactly. As a consequence, the Bayesian formulation reduces to the standard, non-Bayesian use of the surrogate model.

The Eq.~\eqref{eq:bayesian-surrogates-deterministic} strictly holds for deterministic models, where the model output is uniquely determined by the input, i.e., $p(y \mid f, \bm{x}) = \delta(y - f(\bm{x}))\;$. 
Non-deterministic, or statistical, models would demand a prior for and to marginalize over the distribution of model values $\hat{f}$. In other words, the prediction requires integrating over the joint distribution of inputs $\bm{x}$, surrogate parameters $\boldsymbol{\theta}$, and model outputs $\hat{f}$ conditioned on the observed data $\mathcal{D}$. If no surrogate is necessary and a high-fidelity model is used directly, then the dependency on $\vv \theta$ is eliminated, but the dependency on $p(\hat{f})$ remains. A simple example of such a non-deterministic model is a GP, where $\hat{f}$ is treated as a random function.

In fact, for linear surrogate parameters $c_p$, the integrals above can often be solved analytically --- particularly for Gaussian likelihoods combined with conjugate priors, where prior and posterior share the same distributional form. The main computational challenge arises from numerically integrating over the nonlinear parameters $w_p$. As a result, linear methods such as PCEs are tractable (only linear parameters $c_p$ and no non-linear parameters $w_p$), whereas NNs are typically intractable in a fully Bayesian treatment due to the presence of numerous nonlinear parameters $w_p$ (cf. Ranftl et al.~\cite{ranftl2021bayesian} for details).

Previously, in Section~\ref{ssec:propagation}, we concluded that surrogate modeling is a practical necessity for many UQ problems. Here, we have formalized surrogate modeling as an inverse problem within a Bayesian framework. Indeed, the inverse problem concerning the surrogate model is, within this framework, structurally not distinct from the inverse problem for model calibration and parameter estimation in Section~\ref{ssec:inverseI}. We present these types of inverse problems separately merely for their distinct role they are playing in the scheme of UQ. 

It is important to recognize that selecting a surrogate function in Eq.~\eqref{eq:def-surrogate} constitutes a fundamental modeling choice --- much like selecting a likelihood and prior for the probabilistic model or a specific constitutive model that capture physics.
Just as the combination of a constitutive model and a likelihood defines a probabilistic model for experimental data, the combination of a surrogate model and a likelihood defines a probabilistic model for simulation data. The essential distinction is that the surrogate is specifically designed to enable rapid evaluation while maintaining sufficient predictive accuracy.
While most readers are familiar with various constitutive or rheological models, we have not yet discussed specific surrogate architectures. We introduce popular classes of surrogate models along with relevant examples from the literature in Section~\ref{sec:types-of-surrogates}, and then present an illustrative mechanical example of surrogate-based UQ in Section~\ref{sec:example2}

\subsection{Example: Hierarchical inference of Young's modulus from uniaxial tensile tests}\label{sec:example1}
As a first basic example, consider a uniaxial tensile test performed on five biological soft tissue samples obtained from five different patients. For clarity of exposition, we adopt a simple linear-elastic constitutive model for calibration against the experimental data. This allows us to focus entirely on the probabilistic aspects of the inference problem rather than on additional complexities of the physical model.

\textit{Patient-specific level inference}. In the first hierarchical step, we aim to infer patient-specific parameters, i.e., we calibrate the constitutive model for each of the five different samples. We believe this is a natural first step because it accounts for individual variability before attempting to identify population-level trends.
Under the assumption of incompressible linear elasticity, the stress–strain relationship is governed by a single unknown material parameter: the Young’s modulus $E$. In accordance with the introduced notation, we define the model parameter as $\bm{x} = E$ and the control parameter (strain) as $\bm{\xi} = \varepsilon$, while the experimental output (stress) is denoted as $y_{\rm exp}$. For each of the five samples, we denote the patient-specific Young's modulus by $E_i$, $i=1,\dots,5$.

For experiment $i$, the observed data consist of $N=100$ strain--stress pairs,
\begin{align}
    \mathcal{D}_{{\rm exp},i}
    =
    \big(\varepsilon_i^{(n)}, y_{{\rm exp},i}^{(n)}\big)
    _{n=1}^{N}\;,
\end{align}
and the complete dataset is $\mathcal{D}_{{\rm exp}} = (\mathcal{D}_{{\rm exp},i})_{i=1}^{5}$.

We can now define the forward model for experiment $i$ and measurement point $n$ with additive Gaussian measurement noise as
\begin{gather}
    y_{{\rm exp},i}^{(n)}
    =
    f\big(\varepsilon_i^{(n)}; E_i\big) + \eta\;, \nonumber\\ 
    \eta \sim \mathcal{N}(0,\sigma_{\rm meas}^2)\;,
\end{gather}
where the model is given by $f\big(\varepsilon_i^{(n)}; E_i\big) = E_i \, \varepsilon_i^{(n)}$. We assume the noise terms $\eta$ are statistically independent across experiments $i$ and measurement points $n$.\footnote{In practice, each measurement is typically assumed to be normally distributed around the model prediction. While this Gaussian noise assumption is a strong simplification, it is a standard choice in the absence of detailed information regarding noise statistics}
Furthermore, the variance $\sigma_{\rm meas}^2$ is assumed to be known from device calibration and identical for all experiments, since the same experimental setup was used throughout. Repeated testing on the same specimen is not feasible in this example, as irreversible damage during the initial loading cycle --- a common occurrence in biological tissues --- cannot be ruled out.

Under the independence assumption, the likelihood factorizes over experiments and measurement points. Up to proportionality constants, this yields
\begin{multline}
    p(\mathcal{D}_{{\rm exp}} \mid E_i, \sigma_{\rm meas}) \\
    \propto
    \prod_{i=1}^{5}
    \exp\left(
    -
    \frac{1}{2\sigma_{\rm meas}^2}
    \sum_{n=1}^{N}
    \left(
    y_{{\rm exp},i}^{(n)}
    -
    E_i \varepsilon_i^{(n)}
    \right)^2
    \right)\;.
\end{multline}

We assume a flat (improper) prior on each modulus with $p(E_i) \propto 1$, which we assume to be positive, so that the posterior is proportional to the likelihood
\begin{multline}\label{eq:posterior1_exp1}
    p(E_i \mid \mathcal{D}_{{\rm exp}}, \sigma_{\rm meas})
    \propto
    p(\mathcal{D}_{{\rm exp}} \mid E_i, \sigma_{\rm meas})\;.
\end{multline}

Taking the negative logarithm and omitting additive constants yields
\begin{multline}
    -\log p(E_i \mid \mathcal{D}_{{\rm exp}}, \sigma_{\rm meas}) \\
    =
    \frac{1}{2\sigma_{\rm meas}^2}
    \sum_{i=1}^{5}
    \sum_{n=1}^{N}
    \left(
    y_{{\rm exp},i}^{(n)}
    -
    E_i \varepsilon_i^{(n)}
    \right)^2\;.
    \label{eq:exp_1_loglikelihood}
\end{multline}
which demonstrates that maximizing the posterior is equivalent to minimizing this quadratic objective function.

Since each experiment is treated independently, the prefactor $1/\sigma_{\rm meas}^2$ merely rescales the objective function without affecting the location of its minimum. Consequently, the MAP estimate for each experiment $i$ reduces to the classical least-squares estimator
\begin{equation}
    E_i^{\mathrm{MAP}}
    =
    \arg\min_{E_i>0}
    \sum_{n=1}^{N}
    \left(
    y_{{\rm exp},i}^{(n)}
    -
    E_i \varepsilon_i^{(n)}
    \right)^2\;.
\end{equation}
as expected when combining a flat prior with Gaussian noise.

To quantify uncertainty, we approximate the posterior distribution. A flat prior
combined with a Gaussian likelihood admits an analytical posterior for this
problem\footnote{\label{fn:exp1_analytical}Under a flat prior, the posterior is proportional to the likelihood, so its logarithm equals the log-likelihood Eq.~\eqref{eq:exp_1_loglikelihood} up to an additive constant. Expanding the square and completing it in $E_i$ yields $\sum_{n=1}^{N}\big(y_{{\rm exp},i}^{(n)} - E_i \varepsilon_i^{(n)}\big)^2 = \sum_{n=1}^{N}\big(\varepsilon_i^{(n)}\big(E_i - E_i^{\rm MAP}\big)\big)^2 +\,\text{const}$, with $E_i^{\rm MAP} = \sum_{n=1}^{N} y_{{\rm exp},i}^{(n)}\varepsilon_i^{(n)} \big/\sum_{n=1}^{N}\big(\varepsilon_i^{(n)}\big)^2$, where the constant collects all terms independent of $E_i$ and is absorbed into the normalization. Matching this quadratic exponent with that of a Gaussian density identifies the posterior as $\mathcal{N}(E_i^{\rm MAP},\sigma_{E_i}^2)$ with $\sigma_{E_i}^2 = \sigma_{\rm meas}^2 \big/ \sum_{n=1}^{N}\big(\varepsilon_i^{(n)}\big)^2$.},
but we proceed numerically and approximate it locally around the MAP estimate
through a Laplace approximation, since closed-form posteriors are unavailable in
many mechanical problems. This corresponds to a second-order Taylor expansion of
the negative log-posterior around $E_i^{\mathrm{MAP}}$.
For a single experiment $i$, the second derivative of the negative log-posterior is given by
\begin{multline}
\frac{\partial^2}{\partial E_i^2}
\left[
-\log p(E_i \mid \mathcal{D}_{{\rm exp},i}, \sigma_{\rm meas})
\right] \\
=
\frac{1}{\sigma_{\rm meas}^2}
\sum_{n=1}^{N}
\left(
\frac{\partial f(\varepsilon_i^{(n)}; E_i)}{\partial E_i}
\right)^2\;,
\end{multline}
where $\partial f(\varepsilon_i^{(n)}; E_i) / \partial E_i = \varepsilon_i^{(n)}$.

The Laplace approximation therefore yields a Gaussian posterior,
\begin{equation}
    p(E_i \mid \mathcal{D}_{{\rm exp},i}, \sigma_{\rm meas})
    \approx
    \mathcal{N}
    \left(
    E_i^{\mathrm{MAP}},
    \sigma_{E_i}^2
    \right)\;,
\end{equation}
with posterior variance
\begin{equation}
    \sigma_{E_i}^2
    \approx
    \left[
    \frac{1}{\sigma_{\rm meas}^2}
    \sum_{n=1}^{N}
    \left(
    \varepsilon_i^{(n)}
    \right)^2
    \right]^{-1}\;,
\end{equation}
analogously to the analytical solution in Footnote~\ref{fn:exp1_analytical}.

\begin{figure*}[t!]
    \centering
    \includegraphics[scale=1.0]{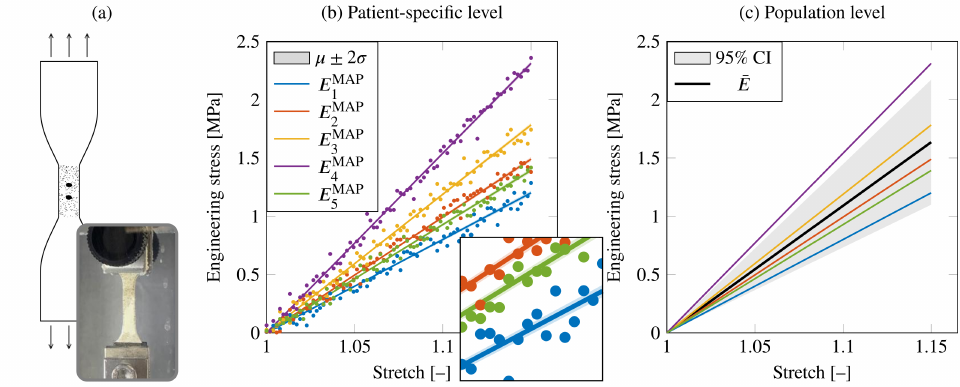}
    \caption{
    (a) Sketch of a uniaxial tensile test, where (b) colored scatter points show the noisy experimental engineering stress--stretch data obtained from five soft biological tissue samples. Solid colored lines represent the individual maximum a posteriori (MAP) estimates of Young’s modulus for each experiment, $E_i^{\rm MAP}$. The shaded colored bands denote the Laplace approximation of the parameter posterior (detaile view), quantifying uncertainty in the estimated modulus arising from measurement noise within each individual dataset.
    (c) The black line shows the arithmetic mean, $\bar{E}$, serving as an estimate of the population represenative Young’s modulus. The gray shaded band indicates the 95\% credible interval (CI) obtained from the Student-$t$ posterior. This interval quantifies uncertainty in the population arithmetic mean arising from variability between experiments and the small number of realizations.
    }
    \label{fig-example1}
\end{figure*}

It should be noted that neither the MAP estimate nor the Laplace approximation is fully Bayesian in the strict sense: the MAP estimate reduces the full posterior to a single point, while the Laplace approximation subsequently approximates the posterior with a Gaussian distribution, rather than computing the full PDF. Nevertheless, both provide computationally tractable uncertainty estimates. The results of this first step are presented in Fig.~\ref{fig-example1}(a) for synthetic experimental data, yielding MAP estimates with Laplace approximations for each individual test. Note that the Laplace bands are narrow, which reflects the relatively large number of measurement points available.

\bigbreak
\textit{Population-level inference}. In the second hierarchical step, we aim to infer a population-representative Young’s modulus, $E_{\rm pop}$, from the five patient-specific parameters $\{E_i\}_{i=1}^{5}$ estimated in the first step.
We can now interpret these patient-specific parameters as independent realizations of an underlying population distribution. Specifically,
\begin{equation}
E_i \sim \mathcal{N}(E_{\rm pop},\sigma_{\rm pop}^2)\;,
\end{equation}
where $\sigma_{\rm pop}^2$ represents the unknown population variance. Since biological variability is assumed to dominate measurement noise in this example, we set $\sigma_{\rm meas}^2 \approx 0$ and treat the inferred $E_i$ as direct observations from the underlying population distribution.

As weakly informative priors we choose, assuming independence between $E_{\rm pop}$ and $\sigma_{\rm pop}$,
\begin{equation}
    p(E_{\rm pop}) \propto 1\;,
    \quad
    p(\sigma_{\rm pop}) \propto \frac{1}{\sigma_{\rm pop}}\;,
\end{equation}
where the prior on $\sigma_{\rm pop}>0$ is the scale-invariant Jeffreys prior. As we will see, these prior assumptions lead to a Student-$t$  posterior distribution for $E_{\rm pop}$. 

If we assume that the five realizations are normally distributed, the likelihood reads
\begin{multline}
    p(E_i \mid E_{\rm pop},\sigma_{\rm pop})
    \\ \propto
    \sigma_{\rm pop}^{-N}
    \exp\left(
    -\frac{1}{2\sigma_{\rm pop}^2}
    \sum_{i=1}^{5}
    (E_i-E_{\rm pop})^2
    \right)\;.
\end{multline}
and applying Bayes’ theorem yields the joint posterior
\begin{align}
    p(E_{\rm pop},\sigma_{\rm pop} \mid E_i)
     &\propto p(E_i \mid E_{\rm pop},\sigma_{\rm pop}) \nonumber\\
     &\hphantom{\propto}\; \times p(E_{\rm pop})p(\sigma_{\rm pop})\;,
\end{align}
which explicitly includes the prior of the unknown population variance $\sigma_{\rm pop}^2$; cf. Eq.~\eqref{eq:posterior1_exp1}, where no such hyperparameter was present.

To proceed analytically, we employ the standard sum-of-squares decomposition
\begin{equation}
    \sum_{i=1}^{5}(E_i-E_{\rm pop})^2
    =
    \sum_{i=1}^{5}(E_i-\bar{E})^2
    +
    5(E_{\rm pop}-\bar{E})^2\;,
\end{equation}
where $\bar{E}=\frac{1}{5}\sum_{i=1}^{5}E_i$ denotes the arithmetic mean of the realizations, and $S=\sum_{i=1}^{5}(E_i-\bar{E})^2$ is the centered sum of squares. The variance $s^2 = (1/(5-1))S$ serves as an unbiased estimator of the population variance $\sigma_{\rm pop}^2$.

With the aim of eliminating the unknown population variance $\sigma_{\rm pop}^2$, we marginalize the joint posterior by integrating out $\sigma_{\rm pop}$
\begin{multline}
    p(E_{\rm pop} \mid E_i)
    \propto
    \int_{0}^{\infty}
    \sigma_{\rm pop}^{-(N+1)} \\
    \times \exp\left(
    -\frac{1}{2\sigma_{\rm pop}^2}
    \left[
    S + 5(E_{\rm pop}-\bar{E})^2
    \right]
    \right)
    \: \mathrm{d}\sigma_{\rm pop}\;.
\end{multline}
Performing this integration (cf.~\cite{ChapterWollner}) and omitting multiplicative constants yields the normalized form
\begin{multline}
    p(E_{\rm pop} \mid E_i)
    \propto
    \left(
    1 + \frac{5(E_{\rm pop}-\bar{E})^2}{S}
    \right)^{-5/2}\;,
\end{multline}
which is recognized as a Student-$t$ distribution centered at the sample arithmetic mean $\bar{E}$ with 4 degrees of freedom.
This result reveals several key properties. First, the posterior mean of $E_{\rm pop}$ coincides with the sample mean
\begin{equation}
    \mathbb{E}[\bar{E}] = E_{\rm pop}\;,
\end{equation}
and second, the estimator of the population variance is unbiased
\begin{equation}
\mathbb{E}[s^2] = \sigma_{\rm pop}^2\;.
\end{equation}
Importantly, however, the population-representative Young’s modulus is not simply the arithmetic average of the five samples. Rather, it is characterized by a full posterior that consistently accounts for both between-sample variability and limited data. The Student-$t$ form reflects this uncertainty: it has heavier tails than a Gaussian, appropriately reflecting the additional uncertainty arising from the small sample size.

Figure~\ref{fig-example1}(b) illustrates the results of this second step, showing the five MAP estimates $E_i^{\text{MAP}}$ obtained from the first hierarchical step, the inferred population arithmetic mean $\bar{E}$, and the corresponding 95\% credible interval for $E_{\rm pop}$. The wide interval demonstrates the considerable uncertainty in estimating the population parameter from only five samples, as expected from the Student-$t$ posterior with 4 degrees of freedom.

\section{Popular surrogate models} \label{sec:types-of-surrogates}
\subsection{General concept}
Let us recall the general concept of surrogate modeling, introduced as an inverse problem in Section~\ref{ssec:inverseII}, where the surrogate model $\hat{f}(\boldsymbol{x}; \boldsymbol{\theta})$ is parameterized by the surrogate parameters $\boldsymbol{\theta}$ and represented as an expansion (cf. Eq.~\eqref{eq:def-expansion}).

With this formulation in mind, we now explore different ways how the surrogate function $\hat{f}$ can be constructed.
For instance, the basis functions $\phi_p$ (Eq.~\eqref{eq:def-expansion}) may be chosen as simple polynomials, in which case the surrogate function represents a polynomial expansion. When prior knowledge about the distribution of $\bm{x}$ is available, it can be advantageous to select these polynomials accordingly, leading to the PCE described in the following Section~\ref{ssec:PCE}. Alternatively, the basis functions $\phi_p$ can be defined through kernel representations, giving rise to GP regression, which can also be viewed as a Bayesian form of machine learning, as shown in the subsequent Section~\ref{ssec:GP}. Finally, by allowing the nonlinear parameters $w_p$ to be trainable as well, we arrive at more flexible surrogate models inspired by modern machine learning, as discussed later in Sections~\ref{ssec:PINN-and-NO} and~\ref{ssec:other-surrogate-models}.

\subsection{Polynomial chaos expansion} \label{ssec:PCE}
A popular choice for surrogate modeling is PCEs~\cite{wiener1938Homogeneous}. Consistent with the notation above, we consider a forward map $f: \bm{x} \mapsto y$, which maps input $\bm{x}$ to output $y$. We approximate this map by a PCE surrogate, denoted as
\begin{equation}
    \hat{f}_{\rm PCE}: \bm{x} \mapsto \hat{y}\;.
\end{equation}
The PCE surrogate is then defined as a finite expansion
\begin{equation} \label{eq:def-pce-linear}
\hat{f}_{\rm PCE}(\bm{x};\bm{\theta}) = \sum_{p=0}^{P} c_p\phi_p(\bm{x})\;,
\end{equation}
where $\phi_p(\bm{x})$ are prescribed (multivariate) polynomial basis functions and $\bm{\theta} = (c_0,\ldots,c_P)^{\rm T}$ denotes the vector of unknown expansion coefficients.
This approach can be viewed as a special, linear case of the generalized model introduced in Eq.~\eqref{eq:def-expansion}, in which the nonlinear parameters $w_p$ are fixed, the basis functions $\phi_p$ are specific types of polynomials, and only the linear coefficients $c_p$ are treated as unknown variables. 

A defining feature of PCEs is the choice of basis. The polynomial basis functions $\phi_p(\bm{x})$ are constructed to be orthonormal with respect to the probability distribution of the input parameters. This orthonormality condition is defined as
\begin{align}\label{eq:def-pce-orthogonality}
    \int \phi_p(\bm{x})\phi_{q}(\bm{x})p(\bm{x})\:\mathrm{d}\bm{x} \stackrel{!}{=} \delta_{p,q}\;,
\end{align}
where $p(\bm{x})$ is the joint probability density of the $D$-dimensional input random vector $\bm{x}$ and $\delta_{p,q}$ denotes the Kronecker delta. Here, $p$ and $q$ index different polynomial basis functions in the expansion. This orthogonality ensures that each basis function captures a distinct, non-overlapping contribution to the variance of the surrogate model $\hat{f}_{\rm PCE}(\bm{x} ; \bm{\theta})$. The advantage of that will become more clear in Section \ref{sec:sensitivity-analysis}, or see \cite{ranftl2021bayesian} for details.

If all input parameters are independent, i.e., the input distribution factorizes as $p(\bm{x}) = \prod_{i=1}^{D_{\bm{x}}} p(x_i)$, the multivariate basis functions admit a tensor-product form,
\begin{equation}
    \phi_p(\bm{x}) = \prod_{k=1}^{D_{\bm{x}}} \psi_{\alpha_k} (x_k)\;,
\end{equation}
where $\psi_{\alpha_k} (x_k)$ is a univariate polynomial of degree $\alpha_k$ in the $k$-th input dimension. Each scalar index $p$ corresponds to a unique multi-index $\bm{\alpha} = (\alpha_1,\ldots,\alpha_D)$ and truncating the expansion amounts to restricting the admissible set of multi-indices. For standard input distributions, the resulting basis functions $\phi_p$ correspond to well-known families of orthogonal polynomials~\cite{xiu2002wiener}: Hermite polynomials for Gaussian inputs, Legendre polynomials for uniform inputs, Laguerre polynomials for Gamma-distributed inputs, etc.

Historically, PCEs were developed for discretizing stochastic differential equations in parameter space, analogously to Galerkin projections~\cite{ghanem1991Stochastic}. This class of methods is often referred to as \emph{intrusive UQ}~\cite{ghanem2017handbook}, as it typically requires direct modifications of the underlying numerical solver (e.g., finite element solver). 

In contrast, the \emph{non-intrusive} approach used here treats the solver as a black box: simulation data are generated first and the PCE is then constructed by solving a regression problem following the generalized model in Eq.~\eqref{eq:def-expansion}, which is linear in the unknown coefficients.

Within the introduced Bayesian framework, this corresponds to an inverse problem in which we infer the surrogate parameters $\bm{\theta} = c_p$ given simulation data $\mathcal{D} = ( \bm{x}^{(n)},y^{(n)})_{n=1}^N$, subject to the orthogonality constraint in Eq.~\eqref{eq:def-pce-orthogonality}. Under these assumptions, the PCE coefficients can be obtained by solving the inference problem in Eq.~\eqref{eq:def-machine-learning-equation}, i.e., the optimization over $\bm{\theta}$. Furthermore, select Bayesian estimates according to Eq.~\eqref{eq:def-machine-learning-equation-Bayesian} admit closed form expression (cf. Ranftl et al.~\cite{ranftl2021bayesian}) as well.

For the PCE surrogate, maximizing the Gaussian likelihood is equivalent to a least-squares fit with fixed nonlinear parameters $w_p$. The optimal coefficient vector $\vv c = (c_0, \cdots, c_P)^{\rm T}$ is then given in closed form by
\begin{align}
\vv c^\ast = (\Phi^{\rm T} \Phi)^{-1}\Phi^{\rm T} \vv y\;,   
\end{align}
where $\vv y = (y^{(1)},\cdots,y^{(N)})^{\rm T}$ and the design matrix satisfies $[\Phi]_{np} = \phi_p(\vv x^{(n)})\;$.
Thus, least squares arises as the maximum likelihood estimate under a Gaussian likelihood, or equivalently as a MAP estimate with a flat prior. The MAP estimate then happens to coincide also with the expected value. In other words, the familiar least-squares fit appears as a special case of Eq.~\eqref{eq:def-machine-learning-equation}. In contrast, introducing a non-uniform prior on $\vv c$ (e.g., a Laplace prior, corresponding to Lasso regularization) eliminates the closed-form solution for Eq.~\eqref{eq:def-machine-learning-equation} and requires numerical optimization methods.

From a Bayesian perspective on regression~\cite{OHagan2013}, the PCE basis functions may appear restrictive. The orthonormality condition (Eq.~\eqref{eq:def-pce-orthogonality}) is simple to satisfy only when the inputs are independent and follow certain standard probability distributions, making the construction of orthonormal PCE bases for experimentally or simulation-derived input densities $p(\bm{x} \mid \mathcal{D})$ often more involved~\cite{Oladyshkin2012,JAKEMAN2019,Torre2019}. Nevertheless, the choice of basis is a modeling decision and non-orthogonal bases may be used when appropriate.

The main advantages of PCEs are interpretability and computational efficiency. When the input distribution $p(\bm{x})$ has a standard form, such as Gaussian, Beta, or Gamma and factorizes into independent components~\cite{Xiu2005,sudret2008global,Crestaux2009}, the variance and conditional variances of the output $y$ admit simple, closed-form solutions. Once the surrogate model is learned, uncertainty propagation becomes essentially free, requiring no numerical integration. This is especially valuable for sensitivity analysis, see Section~\ref{sec:sensitivity-analysis}. These advantages extend to non-standard or non-factorizing distributions, as detailed in~\cite{Oladyshkin2012,JAKEMAN2019,Torre2019}.

The main PCE limitations are (i) poor scalability to high-dimensional inputs and (ii) restricted ability to model correlations. For (i), the number of polynomials and coefficients grows combinatorially with polynomial order and input dimension, limiting standard PCEs to about 20 variables. Various numerical techniques address this, often under additional assumptions or constraints, including hyperbolic truncation~\cite{muhlpfordt2017comments} and sparse, adaptive, or collocation strategies~\cite{minseok2010anchor,foo2010multi,blatman2011Adaptive,doostan2011non,luthen2021Sparse}. However, scalability remains limited without low effective dimension.

To address this challenge, several studies combine PCEs with NNs~\cite{zhang2019Quantifyinga,Schwab2019,cooper2021augmented,Zheng2021,lutjens2021pcepinnsphysicsinformedneuralnetworks,ZHENG2022108732,OLADYSHKIN2023,YAO2023108813,bahmani2025Neural} to leverage NN scalability. Most NN-based PCE methods achieve improved scalability but sacrifice exact analytical expressions for conditional variances, which are the key advantage of PCEs. A notable exception is \emph{DeepPCE}~\cite{exenberger2025deep}, which preserves exact conditional variances while offering NN-like scalability. 

PCEs have long been well-established as surrogate models in (bio)mechanics, predominantly with deterministic hyperparameters; for example, as surrogates for nonlinear constitutive models~\cite{Campos2023a}, for uncertainty propagation of cardiac myofiber orientation and stiffnesses in a computational model of left ventricle deformation~\cite{RodriguezCantano2019a}, or to study the sensitivity of morphological parameters affecting false lumen thrombosis following aortic dissection~\cite{Jafarinia2023a}, to name only a few.

\subsection{Machine learning I: Gaussian process regression} \label{ssec:GP}
In contrast to PCE, which relies on fixed orthogonal polynomial bases and assumes a specific structure in the input distributions, GP regression performs surrogate modeling by defining a probability distribution over functions. Originally established in spatial statistics and UQ under the name \emph{Kriging}~\cite{Krige1952,OHagan1978}, GP regression has experienced a popular revival the machine learning community as a flexible, non-parametric method~\cite{Rasmussen2006}. Here, the term \emph{non-parametric} refers to the fact that we do not assume a specific parametric form for the function itself. Instead, we specify a parametrization of the statistics of, or equivalently the \emph{distribution} over, possible functions. 

Again, we consider a general forward map, $f: \bm{x} \mapsto y$, mapping an input vector $\bm{x}$ to an output $y$, and its surrogate
\begin{equation}
    \hat{f}_{\rm GP}: \bm{x} \mapsto \hat{y}\;.
\end{equation}
Rather than expanding the target function, i.e., the forward model $f$ in a predefined basis, GPs define a probability distribution directly over the space of possible surrogate functions $\hat{f}_{\rm GP}$. This allows the model to infer both the functional form and its associated uncertainty from the data itself, without committing to a particular functional form \emph{a priori}. This does not remove prior assumptions but shifts them. Instead of a basis, one specifies a mean and a covariance (kernel) function. The kernel and its hyperparameters encode assumptions on smoothness, correlation length, and stationarity, which
implicitly define the function space in which the surrogate is inferred.

Per definition, given an input space $\mathcal{X} \subset \mathbb{R}^{D_{\bm x}}$, a GP is a collection of random variables, any finite subset of which follows a joint Gaussian distribution. Colloquially, a GP can be understood as an infinite-dimensional extension of the multivariate Gaussian distribution to functions. Although this concept may appear abstract at first, the resulting equations are remarkably simple.
For a random function $\hat{f}_{\rm GP}: \mathcal{X} \rightarrow \mathbb{R}$  we write 
\begin{equation}
    \hat{f}_{\rm GP} \sim \mathrm{GP}(\mu(\bm{x}), k(\bm{x}, \bm{x}'))\;,
\end{equation}
where $\mu(\bm{x}) = \mathbb{E}\big[\hat{f}_{\rm GP}(\bm{x})\big]$ is the mean function, and $k(\bm{x}, \bm{x}'; \bm{\theta}) = \mathrm{cov}\,[\hat{f}_{\rm GP}(\bm{x}), \hat{f}_{\rm GP}(\bm{x}')]$ is the covariance function parametrized by surrogate parameters $\bm{\theta}$. The covariance function, often called the \emph{kernel} in machine learning, encodes prior assumptions about smoothness, periodicity, or other structural properties of the unknown forward model $f$. The GP framework therefore enables us to express such prior beliefs and to obtain posterior predictive distributions that quantify uncertainty in the surrogate model $\hat{f}_{\rm GP}$ given observed training data.

In surrogate modeling, the GP is typically employed as a \emph{prior} over functions in a Bayesian setting. In GP regression, observations are assumed to be corrupted by Gaussian noise. If this likelihood is also Gaussian, then the resulting posterior is also Gaussian due to conjugacy with the Gaussian prior, ensuring analytical tractability --- the posterior can be computed in closed form. For simplicity, we consider in the following zero-mean priors, i.e., $\mu(\bm{x}) \equiv 0$, meaning we do not assume any systematic trend or offset in the function; a rationale for this choice will be discussed later.

For the covariance function, we require only two basic properties to ensure that the GP is well-defined. First, the kernel must be symmetric, meaning it gives the same value regardless of the order of its inputs: $k(\bm{x}, \bm{x}') = k(\bm{x}', \bm{x})$. Second, it must be positive semi-definite, which guarantees that the covariance matrices constructed from the kernel are valid and correspond to a meaningful probabilistic model\footnote{Note that this property ensures, in particular, that the resulting variances --- diagonal entries of the covariance matrix --- are non-negative, since a negative variance would not define a proper probability distribution. In analogy, $e^{-x^2}$ is normalizable, whereas $e^{x^2}$ is not.}. Formally, for any set of points $(\bm{x}^{(1)}, \ldots,\bm{x}^{(n)}) \subset \mathcal{X}$ and any real coefficients $a_1,\ldots,a_n$, it holds that $\sum_{i=1}^n \sum_{j=1}^n a_i a_j k(\bm{x}^{(i)}, \bm{x}^{(j)}) \geq 0$. 

A default choice for the kernel is the so-called \emph{squared-exponential} kernel, also known as the radial basis function (RBF),
\begin{align} \label{eq:def-SEkernel}
    k(\bm{x}, \bm{x}' ; \vv\theta) = \sigma^2_{f} \exp\left( -\frac{\|\bm{x} - \bm{x}'\|^2}{2\iota^2} \right)\;,
\end{align}
where $\sigma^2_{f}$ is the signal variance or prior variance magnitude and $\iota$ is the correlation length. An entire zoo of symmetric, positive semi-definite kernel functions is available in the literature~\cite{abrahamsen1997gaussian,cuturi2009positivedefinitekernelsmachine,duvenaud2014automatic}. In the following, we will occasionally omit to explicitly denote the dependence of $k(\bm{x}, \bm{x}' ; \vv\theta)$ on $\vv \theta$.

We now compute the posterior predictive.
Given regrouped training data $\mathcal{D} = (\bm{x}^{(n)} , y^{(n)})_{n=1}^N = (\Xset, \vv y)$, with $\Xset=(\bm{x}^{(1)},\cdots, \bm{x}^{(N)})^{\rm T}$ and $\vv y = (y^{(1)},\cdots,y^{(N)})^{\rm T}$, we assume noisy evaluations $y^{(n)} = f(\bm{x}^{(n)}) + \eta$, where $\eta \sim \mathcal{N}(0, \sigma^2)$ is Gaussian i.i.d. noise. We collect the latent (noise-free) function values in the vector $\bm{f} =\big(f(\bm{x}^{(1)}),\ldots,f(\bm{x}^{(N)})\big)^{\rm T} \in \mathbb{R}^N$. The likelihood is therefore $\vv y \mid \vv f, \Xset \sim \mathcal{N}(0, \sigma^2 I)$, where $I\in \mathbb{R}^{N\times N}$ is the identity matrix, and the GP prior is $p(\vv f\mid \Xset,\vv\theta ) = \mathcal{N}(0,K)$. Per Bayes' theorem, $p(\vv f \mid \mathcal{D}, \vv\theta) = \frac{1}{Z} p(\vv y \mid \vv f, \Xset, \cancel{\vv\theta}) p(\vv f\mid\Xset, \vv\theta)$, where $Z$ denotes the marginal likelihood (cf. Eq.~\eqref{eq:GP-logmarginallikelihood}). For a new input $\bm{x}_\ast$, the joint prior of $f_\ast$ and $\bm{f}$ is Gaussian, 
$p(f_\ast , \bm{f} \mid  \bm{x}_\ast, \bm{\theta}) =\mathcal{N}\big(0, 
\begin{bmatrix}
    K & \vv k_\ast\\
    \vv k_\ast^{\rm T} & k(\vv x_\ast,\vv x_\ast)
\end{bmatrix}\big)\;,$
where $K \in \mathbb{R}^{N \times N}$ is the Gram matrix with entries $[K]_{ij} = k(\bm{x}^{(i)}, \bm{x}^{(j)})$, and $\vv{k}_* = \big(k(\bm{x}_*, \bm{x}^{(1)}), \dots, k(\bm{x}_*, \bm{x}^{(N)})\big)^{\rm T} \in \bm{R}^N$ is the covariance vector between the new input and the training inputs.

The predictive distribution is obtained by conditioning this joint Gaussian on the known training values. Using the definition of conditional probability, 
\begin{equation}
p(f_\ast \mid \bm{f},  \bm{x}_\ast, \bm{\theta}) = \frac{p(f_\ast , \bm{f} \mid  \bm{x}_\ast,  \bm{\theta}) }{p(\bm{f} \mid   \cancel{\bm{x}_\ast},  \bm{\theta}) }\;,
\end{equation}
where the denominator does not depend on $\bm{x}_\ast$, since the training values $\bm{f}$ are independent of the new input. Because the joint distribution is Gaussian, this conditioning yields the familiar closed-form expressions for the predictive mean and variance.

The predictive distribution above conditions on the latent training values $\bm{f}$. However, these values are not known exactly; instead, they are distributed according to the posterior $p(\bm{f}\mid\mathcal{D},\bm{\theta})$. Therefore, to obtain the full posterior predictive distribution at a new input $\bm{x}_\ast$ we must marginalize over the uncertainty in $\bm{f}$. This yields
%
\begin{align} \label{eq:gp-aux1}
    p(f_\ast \mid \bm{x}_*, \mathcal{D}, \bm{\theta}) &= \int p(f_\ast \mid \bm{f},  \bm{x}_\ast, \cancel{\mathcal{D}}, \bm{\theta}) p(\bm{f} \mid \mathcal{D}, \bm{\theta}) \:\mathrm{d}\bm{f} \notag \\
    &= \mathcal{N}\big(\mu(\bm{x}_*), \sigma^2(\bm{x}_*)\big)\,,
\end{align}
which, owing to the Gaussian assumptions of the model, results again in a Gaussian distribution. The associated posterior mean and variance are
\begin{align} \label{eq:posterior-mean-var}
\begin{split}
\mu(\bm{x}_*) &= \bm{k}_*^{\rm T} (K + \sigma^2 I)^{-1} \bm{y}\;,\\
\sigma^2(\bm{x}_*) &= k(\bm{x}_*, \bm{x}_*) - \bm{k}_*^{\rm T} (K + \sigma^2 I)^{-1} \bm{k}_*\;.
\end{split}
\end{align}
%

Since the parameter posterior for $\vv\theta^\ast$ is usually intractable, \emph{optimal} hyperparameters $\bm{\theta}$ are often obtained by maximizing the log marginal likelihood. The marginal likelihood is defined as $p(\vv y \mid \Xset, \vv\theta) = \int p(\vv y \mid \vv f, \Xset, \cancel{\vv\theta}) p(\vv f\mid \Xset, \vv\theta)\: \mathrm{d}\vv f$, that is, by integrating out the latent function values $\bm{f}$. Since both the likelihood and the GP prior are Gaussian, this integral can be evaluated analytically by completing the square, resulting in a closed-form expression
%
\begin{align} \label{eq:GP-logmarginallikelihood}
    \log p(\vv y \mid \Xset, \vv\theta) & = -\frac{1}{2} \bm{y}^{\rm T} (K + \sigma^2 I)^{-1} \bm{y} \nonumber \\
    & \hphantom{=}\,\, - \frac{1}{2} \log \det (K + \sigma^2 I) \nonumber \\
    & \hphantom{=}\,\, - \frac{N}{2} \log (2\pi)\;,
\end{align}
with $\boldsymbol{\theta} = (\sigma^2_{f}, \iota)^{\rm T}$, where the covariance matrix $K$ is parameterized by $\boldsymbol{\theta}$. In addition to the kernel parameters, the likelihood variance $\sigma$ is, in principle, also a hyperparameter. As in PCE, the likelihood parameters may likewise be treated as unknown. In GPs, however, $\sigma$ is often estimated from the data or introduced as a technical parameter to stabilize the numerical matrix inversion.
While we focus on zero-mean priors, the prior mean function can also be modeled explicitly, for example by a polynomial expansion such as the PCE introduced in Section~\ref{ssec:PCE}. An extension of GPs to non-zero prior means with linear coefficients is presented in Ranftl et al.~\cite{ranftl2021bayesian}. In machine-learning practice, however, the prior mean is typically set to zero, since the covariance function alone often provides sufficient model flexibility. This property is formalized by the \emph{representer theorem}, which we briefly discuss next.

\bigbreak
%

\textit{Representer theorem}. Simply put, the essence of the {representer theorem} is that the solution of a regularized learning problem in a reproducing kernel Hilbert space --- such as the optimization problems underlying Eq.~\eqref{eq:def-machine-learning-equation} or Eq.~\eqref{eq:GP-logmarginallikelihood} --- admits a representation of the form
\begin{equation} \label{eq:representer-theorem}
f^*(\bm{x}) = \sum_{n=1}^N a_ik(\bm{x}, \bm{x}^{(n)})\;,
\end{equation}
for some coefficients $\bm{a} = (a_1, \dots, a_N)^{\rm T} \in \mathbb{R}^N$. Here, $f^\ast$ denotes the optimal solution of the learning problem within the reproducing kernel Hilbert space defined by the kernel $k$.
In the GP framework, this representation arises directly from the structure of the posterior mean given in Eq.~\eqref{eq:posterior-mean-var}, where $\bm{k}_* = \big(k(\bm{x}_*, \bm{x}^{(1)}), \dots, k(\bm{x}_*, \bm{x}^{(N)})\big)^{\rm T}$ collects the kernel evaluated at the training
inputs, so the posterior mean is a weighted sum of these kernel values. In particular, the GP posterior mean has the form of Eq.~\eqref{eq:representer-theorem} with $\bm{a} = (K + \sigma^2 I)^{-1} \bm{y}$. Thus, the coefficients $a_i$ in Eq.~\eqref{eq:representer-theorem} correspond exactly to the entries of $\bm{a}$ obtained from the GP posterior.
Consequently, it is not strictly necessary to introduce additional surrogate parameters by modeling the prior mean as a separate linear expansion of basis functions, because the representer theorem shows that the posterior mean is already a linear combination of kernel evaluations. Put differently, the GP posterior mean (not the distribution though!) constitutes a generalized linear model of the form~\eqref{eq:def-expansion}, where the basis functions are given by the kernel.

\bigbreak
Nevertheless, in practice it can be advantageous to define a non-zero prior mean function --- for example as a polynomial~\cite{OHagan1978} or a PCE (cf. Section~\ref{ssec:PCE}). Conversely, such models can also be interpreted as PCEs augmented with a GP prior to capture correlations that are not explained by the polynomial trend~\cite{schobi2015polynomial}.

In contrast to PCEs, GPs offer two key advantages: (i) built-in UQ and (ii) explicit modeling of covariance between variables. The first advantage can be viewed as the probabilistic counterpart of deterministic kernel methods such as support vector machines~\cite{cortes1995support} and kernel ridge regression~\cite{vovk2013kernel}, following directly from the predictive variance, the right-hand side of Eq.~\eqref{eq:posterior-mean-var}. The second is achieved through the choice of kernel, which encodes correlation structure.
Importantly, GP uncertainty remains tractable even for non-factorizable input distributions --- a key advantage over PCE --- though uncertainty propagation may require sampling or approximations~\cite{girard2004approximate,marrel2009calculations,wirthl2023global}. Note that this uncertainty represents epistemic uncertainty in the surrogate itself, which is distinct from the propagated aleatoric uncertainty and is not explicitly quantified in PCE.

The major limitations of GPs are: (i) unlike PCE, they do not provide exact solutions for the (conditional) variance of $y$ and require numerical integration (e.g., via Monte Carlo methods) and (ii) limited scalability to high-dimensional problems, despite partial remedies~\cite{hensman2013gaussian,liu2020gaussian,binois2022survey}. 
Limitation (ii) reflects the curse of dimensionality also encountered in PCE and arises from the necessity to invert the covariance matrix during both training and inference. However, computational trade-offs exist: representing a linear expansion as in PCE in terms of kernel functions, as discussed in the representer theorem, or \emph{vice versa}, can yield computational savings depending on the relative sizes of the basis and the dataset~\cite{hensman2013gaussian}.
Lastly, GPs inherently assume that the target variable follows a Gaussian distribution. While this can be restrictive when modeling non-Gaussian outputs, it also provides analytical tractability for Gaussian targets, and enable closed-form posterior distributions.

Similar to PCEs, GPs are a standard surrogate model especially for Bayesian optimization (cf. Section~\ref{ssec:bayesian-optimization}), and here too the GP hyperparameters are typically assumed to be deterministic for uncertainty propagation. Examples include coupled multi-physics models of fluid--structure interaction in biofilms~\cite{Willmann2022a}, sensitivity analyses of output variability in a growth and remodeling model of an aortic aneurysm~\cite{Brandstaeter2021a} or in other biomechanical problems~\cite{Wirthl2023a}, and the quantification of output stress uncertainty under material behavior uncertainty in a nonlinear finite element model of reconstructive surgery~\cite{Lee2018a}. Others used constrained GP models for parameter calibration in a reduced lung model~\cite{Dinkel2024a} and to represent the solution field of a PDE, enabling physics- and data-informed inference with built-in uncertainty quantification via a finite-element discretization of the weak form~\cite{Dalton2026a}. In addition, a comparison of GP and PCE surrogates was performed for hemodynamic pulse-wave propagation modeling~\cite{Paun2025a}.
A Bayesian treatment of the hyperparameters, by contrast, is rare; one of the few examples is found in finite element simulations of impedance cardiography of aortic dissection with multi-fidelity data~\cite{ranftl2019bayesian}.

\subsection{Machine learning II: Neural networks and physics-informed learning} \label{ssec:PINN-and-NO}

\subsubsection{Neural networks} \label{sssec:neural-networks}
Similar to PCE and GP regression, we consider a forward map $f: \bm{x} \mapsto y$, that is, $y = f(\bm{x})$, which we aim to approximate by a surrogate model. In the case of NNs, this surrogate is given by a parametric mapping 
\begin{equation}
    \hat{f}_{\rm NN}: \bm{x} \mapsto \hat{y}\;,
\end{equation}
that defines a learned mapping $\hat{y} = \hat{f}_{\rm NN}(\bm{x} ; \bm{\theta})$.
Thus, NNs map an input $\bm{x} \in \mathbb{R}^{d_0}$ to an output $y \in \mathbb{R}^{d_L}$, with $d_0 \equiv D$ and $d_L\equiv Q$ as used previously. They achieve this through a sequence of layers, where each layer applies a linear transformation followed by a nonlinear activation. The nonlinear transformations are known as \emph{activation functions}. By composing many such transformations, NNs can represent highly complex input–output relationships.

A feedforward network with $L$ layers is defined recursively. We initialize
\begin{equation}
    \mathbf{z}_0 := \bm{x}\;,
\end{equation}
and for each hidden layer $\ell=1,\ldots,L-1$, we compute
\begin{equation}
    \mathbf{z}_\ell = \sigma_\ell(W_\ell \mathbf{z}_{\ell-1} + \mathbf{b}_\ell)\;,
\end{equation}
where $W_\ell \in \mathbb{R}^{d_\ell \times d_{\ell-1}}$ is the weight matrix of layer $\ell$, $\mathbf{b}_\ell \in \mathbb{R}^{d_\ell}$ is the bias vector, and $\sigma_\ell$ is an element-wise nonlinear activation function. Each intermediate variable $\mathbf{z}_\ell$ represents the output of layer $\ell$, obtained by applying a linear transformation followed by a nonlinear actitivation to the previous layer's output $\mathbf{z}_{\ell-1}$. The surrogate is then the output of the final layer,
\begin{equation}\label{eq:def-nn}
    \hat{f}_{\rm NN}(\bm{x} ; \bm{\theta}) := \mathbf{z}_L \;.
\end{equation}
with the full parameter set $\bm{\theta} = (W_\ell , \mathbf{b}_\ell)_{\ell=1}^L$, and usually $\sigma_L$ is the identity function.  
Thus, the network output is obtained by repeatedly transforming the input through all layers.  Training the network consists of finding $\bm{\theta}$ such that $\hat{f}_{\rm NN}(\bm{x})$ closely matches the forward model $f(\bm{x})$ over a set of training examples. This is achieved by minimizing a loss function $\mathcal{L}$, typically the mean-squared error,
\begin{align}
    \mathcal{L}(\bm{\theta}) = \sum_{n=1}^N  \big(y^{(n)} - \hat{f}_{\rm NN}(\bm{x}^{(n)}; \bm\theta)\big)^2\;, 
\end{align}
which measures the discrepancy between the surrogate predictions and reference outputs, i.e., the squared $L_2$-norm. Note that we again seek to find optimal parameters by solving Eq.~\eqref{eq:def-machine-learning-equation}. Minimizing this loss function is equivalent to MAP estimation with a flat prior and the likelihood
\begin{equation}
    p(\mathcal{D} \mid \bm{\theta},f) \propto \exp (-\mathcal{L(\bm{\theta})})\;.
\end{equation}
We observe that Eq.~\eqref{eq:def-nn}, in its vectorized form, shares the same structural form as Eq.~\eqref{eq:def-expansion}, the general expression for the surrogate as an expansion. Specifically, the matrix multiplication $[W_L \vv{z}_{L-1}]_{ij} = \sum_k [W_L]_{ik} [\vv{z}_{L-1}]_{kj} \equiv \sum_k [W_L]_{ik} [\sigma_{L-1}(\vv{x}; \vv{\theta})]_{kj}$ reveals that the activated features $\sigma_{L-1}$ play an analogous role to the basis functions $\phi$ from previous sections. In other words, NNs can be interpreted as generalized \emph{linear} models with learnable basis functions.

In this view, the final weight matrix $W_L$ plays the role of linear coefficients, while the nonlinear basis functions arise from the compositions $\sigma_\ell(W_\ell\mathbf{z}_{\ell - 1} + \mathbf{b}_\ell)$. Unlike traditional expansions such as PCEs or GPs, these basis functions depend on trainable parameters, that greatly increase the expressive power of the model. In other words, this enables NNs to represent functions that lie beyond the expressive capacity of traditional linear models --- including generalized linear models.

This interpretation of NNs as a linear model with complex, trainable basis functions has inspired variants such as the \emph{extreme learning machine}~\cite{huang2006extreme} and \emph{random features approaches}~\cite{rahimi2007random}, where the nonlinear transformations are fixed or randomized and only the final linear layer is optimized. When the loss function is quadratic (e.g., $L_2$-norm), the resulting optimization problem for the linear parameters is convex and sometimes even admits a closed-form solution. A particular advantage of NNs is that they are scalable to high-dimensional inputs~\cite{tripathy2018deep,zabaras2018bayesian}.

NNs, in contrast to both GPs and PCEs, scale very well to high input dimensions. They also exhibit greater capacity for handling nonlinearities and enable new opportunities such as operator learning, which are typically challenging for PCEs or GPs to capture. Strongly nonlinear or even discontinuous functions, in particular, are often difficult to model with PCEs or GPs and NNs offer modeling capabilities that extend significantly beyond those of PCEs or GPs.

However, NNs generally lack the built-in UQ that characterizes GPs and, unlike PCEs, they do not provide analytical expressions for propagated variances or sensitivities. While Bayesian NNs and ensemble-based methods have been proposed to approximate uncertainty~\cite{lakshminarayanan2017simple,wilson2020bayesian}, these approaches come at the cost of substantial additional computational effort and often yield less interpretable results. As such, UQ in NNs and NNs designed for UQ remain an active area of research.

Another limitation is the strong dependence of NNs on architectural and hyperparameter choices, training procedures, and regularization strategies. The mathematical guidance available for designing NN architectures is still limited, often resulting in unstable or inconsistent performance across problems and necessitating extensive heuristic trial and error. Furthermore, NNs typically require much larger amounts of data compared to GPs or PCEs, which can be detrimental in applications such as biomechanics where data are often scarce. These data requirements can be partially mitigated through physics-informed learning or careful design choices, yet they remain an inherent characteristic of most NN–based approaches.

Lastly, NNs are often regarded as black-box models, in contrast to the more interpretable structure of PCEs or the kernel-based formulation of GPs.

\subsubsection{Physics-informed neural networks}\label{sssec:PINN}
While conventional NNs learn purely from data, physics-informed machine learning enriches the learning process by integrating prior physical knowledge directly into a surrogate model~\cite{karniadakis2021Physicsinformed}. In computational mechanics, such prior knowledge often appears in the form of a PDE together with associated boundary and initial conditions.

Let $\Omega_0 \subset \mathbb{R}^{D}$ denote the spatial domain representing the reference (undeformed) configuration with boundary $\partial\Omega_0$, where $\mathbf{X} \in \Omega_0$ is a material point, and let $T > 0$ be the final time. The forward model is governed by the PDE
\begin{align}\label{eq:dev-bvp1}
    \mathcal{F}[u(\mathbf{X}, t)] = q(\mathbf{X}, t)\;, \quad (\mathbf{X}, t) \in \Omega_0 \times (0,T]\;,
\end{align}
where $\mathcal{F}[\cdot]$ is a (possibly nonlinear) differential operator encoding the governing physics, $u(\mathbf{X}, t)$ is the solution field, and $q(\mathbf{X},t)$ is the prescribed forcing or source term on the right-hand side of the PDE.
A simple example is the Laplacian operator, $\mathcal{F} \equiv \nabla^2$.
Depending on the application, $q(\mathbf{X},t)$ may represent body forces, distributed loads, heat generation, reaction terms, or other volumetric inputs to the system.

This PDE is supplemented by the initial and boundary conditions
\begin{align}\label{eq:dev-bvp2}
    u(\mathbf{X}, 0) & = u_{\rm IC}(\mathbf{X})\;, && \mathbf{X} \in \Omega_0\;,\\
    u(\mathbf{X}, t) & = u_{\rm BC}(\mathbf{X}, t)\;, && (\mathbf{X}, t) \in \partial \Omega_0 \times (0, T]\;, \label{eq:dev-bvp3}
\end{align}
where $u$ represents the physical state (e.g., displacement or velocity). Beyond these equations, other forms of prior knowledge can also be incorporated into a learning model, such as conservation laws, symmetry or invariance properties (e.g., incompressibility), canonical variables, or empirical relations. For simplicity, we focus here on the case where the prior knowledge is represented by the PDE system~\eqref{eq:dev-bvp1} to~\eqref{eq:dev-bvp3}.

A physics-informed NN (PINN) aims to learn a surrogate function
\begin{equation}
    \hat{u}: \Omega_0 \times (0,T] \to Y\subseteq\mathbb{R}^Q\;, \quad (\mathbf{X},t) \mapsto \hat{u}(\mathbf{X},t;\vv\theta)\;,
\end{equation}
for $Q$-dimensional spatiotemporal fields, where $\bm{\theta}$ denotes the trainable parameters. The surrogate $\hat{u}(\mathbf{X},t;\mathbf{\theta})$ approximates the true solution $u(\mathbf{X},t)$ of the forward model (Eq.~\eqref{eq:dev-bvp1}). The key distinction is that the NN is trained to directly satisfy the PDE constraint $\mathcal{F}[\hat{u}(\mathbf{X},t;\mathbf{\theta})] = q(\mathbf{X}, t)$ at collocation points, rather than solely relying on input-output data pairs. Note that what was denoted as $y$ previously is denoted as $u$ here to align with standard notation in the PDE literature.

The key idea of PINNs is to train $\hat{u}$ not only to match available data but also to satisfy the PDE, boundary conditions, and initial condition. This is achieved by minimizing the composite loss function $\mathcal{L}(\bm{\theta})$ with hyperparameters $\bm{\theta}$ that penalizes discrepancies between the surrogate solution $\hat{u}$ and the physical constraints
\begin{align} \label{eq:pinn-loss-term}
\mathcal{L}(\bm{\theta}) 
& = \mathcal{L}_{\rm data} 
+ \lambda_{\rm PDE}\,\mathcal{L}_{\rm PDE} \nonumber\\
& \hphantom{=}\,\, + \lambda_{\rm BC}\,\mathcal{L}_{\rm BC} 
+ \lambda_{\rm IC}\,\mathcal{L}_{\rm IC}\;,
\end{align}
where the $\lambda_{(\cdot)}$'s are tunable weighting parameters, and  each term enforces a different constraint: (i) agreement with available data, (ii) residual of the PDE, (iii) boundary conditions, and (iv) the initial condition.
Explicitly, 
\begin{align}\label{eq:def-pinn-lossfunction}
\begin{split}
\mathcal{L}(\theta) & = \frac{1}{N_{\text{data}}} \sum_{i=1}^{N_{\text{data}}} \left| \hat{u}(\mathbf{X}^{(i)}, t^{(i)}) - u^{(i)} \right|^2\\
& \hphantom{=}\,\, + \frac{\lambda_{\text{PDE}}}{N_{\text{PDE}}}\sum_{j=1}^{N_{\text{PDE}}} \left| \mathcal{F}[\hat{u}(\mathbf{X}^{(j)}, t^{(j)})] - q(\mathbf{X}^{(j)}, t^{(j)}) \right|^2\\
& \hphantom{=}\,\, + \frac{\lambda_{\text{BC}}}{N_{\text{BC}}} \sum_{m=1}^{N_{\text{BC}}} \left| \hat{u}(\mathbf{X}^{(m)}, t^{(m)}) - u_{\rm BC}(\mathbf{X}^{(m)}, t^{(m)}) \right|^2\\
& \hphantom{=}\,\, + \frac{\lambda_{\text{IC}}}{N_{\text{IC}}}\sum_{n=1}^{N_{\text{IC}}} \left| \hat{u}(\mathbf{X}^{(n)}, 0) - u_{\rm IC} (\mathbf{X}^{(n)}). \right|^2\;,
\end{split}
\end{align}
where the collocation points $N_{\rm PDE}$, $N_{\rm BC}$, $N_{\rm IC}$ and data points $N_{\rm data}$, are \emph{sampled} within the domain, on the boundary, and on the initial surface. Taking a NN for an ansatz for $\hat{u}$ and minimizing $\mathcal{L}(\bm{\theta})$ yields a surrogate  whose predictions supposedly both match data and approximately satisfy the underlying physics. The most important point of PINNs is that the physics-related loss-terms in Eq. \eqref{eq:pinn-loss-term} effectively act as a regularizer that significantly reduces the amount of data, $N_{\rm data}$, necessary for convergence. This approach can be taken to the extreme by training an \emph{unsupervised} PINN that omits the data-loss term in Eq.~\eqref{eq:pinn-loss-term} entirely and relies purely on the physical equations~\cite{zabaras2019physics}. 

While PINNs enforce physics pointwise at collocation points, the residuals may remain large in regions that are insufficiently sampled. Moreover, training is often sensitive to the weighting parameters $\lambda_{\rm PDE}$, $\lambda_{\rm BC}$, and $\lambda_{\rm IC}$, which has motivated various adaptive weighting strategies~\cite{wang2021understanding,wang2022respecting,basir2022physics,wang2023expert}. 

Finally, the notion of \emph{physics-informed learning} here hinges on the loss function (Eq.~\eqref{eq:pinn-loss-term}), not on the functional form of $\hat{u}$. That is, the representation $\hat{u}$ need not be a NN. Other surrogate families --- such as PCEs~\cite{novak2024physics}, kernel methods, or GPs --- can also be made physics-informed by interpreting the PDE residuals and boundary and initial conditions as contributions to the likelihood or prior~\cite{raissi2017machine,raissi2018hidden,cross2024spectrum}.

\subsubsection{Neural operators}\label{sssec:NO}
Neural operators (NOs) extend surrogate modeling beyond learning pointwise mappings. While standard NNs learn functions $\hat{f}: \bm{x} \mapsto \hat{y}$ and PINNs learn mappings $\mathbf{X} \to \hat{u}$, NOs learn \emph{operators} --- that is, mappings between infinite-dimensional function spaces. Rather than mapping points to points, NOs map entire \emph{input functions} to \emph{output functions}.

Formally, the true operator is denoted by $\mathcal{F}_{\rm NO}: u \mapsto v$, where $u$ is an input function, e.g., boundary conditions, forcing terms, or material coefficients, and $v$ is the corresponding output function, e.g., a PDE solution field. A neural operator then serves as a surrogate
\begin{equation}
    \hat{\mathcal{F}}_{\rm NO}: u \mapsto \hat{v}\;,
\end{equation}
approximating the input--output relationship encoded by $\mathcal{F}_{\rm NO}$.

The most widely used neural-operator architecture is the deep operator network (DeepONet)~\cite{lu2021learning}. Note that while our notation here implies that $\mathbf{X}$ is a spatial coordinate like in Section~\ref{sssec:PINN}, and often indeed it is, however it can also be a stochastic parameter $\vv x$.

A DeepONet consists of two subnetworks that work together to represent the operator:
\begin{itemize}
    \item \textit{Branch network}. The branch network encodes the input function $u$. The function is sampled at a fixed set of sensor points and passed through the branch network, which outputs a latent feature vector $b(u) \in \mathbb{R}^L$.
    \item \textit{Trunk network}. In contrast, the trunk network encodes the evaluation location $\mathbf{X}$. For each spatial (or spatio-temporal) coordinate, the trunk network outputs a latent feature vector $t(\mathbf{X}) \in \mathbb{R}^L$.
\end{itemize}
The surrogate operator prediction at a location $\mathbf{X}$ is then given by
\begin{equation}
    \hat{v}(\mathbf{X}; \vv\theta) = \hat{\mathcal{F}}_{\rm NO}(u)(\mathbf{X}) = b(u)^{\rm T} t(\mathbf{X})\;,
\end{equation}
where the trainable parameters of both subnetworks are collected in $\bm{\theta}$, and $L$ is the dimension of the latent feature space. Intuitively, the branch network thus captures how the input function influences the output, while the trunk network modulates this dependence at each evaluation point $\mathbf{X}$. In relation to the generalized model Eq.~\eqref{eq:def-surrogate} or NNs (Section~\ref{sssec:neural-networks}) before, the elements of $t(\mathbf{X})$ can also be considered learnable basis functions with coefficients $b(u)$.


Given a dataset of input--output function pairs $( u^{(i)}, v^{(i)})_{i=1}^{N_u}$, each sampled at evaluation points $(\mathbf{X} ^{(j)})_{j=1}^{N_x}$, the surrogate operator $\hat{\mathcal{F}}_{\rm NO}$ is trained by minimizing the data misfit
\begin{equation}
    \mathcal{L}_{\text{data}}(\bm{\theta})
    = \frac{1}{N_u N_x}
      \sum_{i=1}^{N_u}
      \sum_{j=1}^{N_x}
      \big|
          \hat{\mathcal{F}}_{\rm NO}(u^{(i)})(\mathbf{x}_j)
          - v^{(i)}(\mathbf{x}_j)
      \big|^2\;.
\end{equation}
Here, $N_u$ is the number of distinct input functions, $N_x$ denotes the number of evaluation (sensor) points per function, and $v^{(i)}(x_j)$ denotes the true output function corresponding to $u^{(i)}$ at location $\mathrm{x}_j$. 

In computational physics and biomechanics, the input function $u$ may represent a boundary displacement, a spatially varying material parameter, a forcing profile, or any other function that parametrizes the governing PDE. The trained NO $\hat{\mathcal{F}}_{\rm NO}$ can then predict the corresponding solution function $\hat{v}(\mathbf{x};\vv\theta^\ast)$ approximately, e.g., instead of a PDE solver.

NOs can also be trained in a physics-informed manner analogous to PINNs~\cite{SifanWang2021}, where additional loss terms enforce PDE residuals, boundary conditions, or conservation laws. However, because these constraints must be evaluated across many input functions and spatial points, the associated computational cost can be significantly higher. Efficient training schemes including separable architectures, low-rank factorizations, and adaptive collocation strategies are active research topics~\cite{yu2024separable,MANDL2025117586}.

Examples of NO as a surrogate in biomechanics are diverse. They map the aortic wall microstructure to pressure-volume curves and damage progression during aortic dissection progression~\cite{Yin2022a}, infer patient-specific mechanobiological insult profiles driving aneurysm formation from clinical maps of aortic dilatation and distensibility~\cite{Goswami2022a}, reconstruct full-field brain displacement fields from multimodal magnetic resonance elastography data~\cite{Agarwal2025a}, relate aortic flow-rate waveforms and heart rate to the corresponding pressure waveform~\cite{Hong2024a}, and denoise displacement fields in ultrasound elastography~\cite{Zhu2024a}. In material modeling, NO surrogate were used to link crack configuration and loading steps to global displacement and damage fields~\cite{Goswami2022b}, experimental load-displacement field measurements to a nonlocal constitutive law together with the fiber-orientation field and full-field stress predictions~\cite{Jafarzadeh2025a}, and loading protocols to the resulting displacement field~\cite{You2022a}. Further examples and perspectives are given in~\cite{Ahmadi2026a,ChapterGoswami}.

\subsection{Alternative surrogate modeling strategies}\label{ssec:other-surrogate-models}
While PCEs, GPs, and PINNs represent some of the most widely used surrogate modeling techniques today, the broader literature on surrogate modeling, particularly within UQ and machine learning, is considerably more diverse. In the following sections, we will briefly introduce several additional approaches that have gained attention in recent years and discuss their respective advantages, challenges, and areas of application.

\subsubsection{Tree-based regressors and bootstrapping}\label{sssec:Tree}
Yet another important class of machine learning models applicable to regression tasks are tree-based and bootstrapping methods, including decision trees~\cite{breiman2017classification}, random forests~\cite{breiman2001random}, and gradient boosting~\cite{FRIEDMAN2002367}. These approaches aim to learn (ensembles of) nonlinear mappings
\begin{equation}
    \hat{f}_{\rm Tree}^{(j)}\colon \bm{x} \mapsto \hat{y}\;,
\end{equation}
that approximate the relationship between an input vector $\bm{x}\in\mathbb{R}^{D_{\bm{x}}}$ and a scalar or vector-valued response $y\in\mathbb{R}^{D_y}$.
Regression trees iteratively partition the input space into regions and fit simple predictive rules locally. A random forest then is an entire (random) ensemble of such trees $j=1,\cdots, J$. 
The sample functions, sometimes called \emph{weak learners}, or rather the predictions thereof, are then aggregated into a single prediction --- such as a mean --- through different bootstrapping methods~\cite{efron1992bootstrap}, typically bagging or boosting. The latter may be guided by the gradients of the weak learners, hence the name \emph{gradient boosting}.

Although these approaches have been extensively developed and analyzed within the machine learning community, and efficient open-source implementations such as XGBoost~\cite{Chen2016_XGBoost} and CatBoost~\cite{prokhorenkova2019catboostunbiasedboostingcategorical} are readily available, their use for surrogate modeling and UQ in computational science remains comparatively limited, with few examples in (bio)mechanics~\cite{ChapterRanftl2}. 

These models can handle nonlinear, non-smooth, and even discontinuous system responses effectively; situations where smooth-function approximators such as GPs or PCEs struggle. Conversely, tree-based models often lack the smoothness desiderata of a given physical system. Moreover, ensemble methods naturally provide measures of predictive uncertainty per the nature of bootstrapping, which can be exploited for uncertainty estimation in surrogate modeling. To maintain focus, we refer the reader to a more detailed discussion of tree-based ensemble methods~\cite[Secs.~2.2.2 and~2.2.3]{ChapterRanftl2}.

\subsubsection{Reduced-order modeling}\label{sssec:ROM}
So far, we have primarily discussed data-driven surrogate models, which learn mappings from input to output while treating the underlying physical system as a black box.
An exception to this paradigm are the physics-informed learning approaches introduced in Section~\ref{ssec:PINN-and-NO}, where physical constraints are explicitly incorporated into the learning process.

In contrast, we have not yet discussed \emph{reduced-order modeling} (ROM) and intrusive UQ, methods that require reformulation of the governing equations~\cite{ghanem2017handbook}. These approaches tackle the problem from the side of physics-based model reduction rather than purely data-driven approximation.
Similar to surrogate models, ROMs aim to reduce the computational complexity, dimensionality or number of degrees of freedom of the governing equations in Eq.~\eqref{eq:def-uncertainty-quantification-deterministic} to obtain computationally efficient approximations of the true solution. Traditionally, ROMs are derived by projecting the governing equations onto a reduced basis, thereby retaining the essential physical structure of the system while substantially decreasing computational cost~\cite{kerschen2005method,lassila2014model,rathore2025projection}.

More recently, data-driven ROMs have emerged by combining projection-based model reduction with machine learning techniques~\cite{kutz2016dynamic}. Such approaches have gained increasing relevance in biomechanics~\cite{chinesta2023reduced,siena2023fast,balzotti2024reduced}, where high-fidelity simulations of complex dynamical systems are computationally demanding.
As a result, the boundary between \emph{physics-informed surrogates} and \emph{data-driven ROMs} has become increasingly blurred, often depending on the degree to which physical modeling and data-driven inference are fused. We have already touched upon these hybrid approaches in Section~\ref{ssec:PINN-and-NO}; however, a comprehensive overview would warrant its own treatment and therefore lies beyond the scope of this review article. For readers interested in UQ from the ROM perspective, we refer to the excellent literature in~\cite{chen2015reduced,chinesta2023reduced}.

\subsubsection{Multi-fidelity modeling}\label{sssec:MF}
\begin{figure*}
    \centering
    \includegraphics[width=1\linewidth]{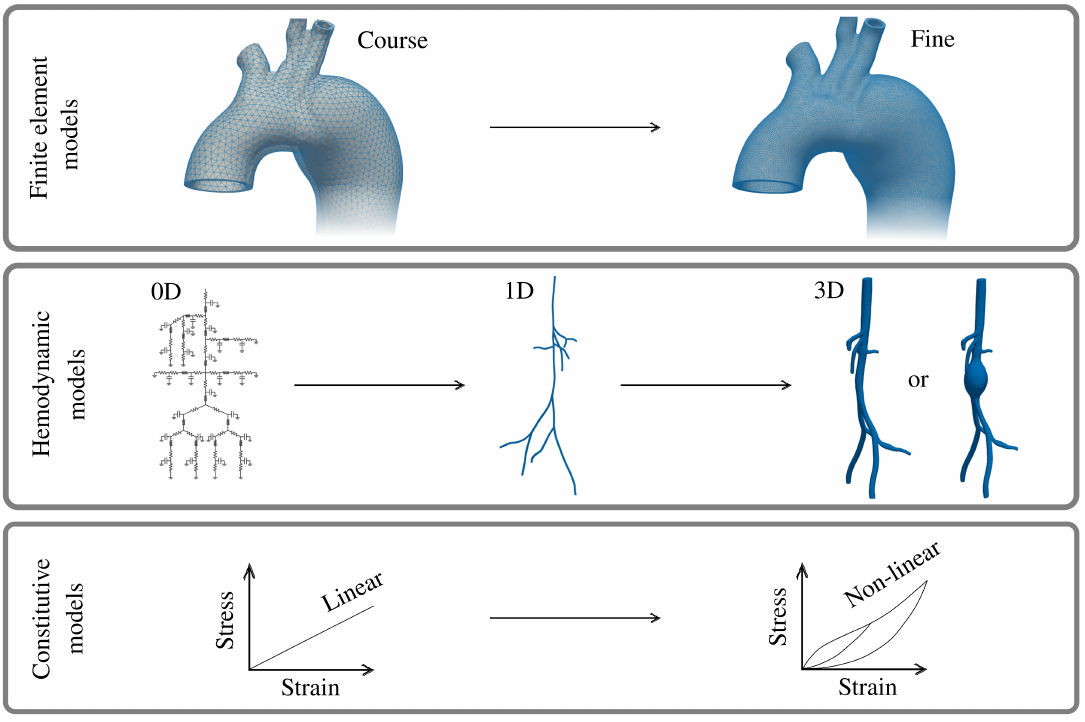}
    \caption{
    Schematic overview of multi-fidelity modeling using physical and numerical models of varying fidelity: (a) patient-specific finite element models of aortic dissection (patient model based on~\cite{Baeumler2024a}; no permission needed) with increasing spatial resolution, ranging from coarse to fine meshes; (b) hemodynamic models of increasing dimensionality, progressing from 0D (lumped-parameter) to 1D and fully resolved 3D representations (modified from~\cite{Fleeter2020a}; with permission from Elsevier); and (c) constitutive models spanning from linear to nonlinear material behavior.}
    \label{fig-multi_fidelity}
\end{figure*}
Multi-fidelity (MF) models combine simulations or data sources of varying accuracy and cost. The idea is to fuse high-fidelity (HF) data, which are accurate but expensive, with low-fidelity (LF) data, which are cheaper but less accurate. To place these ideas in context, we briefly sketch MF modeling here; see, e.g.,~\cite{ChapterSchiavazzi} for a dedicated introduction, implementation aspects, and applications.

An illustrative example is a HF model of a boundary-value problem with a fine spatial discretization or higher-order finite elements, while the LF model could be the same boundary-value problem solved on a coarser mesh or using lower-order elements~\cite{Biehler2015a,ranftl2019bayesian}, see Fig.~\ref{fig-multi_fidelity}. Similarly, in fluid mechanics, the full three-dimensional Navier–Stokes equations may serve as the HF model, whereas corresponding LF representations could be based on the Reynolds-averaged Navier–Stokes equations, one-dimensional flow models, or even zero-dimensional lumped-parameter networks~\cite{tran2017automated,fleeter2020multilevel}.

For vector-valued inputs and outputs, we denote the HF and LF forward models as
\begin{align}
    y_{\rm HF} = f_{\rm HF}(\bm{x})\;, \quad y_{\rm LF} = f_{\rm LF}(\bm{x})\;.
\end{align}
Rather than directly learning the expensive relation
\begin{equation}
    \hat{f}_{\rm HF}: \bm{x} \mapsto y_{\rm HF}(\bm{x}),
\end{equation}
a MF surrogate decomposes this mapping into two stages
\begin{equation}
    \hat{f}_{\rm MF}: \bm{x} \mapsto y_{\rm LF} \mapsto \hat{y}_{\rm HF}\;,
\end{equation}
where $y_{\rm LF}(\bm{x})$ serves as an intermediate representation. The first stage provides a cheap approximation, and the second stage learns a corrective relationship from $y_{\rm LF}(x)$ to $\hat y_{\rm HF}(\bm{x})$. Thus, the composed surrogate still maps $\bm{x}$ to $\hat{y}_{\rm HF}(\bm{x})$, but does so more efficiently by leveraging the intermediate LF prediction.

In the Bayesian framework~\cite{Kennedy2000}, the relationship between HF and LF models can be expressed through a parameterized function
\begin{align} \label{eq:MF-general}
    y_{\rm HF} = \hat{f}_{\rm MF}\big(y_{\rm LF}(\bm{x}) ; \bm{\theta}\big)\;,
\end{align}
where $\hat{f}_{\rm MF}$ denotes the MF surrogate as a function of the LF prediction, and $\bm{\theta}$ represents model parameters.

Alternatively, one may directly model the cross-covariance
\begin{align}
    \mathrm{cov}\big[y_{\rm HF}(\bm{x}), y_{\rm LF}(\bm{x}) \mid \bm{\theta} \big]\;,
\end{align}
to directly infer the correlation structure from data. A particularly common choice is the additive model
\begin{align} \label{eq:MF-additive}
    y_{\rm HF}(\bm{x}) \approx \rho_{\rm LF} \cdot y_{\rm LF}(\bm{x}) + \delta_{\rm MF}(\bm{x})\;,
\end{align}
where $\rho_{\rm LF}$ is a (learnable) correlation parameter while $\delta_{\rm MF}(\bm{x})$ is the discrepancy between HF and LF. A popular Bayesian construction assumes GP priors
\begin{align}
    y_{\rm LF}(\bm{x}) & \sim \mathrm{GP}\big(0, k_{\rm LF}(\bm{x}, \bm{x}' ; \bm{\theta}_{\rm LF})\big)\;, \\ 
    \quad \delta_{\rm MF}(\bm{x}) & \sim \mathrm{GP}(0, k_{\rm MF}\big(\bm{x}, \bm{x}' ; \bm{\theta}_{\rm MF})\big)\;,
\end{align}
where $k_{\rm LF}$ and $k_{\rm MF}$ are the covariance functions associated with the LF model and the discrepancy between the low- and high-fidelity models, while $\bm{\theta}_{\rm LF}$ and $\bm{\theta}_{\rm MF}$ are the hyperparameters of the kernels, respectively. Under this construction, $y_{\rm HF}$ is itself a GP, enabling semi-analytic posterior expressions~\cite{ranftl2019bayesian}.

This formulation can be viewed as an extension of the general UQ framework (cf. Eq.~\eqref{eq:def-uncertainty-quantification}), where the HF model output $y_{\rm HF}$ is substituted by the MF surrogate defined in Eqs.~\eqref{eq:MF-general} to~\eqref{eq:MF-additive}. 
Consequently, the additional parameters $\{\rho_{\rm LF}, \bm{\theta}_{\rm LF}, \bm{\theta}_{\rm MF}\}$ enter the inference process in Eq.~\eqref{eq:def-uncertainty-quantification}. We marginalize over them in a fully Bayesian treatment by extending the corresponding integration.

In the literature, numerous examples demonstrate the use of multi-fidelity modeling in biomechanics to accelerate computational time, particularly for UQ. Biehler et al.~\cite{Biehler2015a} applied this strategy to a patient-specific abdominal aortic aneurysm model by correcting the low-fidelity structural solution with a probabilistic discrepancy term inferred from a few high-fidelity simulations based on finer finite element meshes.
Similarly, Ranftl et al.~\cite{ranftl2019bayesian} quantified uncertainty in impedance cardiography for aortic dissection using multi-resolution finite element models combined with a normal GP surrogate. 
In the context of soft tissue growth and remodeling, Lee et al.~\cite{Lee2020a} propagated uncertainty in mechanical and biological parameters through a multi-fidelity GP surrogate trained on finite element models of tissue expansion with varying mesh refinement, enabling spatio-temporal UQ at a fraction of the high-fidelity cost.
In the vascular domain, several works compare multilevel multi-fidelity estimators across hierarchical zero-, one-, and three-dimensional hemodynamic models~\cite{Fleeter2020a} or link zero-dimensional Windkessel models to patient-specific three-dimensional vascular geometries by treating model discrepancy as random noise~\cite{Choi2025a}, whereas in cardiac modeling, parameter-mapping strategies have been proposed to enable fast zero-dimensional approximations of high-fidelity three-dimensional simulations~\cite{Mollero2018a}.
For flow and coupled fluid--structure interaction problems, Nitzler et al.~\cite{Nitzler2022a} proposed a Bayesian multi-fidelity UQ framework that learns a nonlinear output-to-output dependency between fidelity levels via a GP, augmented by informative input features to improve accuracy in the small-data regime.
Rounding out this overview, Nitzler et al.~\cite{Nitzler2026a,Nitzler2026b} addressed high-dimensional spatial Bayesian inversion by inferring spatially varying permeability fields in poro-elastic media, combining a cheap single-physics porous flow model as low-fidelity surrogate with a Gaussian Markov random field prior and stochastic variational inference to approximate the expensive coupled high-fidelity posterior.

\subsection{Qualitative comparison of surrogate models} \label{sssurrogate-selection}
Each model class (type) of surrogate models presented in Section~\ref{sec:types-of-surrogates} has quantitative measures for selecting models within that class. However, choosing a particular model class in the first place requires a careful analysis of the requirements and goals of the problem, as well as the advantages and limitations of the different model classes in question, as previously discussed. 

In summary, GPs and PCEs are well-established and well-understood surrogate modeling techniques for UQ, although several limitations and unresolved advanced problems remain. In contrast, NNs promise to overcome some of these limitations but are generally far less well understood and face a different set of challenges. Indeed, NNs pose a broader and potentially more impactful range of open questions regarding both methodology and fundamental theory.
The best advise is to select the surrogate model that best fits the requirements and characteristics of their specific problem. For an overview, a simplified tabular comparison of the main aspects of GPs, PCEs, and NNs in the context of surrogate modeling for UQ, presented in Table~\ref{tab:comparison-surrogates}.
\begin{table}[t!]
\centering
\caption{
Qualitative comparison of Gaussian processes (GPs), polynomial chaos expansions (PCEs), and neural networks (NNs) for surrogate modeling in uncertainty quantification (\textit{high}, \textit{medium}, and \textit{low}).
}
\label{tab:comparison-surrogates}
\begin{tabular}{lccc}
\hline
Aspect & GP & PCE & NN \\
\hline
Nonlinearity representation       & \textit{High}   & \textit{Medium} & \textit{High}   \\
Handling of discontinuities       & \textit{Low}    & \textit{Medium} & \textit{High}   \\
Computational scalability         & \textit{Low}    & \textit{Medium} & \textit{High}   \\
Data efficiency                   & \textit{High}   & \textit{Medium} & \textit{Low}    \\
Uncertainty quantification        & \textit{High}   & \textit{Medium} & \textit{Medium} \\
Interpretability                  & \textit{Medium} & \textit{High}   & \textit{Low}    \\
\hline
\end{tabular}
\end{table}

\subsection{Heuristic accuracy metrics and selection criteria}
The performance and accuracy of surrogate models are commonly assessed using a combination of heuristic pointwise error metrics, such as the mean squared error (MSE) and the root-mean-squared error (RMSE). Therein, one has to distinguish between metrics computed over the training dataset and those evaluated on a test or validation dataset. In particular, predictive accuracy is typically quantified via sample-based error norms computed over a validation set $\mathcal{D}_{\mathrm{val}} = (\bm{x}^{(n)}, y^{(n)})_{n=1}^{N_{\mathrm{val}}}$.

Collecting the data into a vector $\bm{y} =(y^{(1)},\cdots,y^{(N)})^\mathrm{T}$ and the corresponding predictions of the surrogate at the input points for a given set of optimal hyperparameters, $\bm{\theta}^\ast$, into a vector $\hat{\bm{y}} = \hat{f}(\underline{\vv x} ;\vv\theta^\ast) \:\coloneqq\: \big(\hat{f}(\vv x^{(1)};\vv\theta^\ast),\cdots,\hat{f}(\vv x^{(N)};\vv\theta^\ast)\big)^\mathrm{T}$, we can define commonly used error metrics as
\begin{align} \label{eq:error-metrics1}
\mathrm{MSE} &= \frac{1}{N_{\mathrm{val}}} \sum_{n=1}^{N_{\mathrm{val}}} \left( y^{(n)} - \hat{f}(\bm{x}^{(n)}; \vv\theta^\ast) \right)^2\,,\\ 
\qquad \mathrm{RMSE} & = \sqrt{\mathrm{MSE}}\;, \\
\begin{split}
\epsilon_{\mathrm{rel}} & = \frac{L_2}{\left\| \bm{y} \right\|_2} = \frac{\left\| \bm{y} - \hat{\bm{y}} \right\|_2}{\left\| \bm{y} \right\|_2}\\
 &= \frac{1}{\left\| \bm{y} \right\|_2}  \left(
\sum_{n=1}^{N_{\mathrm{val}}}
\left( y^{(n)} - \hat{f}(\bm{x}^{(n)}; \vv\theta^\ast) \right)^2
\right)^{1/2} \;,
\end{split}
\end{align}

In fact, the test set is typically drawn from the same data distribution as the training set, but it contains different data points. The difference between training error and test error is called the \emph{generalization gap}. It measures how well a model performs on unseen data and helps detect overfitting or model misspecification. In contrast, \emph{out-of-distribution} evaluation measures performance on test data that does \emph{not} originate from the same distribution as the training data.
The error metrics in Eq.~\eqref{eq:error-metrics1} are sample-based discrete approximations of an expectation, i.e., an integral over the input space. Therefore, the error estimates are themselves random variables, and their accuracy depends on the number and selection of samples.

To obtain reliable error estimates, datasets are often \emph{randomly} split into training and test sets, according to some \emph{split ratio}. However, in surrogate modeling data are typically scarce. For example, with $N=40$ samples, a 90/10 split leaves only 4 data points for testing --- too few for meaningful error estimation. A 10/90 split improves testing but leaves too little data for training. Alternatively, a 50/50 split may mitigate either issue whilst leaving both insufficient, hence we face a challenging situation.

Resampling methods such as $\tau$-fold \emph{cross-validation} address this problem by splitting the dataset into $\tau$ disjoint subsets. The model is trained on $\tau-1$ subsets and tested on the remaining one. This is repeated $\tau$ times so that each subset serves once as the test set, and the final error is the average over all $\tau$ runs.
A special case is \emph{leave-one-out cross-validation}, short LOOCV, where $\tau = N$, i.e., the number of folds equals the total number of samples. In this setting, the model is trained on $N-1$ samples and tested on the single remaining sample, and this is repeated once for every data point. This is the standard error measure in PCE modeling. In contrast, NNs are often evaluated using an $L_2$ error with a simple train–test split, since they typically require large datasets where additional resampling is less critical.

For probabilistic surrogates, a common metric is the \emph{negative log predictive density} (NLPD),
\begin{align}
\mathrm{NLPD} =
-\frac{1}{N_{\mathrm{val}}}
\sum_{n=1}^{N_{\mathrm{val}}}
\log p\left(
y^{(n)} \mid \bm{x}^{(n)}, \mathcal{D}
\right) \;,
\end{align}
which measures how well the predicted probability distribution explains the observed test data. For GPs, it is the standard test metric and corresponds to the average negative log marginal likelihood (Eq.~\eqref{eq:GP-logmarginallikelihood}) evaluated on the test set.

Probabilistic surrogates also provide predictive uncertainty, for example through credible intervals. A simple consistency check is to compare the expected probability mass of an uncertainty band with the observed frequency in the test data. 
For instance, under a Gaussian prediction, about 68\% of the data should lie within one standard deviation of the mean, i.e., $\Pr\left(\lvert y^{(n)} - \mu(\vv x^{(n)})\rvert < \sigma\right) \approx 0.68$. Deviations from this indicate miscalibrated uncertainty.

Evaluation metrics are mainly used to assess a trained model' performance, but they are often also used for heuristic model selection --- for example, choosing the model with the lowest MSE.
However, surrogate modeling for UQ is more subtle than standard machine learning. In typical machine learning benchmarks, large datasets are available and predictive error on held-out data is the main criterion. In UQ, what matters most is accuracy in regions with high \emph{probability mass}.\footnote{For example, consider a prior $p(\vv x)$ with two sub-regions $\mathcal{X}_1, \mathcal{X}_2 \subset \mathcal{X}$ of similar size, i.e., $\int_{\mathcal{X}_1} \mathrm{d}\vv x \approx \int_{\mathcal{X}_2} \mathrm{d}\vv x $.
Suppose, $\mathcal{X}_1$ carries only about $10^{-6}$ of the total probability mass, while $\mathcal{X}_2$ carries about $10^{-1}$. 
Then, an accurate surrogate in $\mathcal{X}_2$ is far more important than in $\mathcal{X}_1$, because UQ typically involves integrals of the form $\int_{\mathcal{X}} \big(y(\vv x)\big)^\varrho p(\vv x) \mathrm{d}\vv x$, where $\varrho$ is a parameter that specifies the moment of interest (e.g., $\varrho = 1$ for the mean, $\varrho = 2$ for the second moment). Regions with negligible probability mass contribute very little to such integrals, unless $y(\vv x)$ grows so large in those regions that it compensates for the small probability.
The same reasoning applies to sensitivity analysis, inverse problems, and related tasks.}

These heuristic metrics ignore the prior and are therefore not fully suitable for performance evaluation or model selection. While surrogate accuracy in UQ was discussed in Section~\ref{sssec:Bayesian-surrogate-modeling}, the following Section~\ref{ssec:model-selection} addresses principled model selection from a Bayesian perspective.


\subsection{Example: Surrogate-based uncertainty propagation of arterial properties in a pressurized symmetric cylinder}\label{sec:example2}
As a second example, we aim to propagate uncertainties of a material parameter through a computational model, our forward model $f$. In particular, we consider a finite element simulation of an arterial segment with initial inner radius $R_{\rm in}$ and outer radius $R_{\rm out}$, subjected to a constant physiological inner pressure $P_{\rm in} = 100$\,mmHg. Technically, $P_{\rm in}$ is a control parameter; it is held fixed and therefore suppressed in the following. For the sake of argument, the arterial wall mechanics are described by linear elasticity under an incompressibility constraint, with Young's modulus $E$ as the uncertain parameter, i.e., $\bm{x} = E$. As output, we consider the deformed outer radius $r_{\rm out}$, i.e., $y = r_{\rm out}$. The input-output relationship can be written as
\begin{equation}
    r_{\rm out} = y = f(E)\;.
\end{equation}
We assume that simulation data are available in the form 
\begin{equation}
    \mathcal{D} = \big( E^{(n)} , y^{(n)} \big)_{n=1}^N\;,
\end{equation}
where $y^{(n)} = f(E^{(n)})$ denotes the simulated outer radius at a given Young's modulus $E^{(n)}$, for a total of $N$ data pairs. 

We assume that a previous Bayesian inverse problem based on experimental mechanical tests, such as the one shown in Section~\ref{sec:example1}, has provided us with a log-normal prior for the Young's modulus with parameters $\mu_{\ln E} = 4.0$ (corresponding to approximately $54.6$\,kPa) and $\sigma_{\ln E} = 0.3$. This choice ensures physical consistency by strictly enforcing $E > 0$.

To avoid the high computational cost of repeatedly evaluating the forward model, we construct a PCE as a surrogate for the finite element simulation. Because $E$ is log-normally distributed, \textit{Hermite polynomials} $\phi_p$ are the appropriate orthogonal basis, as they are orthogonal with respect to the Gaussian measure in the log-transformed space. Since $E$ is log-normal, $\ln E$ is Gaussian with mean $\mu_{\ln E}$ and standard deviation $\sigma_{\ln E}$, so that standardizing $\ln E$ yields a standard Gaussian variable
$\tilde{E} \sim \mathcal{N}(0,1)$ that serves as a proxy for the input,
\begin{equation}
    \tilde{E}(E) = \frac{\ln E - \mu_{\ln E}}{\sigma_{\ln E}}\;.
\end{equation}
The PCE approximation is then expressed as
\begin{equation}
    \hat{f}_{\text{PCE}}(E) = \sum_{p=0}^{P} c_p \phi_p(\tilde{E})\;,
\end{equation}
where $\phi_p$ denotes the $p$-th orthogonal Hermite polynomial and the polynomial order $P$ is a hyperparameter. To determine the expansion coefficients, we sample $N=50$ training points $(E^{(n)})_{n=1}^N$ directly from the log-normal prior using Latin hypercube sampling. For each sampled value, the forward model is evaluated to obtain the corresponding outer radius $y^{(n)} = f(E^{(n)})$, and the basis functions $[\bm{\Phi}]_{pn} = \phi_p(\tilde{E}(E^{(n)}))$ are used to form the design matrix $\bm{\Phi}$. The coefficients $c_p$ are then determined by solving the optimization problem in Eq.~\eqref{eq:def-machine-learning-equation} via least squares (cf. Section~\ref{ssec:PCE}), corresponding to the classical deterministic construction of PCEs. As shown in Section~\ref{sec:sensitivity-analysis}, the PCE yields an analytic expression for the variance of the model response
\begin{align}
    \text{Var}[y] \approx \text{Var}[\hat{f}_{\text{PCE}}(E)] = \sum_{p=1}^P c_p^2\;.
\end{align}
From a Bayesian perspective, this procedure does not constitute a fully probabilistic surrogate model and does not explicitly quantify surrogate uncertainty. A fully Bayesian treatment would instead formulate the surrogate construction itself as an inference problem, assigning probability distributions to the expansion coefficients, as exemplified in a later example (cf. Section~\ref{sec:example4}).

\begin{figure*}
    \centering
    \includegraphics[scale=1.0]{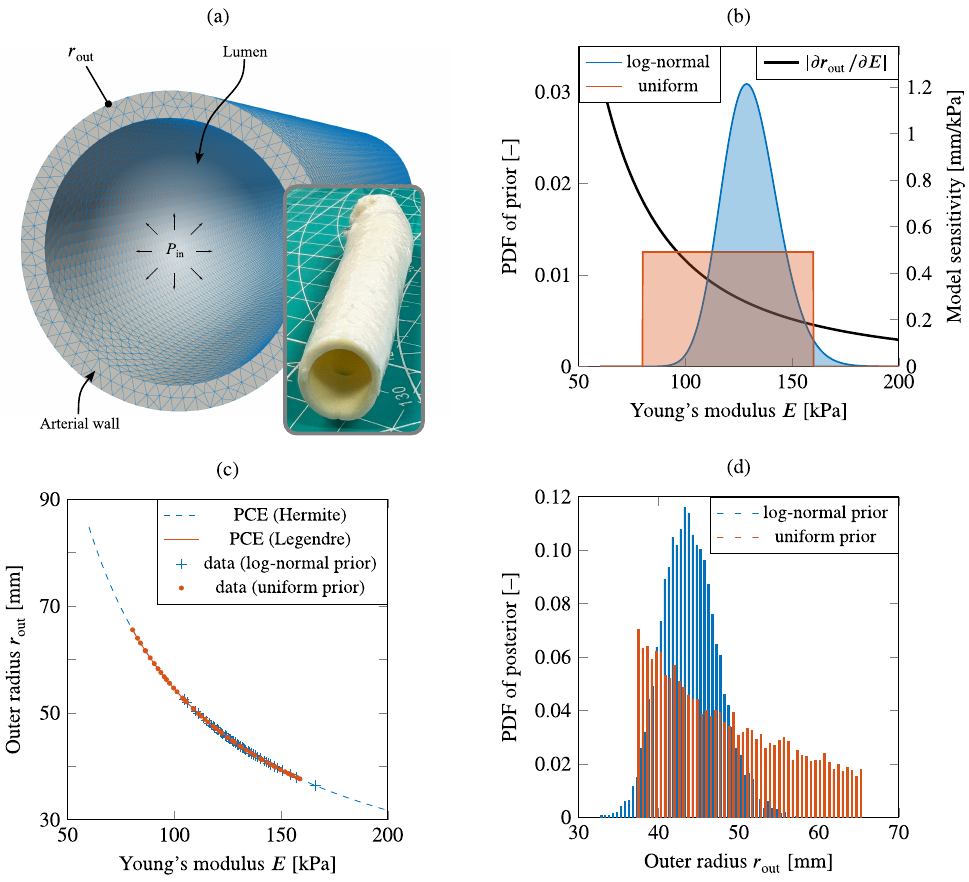}
    \caption{Surrogate-based uncertainty propagation through a finite element simulation of an idealized arterial wall under constant internal pressure, comparing the influence of informative (log-normal) vs. non-informative (uniform) priors on outer radius uncertainty.
    (a) Sketched boundary value problem of an idealized arterial wall with inner radius $r_{\text{in}}$, outer radius $r_{\text{out}}$, and applied internal pressure $P_{\rm int}$.
    (b) PCE surrogate models with Hermite basis (log-normal prior) and Legendre basis (uniform prior) of 5th-order, respectively, both achieve relative error $< 0.001\%$ compared to the simulation data. The Legendre PCE is restricted to the interval $[80, 160]$\,kPa as the uniform prior provides no information beyond this range. Training data are shown as markers.
    (c) Log-normal (median 130 kPa, shape parameter 0.1) and uniform priors ($80-160$\,kPa); overlaid is the model sensitivity $|\partial r_{\text{out}}/\partial E|$, which increases at lower stiffness values.
    (d) Propagated output distributions for outer radius obtained from $10,000$ Monte Carlo samples evaluated through PCE surrogates trained on $50$ Latin Hypercube samples. The non-informative uniform prior produces 25\% larger uncertainty (standard deviation $0.015$ vs. $0.012$\,mm), demonstrating how epistemic input uncertainty amplifies through the nonlinear forward model into predictive uncertainty.}
    \label{fig:example2}
\end{figure*}

Together with the log-normal prior on $E$, the PCE surrogate enables efficient computation of the posterior distribution of the output, the outer radius. The posterior is approximated using Monte Carlo sampling. Exemplary results based on synthetic simulation data are shown in Fig.~\ref{fig:example2}, illustrating how uncertainty in the Young's modulus propagates into predictions of the outer radius under physiological loading conditions. For comparison, the results also show the posterior obtained with a uniform prior on the Young's modulus, demonstrating the sensitivity of the inference to the choice of prior. In this case, Legendre polynomials were used as the PCE basis functions to enable a fair comparison.

\section{Bayesian model selection}\label{ssec:model-selection}
Beyond qualitative comparisons, Bayesian model selection identifies the candidate model that best explains the observed data, formulated as
\begin{quotation}
    \textit{Identification of the most plausible model among competing hypotheses based on their posterior, thereby balancing model fit and complexity through the Bayesian evidence.}
\end{quotation}

\subsection{General concept}
In mechanics, we often encounter situations where several competing models may explain the same experimental data. For example, one may compare different surrogate models against a physics-based computational model or evaluate alternative constitutive laws for describing the experimental behavior of cardiac tissue, possibly using a multimodal dataset. Bayesian probability theory offers a principled mechanism for determining which model best explains the observations while accounting for uncertainty. This is the goal of Bayesian \textit{model selection}~\cite{sivia2006}.

To formalize this, we must make the dependence on the chosen model explicit. The probabilities must be conditioned not only on the model parameters, such as the parameters of a surrogate model, but also on the model hypothesis itself. Indeed, all probabilities until now were implicitly to be understood in the context of a given model hypothesis (cf. Section~\ref{sec:logic}); we had merely swept this information under the rug and not denoted this dependence explicitly.

For the $m$-th candidate model, we now make this dependence explicit by conditioning all probabilities on the model hypothesis. The likelihood function, which previously depended only on the parameters, is now written as
\begin{equation}\label{eq-likelihood_model_selection}
    p(\mathcal{D} \mid \bm{\theta}_m) \equiv p(\mathcal{D} \mid \bm{\theta}_m, \mathcal{M}_m)\;,
\end{equation}
where $\mathcal{M}_m$, $m = 1,\dots,N_m$, denotes the model hypotheses and $\bm{\theta}_m$ its associated parameter vector. In the context of Section~\ref{ssec:inverseII}, each hypothesis $\mathcal{M}_m$ specifies a particular surrogate $\hat{f}_m$ and therefore comes with its own parameterization. These parameters may represent, for example, the hyperparameters of a GP or the weights of a NN. Of course, each hypothesis $\mathcal{M}_m$ can also represent a physics-based forward model $f_m$, where the parameters correspond to the material parameters in a constitutive law (cf. Section~\ref{ssec:inverseI}).

While the likelihood (Eq.~\eqref{eq-likelihood_model_selection}) evaluates the data given specific parameter values and the $m$-th candidate, model selection requires assessing how well each model explains the data regardless of the particular parameter values chosen.
To quantify this, we perform a first marginalization over the parameters and introduce the \emph{marginal likelihood}, also called the \emph{model evidence},
\begin{align} \label{eq:marginal-likelihood-for-model}
    p(\mathcal{D} \mid \mathcal{M}_m)
    & = \int p(\mathcal{D} \mid \boldsymbol{\theta}_m, \mathcal{M}_m)
         \nonumber\\
    & \hphantom{=}\,\, \times p(\boldsymbol{\theta}_m \mid \mathcal{M}_m)\:
    \mathrm{d}\boldsymbol{\theta}_m\;,
\end{align}
which measures how well the hypothesis $\mathcal{M}_m$ explains the data on average, after integrating over all admissible parameter values. 
This quantity, Eq.~\eqref{eq:marginal-likelihood-for-model}, previously appeared as a normalization constant when solving the inverse problem of parameter inference (cf. Sections~\ref{ssec:inverseI} and~\ref{ssec:inverseII}) and could be neglected since it does not affect the posterior distribution over parameters when the model is fixed. However, it becomes the central object for model \emph{comparison}.

Having computed the model evidence for each candidate model in Eq.~\eqref{eq:marginal-likelihood-for-model}, we can now apply Bayes' theorem to determine which model is most probable given the data. Treating the model hypothesis $\mathcal{M}_m$ itself as the unknown quantity, Bayes' theorem yields the posterior of the model hypothesis,
\begin{align}
    p(\mathcal{M}_m \mid \mathcal{D})
    =
    \frac{
        p(\mathcal{D} \mid \mathcal{M}_m)p(\mathcal{M}_m)
    }{
        p(\mathcal{D})
    }\;,
\end{align}
which quantifies how plausible the model $\mathcal{M}_m$ is in light of the data, irrespective of the specific parameter values. Two key factors appear: (i)~the model evidence $p(\mathcal{D} \mid \mathcal{M}_m)$, which plays the role of the likelihood and reflects how well the model explains the observations, and (ii)~the prior model $p(\mathcal{M}_m)$, encoding how plausible the model was before seeing any data.

Treating the model itself as uncertain requires, according to the Bayesian paradigm, a second marginalizing, now over all model hypotheses. By the sum rule of probability,
\begin{equation}
    p(\mathcal{D})
    =
    \sum_{m=1}^{N_m}
    p(\mathcal{D} \mid \mathcal{M}_m)p(\mathcal{M}_m)\;.
\end{equation}
However, computing this summation is rarely feasible in practice.
Each term requires evaluating the model evidence via the integral in Eq.~\eqref{eq:marginal-likelihood-for-model}, which can be computationally prohibitive. Even when the marginal likelihood can be evaluated analytically for given hyperparameters (as in GP regression; cf. Section~\ref{ssec:GP}), one often optimizes over hyperparameters rather than marginalizing over them, avoiding a second demanding integration. For this reason, model comparison often relies on approximations.

For practical comparison of two models, $\mathcal{M}_m$ and $\mathcal{M}_{m'}$, it is typically sufficient to consider the ratio of their posteriors $o$, defined as
\begin{align}
    o
    =
    \underbrace{
        \frac{p(\mathcal{M}_m \mid \mathcal{D})}
             {p(\mathcal{M}_{m'} \mid \mathcal{D})}
    }_{\text{odds ratio}}
    =
    \underbrace{
        \frac{p(\mathcal{D} \mid \mathcal{M}_m)}
             {p(\mathcal{D} \mid \mathcal{M}_{m'})}
    }_{\text{Bayes factor}}
    \;
    \underbrace{
        \frac{p(\mathcal{M}_m)}{p(\mathcal{M}_{m'})}
    }_{\text{prior odds}}\;,
\end{align}
where the overall evidence $p(\mathcal{D})$ cancels out, making pairwise comparison tractable and often sufficient for practical applications.
In fact, the three appearing terms have clear interpretations: the Bayes factor quantifies how strongly the \emph{data} support one model over another, while the prior odds capture any \emph{prior preference} we may have between the models. Together, they determine the posterior odds ratio, which represents our updated belief about which model is more probable after seeing the data\footnote{As a special case, if two mutually exclusive and exhaustive models are considered, i.e., only one hypothesis can be true, the posteriors must satisfy
$
p(\mathcal M_m \mid {\mathcal{D}}) + p(\mathcal M_{m^\prime}\mid{\mathcal{D}}) = 1.
$
In this binary case, the odds ratio $o$ directly quantifies how much more probable one model is compared to the other after observing the data.}.

But how should we interpret the magnitude of the odds ratio $o$? Since the odds ratio can span many orders of magnitude, a logarithmic scale is commonly used. Although interpretation remain a matter of debate, Kass and Raftery~\cite{kass1995bayes} propose the following widely adopted classification:
\begin{align*}
    &\text{Hardly significant}\quad  & 0<\lvert & \log_{10}  (o)\,\rvert < 0.5\;,  \\
    &\text{Positive evidence}\quad & 0.5 < \lvert & \log_{10} (o)\,\rvert < 1\;,  \\
    &\text{Strong evidence}\quad & 1 < \lvert & \log_{10} (o)\,\rvert < 2\;,  \\
    &\text{Overwhelming evidence}\quad &  \lvert &\log_{10} (o)\,\rvert >2\;.
\end{align*}
For instance, an odds ratio of 100 ($\log_{10} (o) = 2$) means that one model is 100 times more probable than the other, providing overwhelming evidence. An odds ratio of 10 ($\log_{10} (o) = 1$) indicates that one model is 10 times more probable, already showing strong support. Conversely, an odds ratio of $\approx 1$ ($\log_{10} (o) \approx 0$) suggests both models explain the data equally well, making it difficult to justify preferring one over the other.

While the Bayes factor is determined by the data and can often be computed or approximated, Bayesian reasoning also allows us to incorporate prior beliefs about the models themselves through the prior odds
\begin{equation}
    \frac{p(\mathcal{M}_m)}{p(\mathcal{M}_{m'})}\;,
\end{equation}
which is generally less straightforward than computing the Bayes factor. 
A common assumption, particularly in surrogate modeling where no strong prior preference exists, is that all candidate models are equally likely \emph{a priori}, i.e., before seeing any data. In such cases, the odds ratio reduces to the Bayes factor alone. 
However, when data are limited and cannot strongly discriminate between models, the choice of prior odds becomes more influential in determining the posterior model probabilities. Further discussions and advanced techniques can be found in the specialized literature~\cite{o1995fractional, sivia2006, VonToussaint2011,vonderLinden2014}.

While our discussion has emphasized surrogate models as competing hypotheses, model selection is equally relevant to physics-based forward models in mechanics, particularly within constitutive modeling. Bayesian model selection has been applied in discriminating among constitutive descriptions for soft biological tissues~\cite{Madireddy2015a, Aggarwal2023a} and various engineering materials~\cite{Ritto2015a, Mototake2020a, Battalgazy2025a}, with comparisons to sparse regression algorithms for automated model discovery~\cite{UrreaQuintero2026a}. Similar Bayesian frameworks have been employed to evaluate competing fatigue damage progression laws in composites~\cite{Chiachio2015a} and to adjudicate between computational fracture models~\cite{Hamdia2019a}. Furthermore, researchers have aimed to identify the optimal constitutive models for predicting pulmonary hemodynamics~\cite{Paun2020a}, to select among phenomenological tumor-growth models~\cite{Oden2013a}, or to differentiate between rheological models for yield-stress fluids~\cite{Rinkens2026a}, to name a few.

\subsection{Example: Derivation of the Bayesian information criterion}\label{sssec:model-selection-BIC}
Since the evidence integral in Eq.~\eqref{eq:marginal-likelihood-for-model} is generally intractable, we next derive a closed-form approximation for practical model comparison. The following argument provides a controlled sequence of approximations, following the spirit of~\cite{VonToussaint2011, vonderLinden2014}.

\bigbreak
 \noindent \textit{Step 1: Approximating a flat prior.} We assume that the prior $p(\bm{\theta}_m)$ is sufficiently broad compared to the likelihood, so that within the region where the likelihood is significant, the prior is effectively constant
\begin{equation}
    p(\bm{\theta}_m)
    \approx
    \frac{1}{\Lambda_m^{P_m}}p(\bar{\bm{\theta}}_m)\;,
\end{equation}
where $\Lambda_m$ is the characteristic prior width per dimension arising from normalization, $P_m$ is the number of parameters in model $\mathcal{M}_m$, and $\bar{\bm{\theta}}_m$ is a representative prior value, here taken as the mode of the likelihood. Substituting this into the evidence integral gives
\begin{align}
    p(\mathcal{D}\mid \mathcal{M}_m)
    & \approx \frac{1}{\Lambda_m^{P_m}}p(\bar{\bm{\theta}}_m) \nonumber\\
    & \hphantom{=}\,\, \times\int p(\mathcal{D}\mid\bm{\theta}_m,\mathcal{M}_m)\: \mathrm{d}\bm{\theta}_m\;.
\end{align}

\bigbreak
\noindent \textit{Step 2: Approximating the likelihood near its maximum.} Next, we assume that for large sample size $N$, the likelihood becomes sharply peaked around the maximum likelihood estimate
\begin{equation}
    \bm{\theta}^\ast_m
    = \arg\max_{\bm{\theta}_m}
      p(\mathcal{D}\mid\bm{\theta}_m,\mathcal{M}_m)\;.
\end{equation}
To leading order, we approximate the likelihood as approximately constant within a small region of volume $\Delta_m^{P_m}$, and negligible outside. Under this approximation,
\begin{equation}
    \int p(\mathcal{D}\mid\bm{\theta}_m,\mathcal{M}_m)\, \mathrm{d}\bm{\theta}_m
    \approx
    p(\mathcal{D}\mid\bm{\theta}^\ast_m,\mathcal{M}_m)\, \Delta_m^{P_m}\;.
\end{equation}

\bigbreak
\noindent \textit{Step 3: Combining both approximations}. Combining the results from Steps 1 and 2, and taking the logarithm on both sides, yields
\begin{align}
    \log p(\mathcal{D}\mid \mathcal{M}_m)
    & \approx \log p(\mathcal{D}\mid\bm{\theta}^\ast_m,\mathcal{M}_m) \nonumber\\
    & \hphantom{=}\,\, - P_m \log\left( \frac{\Lambda_m}{\Delta_m} \right)\;.
\end{align}

\bigbreak
\noindent \textit{Step 4: Scaling assumptions with sample size.} To make further progress, we relate the likelihood width $\Delta_m$ to the number of data points $N$. For many standard likelihoods --- in particular for Gaussian likelihoods --- the width of the region where the likelihood is significant shrinks proportionally to $1/\sqrt{N}$ as more data accumulates
\begin{equation}
    \Delta_m = \frac{\sigma}{\sqrt{N}}\;,
\end{equation}
for some $O(1)$ proportionality constant $\sigma$. If the prior information is taken to be on the same order as the information contained in one data point, i.e., $\Lambda_m \approx \sigma$, then,
%
\begin{align}
    \log p(\mathcal{D}\mid \mathcal{M}_m) 
    & \approx \log p(\mathcal{D}\mid\bm{\theta}^\ast_m,\mathcal{M}_m) \nonumber\\
    & \hphantom{=}\,\, - \frac{P_m}{2}\, \log N\;.
\end{align}

\bigbreak
This motivates the standard definition of the \emph{Bayesian information criterion} (BIC) as
\begin{equation}
    \mathrm{BIC}_m = -2\log p(\mathcal{D}\mid\bm{\theta}^\ast_m,\mathcal{M}_m) + P_m \log N\;,
\label{eq:BIC}
\end{equation}
which provides a crude but practical approximation to the model evidence and is widely used when exact evidence computation is infeasible. Since model selection seeks to maximize the evidence or, equivalently, to minimize its negative logarithm, we choose the model with the smallest BIC value.

The key reason for the practicality of BIC is that it replaces the intractable high-dimensional evidence integral with a simple closed-form expression that depends only on the maximum likelihood value, the number of model parameters, and the size of the dataset. This is achieved through two simplifying assumptions: (i) the prior is effectively constant in the region where the likelihood is significant, and (ii) the likelihood is sharply peaked around its maximum and can be approximated by a hypercube. As a result, BIC is far easier to compute than the full evidence.

Note that the BIC value \emph{decreases} as the maximum likelihood increases, but \emph{increases} with the number of parameters $P_m$. This creates a natural trade-off: models with more parameters can achieve better fits to the data (lower first term), but are penalized for their increased complexity (higher second term). This automatic model complexity penalization is a manifestation of \emph{Occam's razor}, meaning that if two models explain the data equally well (as measured by the maximum likelihood), the simpler model (with fewer parameters) is preferred because it has the smaller BIC value.

The literature offers a wide variety of related, more or less controlled approximations and model selection criteria, such as the  \textit{Akaike information criterion}~\cite{akaike2003new}. The interested reader is referred to~\cite{Stoica2004} for a broader overview of such methods.

\subsection{Example: Kernel selection for Gaussian processes}\label{sssec:model-selection-GP}
As a first example of Bayesian model selection, we consider comparing several GP surrogate models that differ only in their choice of covariance kernel. But why does the kernel choice matter? Different kernels encode different prior assumptions about the underlying function: smoothness (squared exponential), controlled roughness (Matérn), periodicity (periodic kernels), multi-scale behavior (rational quadratic), or additive/compositional structure. Bayesian model selection provides a principled framework to let the data inform this choice. Each kernel $k_m(\cdot,\cdot ; \bm{\theta}_m)$, together with its hyperparameters $\bm{\theta}_m$, defines a distinct model hypothesis $\mathcal{M}_m$.

Let us first revisit the GP framework as introduced in Section~\ref{ssec:GP}. For dataset $\mathcal{D} = (\Xset,\bm{y})$, assuming a zero-mean GP prior with kernel $k_m(\cdot,\cdot;\bm{\theta}_m)$ and i.i.d. Gaussian observation noise $\eta^{(n)}\sim \mathcal{N}(0,\sigma^2)$, the likelihood for the $m$-th model takes the multivariate Gaussian form
\begin{equation}
    \bm{y} \mid \Xset, \bm{\vv\theta}_m, \mathcal{M}_m
    \sim
    \mathcal{N}\Big(
        \mathbf{0},
        K_m(\bm{\theta}_m) + \sigma^2 I
    \Big)\;,
\end{equation}
where $K_m(\bm{\theta}_m)$ is the Gram matrix with entries $[K_m]_{ij} = k_m(\mathbf{x}^{(i)},\mathbf{x}^{(j)};\bm{\theta}_m)$. Inserting this into Eq.~\eqref{eq:marginal-likelihood-for-model}, the model evidence becomes
\begin{align}
\label{eq:gp-evidence}
    p(\mathcal{D}\mid\mathcal{M}_m)
    & = \int
    \frac{\exp\Big(
        -\tfrac{1}{2}\,\bm{y}^{\rm T}
        (K_m(\bm{\theta}_m) + \sigma^2 I)^{-1}\bm{y}
    \Big)}
    {\sqrt{(2\pi)^N \big|K_m(\bm{\theta}_m) + \sigma^2 I\big|}}\nonumber\\
    & \hphantom{=}\,\, \times p(\bm{\theta}_m) \; \mathrm{d}\bm{\theta}_m\;.
\end{align}
Crucially, the choice of kernel $k_m(\cdot,\cdot;\bm{\theta}_m)$ determines the entire GP model hypothesis $\mathcal{M}_m$. Different kernels yield different Gram matrices $K_m(\bm{\theta}_m)$, which in turn affect both the data-fit term $\bm{y}^{\rm T} (K_m+\sigma^2 I)^{-1}\bm{y}$, measuring how well the data conform to the GP model's assumed covariance structure, and the complexity penalty $\lvert K_m+\sigma^2 I\rvert$, which penalizes overly flexible models that can accommodate a wide range of functions.

Computing the integral in Eq.~\eqref{eq:gp-evidence} exactly rarely admits a closed-form solution and is often intractable. A common approximation is to assume that the posterior over hyperparameters is sharply peaked at its MAP estimate, i.e., $p(\bm{\theta}_m \mid \mathcal{D}, \mathcal{M}_m) \approx \delta(\bm{\theta}_m - \bm{\theta}_m^\ast)$,
where
\begin{equation}
    \bm{\theta}_m^\ast = \arg\max_{\bm{\theta}_m} p(\bm{\theta}_m \mid \mathcal{D}, \mathcal{M}_m)\;.
\end{equation}
Under this approximation, the model evidence becomes
\begin{equation}
    p(\mathcal{D}\mid\mathcal{M}_m)
    \approx
    \frac{\exp\big(
        -\tfrac{1}{2}\bm{y}^{\rm T}
        (K_m(\bm{\theta}_m^\ast)+\sigma^2 I)^{-1}\bm{y}
    \big)}
    {\sqrt{(2\pi)^N
            \lvert K_m(\bm{\theta}_m^\ast)+\sigma^2 I\rvert}}
    p(\bm{\theta}_m^\ast)\;,
\end{equation}
Taking the logarithm and assuming a flat prior on $\bm{\theta}_m$, we can reduce this to the log marginal likelihood in Eq.~\eqref{eq:GP-logmarginallikelihood} --- precisely the GP hyperparameter optimization objective from Section~\ref{ssec:GP}. 

This derivation explains why the log marginal likelihood serves as a legitimate criterion for GP model selection and clarifies the underlying approximation: we evaluate at the MAP hyperparameter estimate rather than integrating over all possible values. Kernel comparison thus reduces to computing Eq.~\eqref{eq:GP-logmarginallikelihood} for each kernel following hyperparameter optimization and forming their ratio as the Bayes factor. In contrast, it would be unclear how to generalize ad-hoc criteria such as least squares fit or loss functions (cf. Sections~\ref{ssec:PCE},~\ref{sec:example1} and~\ref{sssec:neural-networks}) to GP model comparison without Bayesian probability theory as a foundation.

In practice, however, the posterior over hyperparameters is often \emph{not} sharply peaked and may even multi-modal, limiting the validity of the MAP approximation. For low-dimensional hyperparameter spaces, the integral in Eq.~\eqref{eq:gp-evidence} can sometimes be evaluated numerically. This becomes infeasible for high-dimensional parameter spaces, as encountered in NNs. In such cases, a common approach is to approximate the evidence locally near the MAP point using a Laplace approximation or related variants~\cite{javid2020compromise, wilson2020bayesian}.

Due to these difficulties, many heuristic or approximate model–selection criteria exist, such as cross-validation~\cite{stone1974cross, stone1977asymptotic}, various \emph{information criteria}~\cite{hastie2009elements}, or other approximations of the model evidence. In the following examples, we will examine one such \emph{controlled} approximation in more detail.

\subsection{Example: Analytical model evidence for polynomial chaos expansions}\label{sec:model-selection-PCE}
How can a flat prior on the PCE coefficients penalize model complexity at all? For certain classes of PCEs the marginal likelihood (cf. Eq.~\eqref{eq:marginal-likelihood-for-model}) admits a closed form, which makes them an ideal showcase for how Bayesian model selection trades model complexity against parsimony. We construct the evidence term by term, isolate the factors that act as an effective prior on complexity, and then demonstrate the mechanism on a nested sequence of polynomials.

As shown in~\cite{ranftl2021bayesian}, the model evidence for a PCE $\mathcal{M}_m : \mathbb{R}^{D} \to \mathbb{R}$ --- or, more generally, for a linear surrogate model as in Eq.~\eqref{eq:def-expansion} --- admits a closed-form expression under three assumptions. First, the likelihood is Gaussian with known or unknown variance, equivalent to a Student-$t$ marginal likelihood. Second, the prior on the expansion coefficients is flat. Third, the basis functions are fixed, and only their number $P_m$ changes with model index $m$.

Before stating the evidence, we introduce the notation. The design matrix $M_m$ collects the basis functions evaluated at the data, with entries $[M_m]_{i,q} = \phi_q(\mathbf{x}^{(i)})$, and $H_m = M_m^{\mathrm{T}} M_m$ is the associated Gram matrix (cf. Section~\ref{ssec:PCE}). The minimal residual sum of squares is $\chi_m^2 = \mathbf{y}^{\mathrm{T}}(I - M_m H_m^{-1} M_m^{\mathrm{T}})\mathbf{y}$, where the term in the brackets is the orthogonal projector onto the complement of the column space of $M_m$. This $\chi_m^2$ coincides with the value obtained from either the posterior mean or the maximum likelihood estimate of the coefficients, and $N$ denotes the number of data points.

With known variance, the matter boils down to a simple marginalization of the coefficients $c_p$, a Gaussian integral to be solved by completing the square. In surrogate modeling, however, the variance of the surrogate is \emph{not} known \emph{a priori}. Indeed, that seems a rather incredulous assumption!
With unknown variance, one must additionally marginalize over the unknown variance. Placing Jeffreys' prior on the scale, $p(\sigma) \propto 1/\sigma$, the model evidence becomes
\begin{align}\label{eq:evidence-PCE}
    p(\mathcal{D}\mid \mathcal{M}_m)
    & = \Omega_{P_m}\;
      |H_m|^{-\frac{1}{2}}
      \big(\chi_m^2\big)^{-\frac{N-P_m}{2}} \nonumber\\
      & \hphantom{=}\,\, \times 
      \frac{\Gamma\left(\frac{P_m}{2}\right)
        \Gamma\left(\frac{N-P_m}{2}\right)}
        {\Gamma\left(\frac{N}{2}\right)}\;,
\end{align}
where $P_m$ denotes the number of PCE basis functions in model $\mathcal{M}_m$, $\Gamma(\cdot)$ is the Gamma function, and $\Omega_{P_m} = 2\pi^{P_m/2}/\Gamma(P_m/2)$ is the solid angle of the unit sphere in $P_m$ dimensions.

The factors in Eq.~\eqref{eq:evidence-PCE} split into two groups that play different roles. The \emph{data-dependent} factors $|H_m|^{-1/2}$ and $\big(\chi_m^2\big)^{-(N-P_m)/2}$ reward a model that fits the data and occupies a favorable region of parameter space. The \emph{data-independent} factors, namely the solid angle $\Omega_{P_m}$ and the two Gamma functions $\Gamma\left(P_m/2\right)$ and $\Gamma\left((N-P_m)/2\right)$, depend only on the counts $N$ and $P_m$. Keeping these two groups apart is the key to the analysis, since the data-independent group alone carries the automatic complexity penalty we isolate below.

\begin{figure*}
    \centering
    \includegraphics[width=\linewidth]{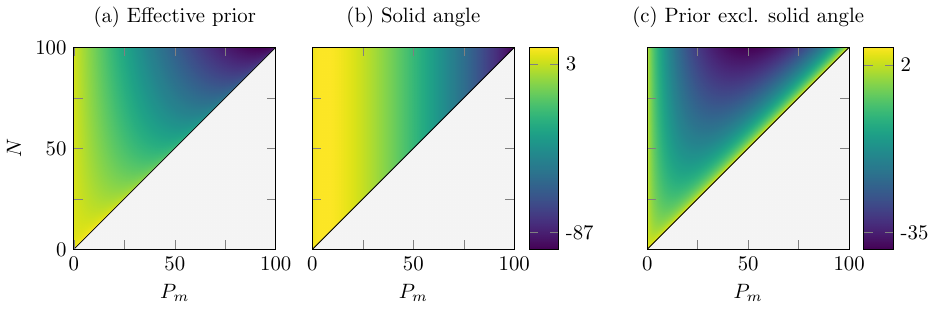}
    \caption{
    Illustration of the (a) effective prior structure as a function of the number of observations $N$ and surrogate parameters $P_m$, as introduced in Eq.~\eqref{eq:evidence-PCE}, showing the combined influence of solid angle and parameter prior. (b) Solid angle effect demonstrating geometric constraints from the likelihood. (c) Intrinsic prior term excluding the solid angle contribution. The boundary $N = P_m$ (black line) separates valid from invalid parameter regions. Colormap in (b) also applied to (a), y-label in (a) also applies to (b,c).
    }
    \label{fig-example_3}
\end{figure*}

This answers the opening question. Even with a flat prior on the coefficients, the evidence integral induces a preference for simpler models through its geometric factors, so these terms act as an \emph{effective prior} on model complexity. The effect depends on both the number of observations $N$ and the model dimension $P_m$, as visualized in Fig.~\ref{fig-example_3}. Figure~\ref{fig-example_3}(b) shows that the solid angle contribution alone biases strongly toward low-dimensional models, independently of any explicitly chosen prior.

Consider a nested sequence of polynomial models in one input dimension, $\bm{x}\in\mathbb{R}$ \textcolor{red}{or $\bm{x}\coloneqq x \in\mathbb{R}$?}. Each model adds one polynomial to the previous basis,
\begin{gather}
    \phi_1 = \{1\}\;,\quad
    \phi_2 = \{1,x\}, \quad \phi_3 = \{1,x,x^2\}\;, \nonumber\\
    \dots\;,\quad
    \phi_{P_m} = \{1,x,\dots,x^{P_m-1}\}\;,
\end{gather}
so that $\mathcal{M}_m \subset \mathcal{M}_{m+1}$ for all $m$.

To illustrate the automatic complexity penalization in Eq.~\eqref{eq:evidence-PCE}, consider its data-independent factors, the solid angle $\Omega_{P_m}$ and the two Gamma functions $\Gamma(P_m/2)$ and $\Gamma((N-P_m)/2)$. We advance along a coarse subsequence of the nested hierarchy in steps of two, $P_{m+1}=P_m+2$.\footnote{The even step ensures that for even $N$ the half-integer arguments of the Gamma functions in Eq.~\eqref{eq:evidence-PCE} collapse to factorials, so the evidence ratio takes an elementary form. A unit step leaves half-integer Gamma functions, which remain computable but obscure the mechanism this example isolates.}

Consider first the solid angle alone, isolated in Fig.~\ref{fig-example_3}(b). For the step $P_m\to P_m+2$ it contributes
\begin{equation}\label{eq:solid-angle-ratio}
    \frac{\Omega_{P_m+2}}{\Omega_{P_m}}
    = \pi\frac{\Gamma(\frac{P_m}{2})}{\Gamma(\frac{P_m}{2+1})}
    = \frac{2\pi}{P_m}\;,
\end{equation}
where we used $\Gamma(P_m/2+1)=(P_m/2)\,\Gamma(P_m/2)$. Its logarithm scales as $-\log P_m$, a geometric bias toward low-dimensional models that persists even under the flat coefficient prior.

Including the Gamma functions changes this scaling. Their ratio contributes
\begin{equation}\label{eq:gamma-ratio}
    \frac{\Gamma(\frac{P_{m+1}}{2})\Gamma\big(\frac{N-P_{m+1}}{2}\big)}
         {\Gamma(\frac{P_{m}}{2})\Gamma\big(\frac{N-P_{m}}{2}\big)}
    = \frac{P_m}{N-P_m-2}\;,
\end{equation}
whose factor $P_m$ cancels the one from the solid angle. The full data-independent prefactor is therefore
\begin{equation}\label{eq:prefactor}
    \frac{\Omega_{P_m+2}}{\Omega_{P_m}}\frac{P_m}{N-P_m-2} = \frac{2\pi}{N-P_m-2}\;.
\end{equation}
Reinstating the data-dependent factors from Eq.~\eqref{eq:evidence-PCE}, namely $\sqrt{|H_m|/|H_{m+1}|}$ and the $\chi^2$ terms, gives the full evidence ratio
\begin{align}\label{eq:evidence-ratio-PCE}
    \frac{\p{\mathcal{M}_{m+1}}{\mathcal{D}}}{\p{\mathcal{M}_{m}}{\mathcal{D}}} & = \frac{2\pi}{N-P_m-2}\sqrt{\frac{|H_m|}{|H_{m+1}|}}\nonumber\\
    &\hphantom{=}\,\,\times\biggl(\frac{\chi_m^2}{\chi_{m+1}^2}\biggr)^{(N-P_m)/2}\chi_{m+1}^2\;.
\end{align}
Keeping only this prefactor and dropping the fit terms $H$ and $\chi^2$, the logarithmic contribution scales as
\begin{align}
    \log \frac{\p{\mathcal{M}_{m+1}}{\mathcal{D}}}{\p{\mathcal{M}_{m}}{\mathcal{D}}}
    \stackrel{\propto}{\sim} -\log\!\big(N-P_m-2\big)\;.
\end{align}
In other words, moving to a higher polynomial order multiplies the posterior odds by the factor $2\pi/(N-P_m-2)$. So, the more complex model is penalized unless the extra parameters improve the fit at least by the same factor. This penalty grows only slowly, like $\log(N-P_m-2)$, with the number of observations $N$, and it does not depend on how well either model fits the data. Over a whole sequence of models these factors multiply, so the penalty adds up as the models grow more complex. 
This fundamental effect again illustrates \emph{Occam's razor}: If two models explain the data equally well, the simpler model is more likely to be true.

The same complexity penalty appears when we replace the exact evidence with its BIC approximation. Following the reasoning in the previous example, we obtain
\begin{align}
\frac{p(\mathcal{M}_{m+1}\mid\mathcal{D})}
     {p(\mathcal{M}_m\mid\mathcal{D})}
& \approx
\underbrace{
\frac{p(\mathcal{D}\mid\boldsymbol{\theta}^\ast_{m+1},\mathcal{M}_{m+1})}
     {p(\mathcal{D}\mid\boldsymbol{\theta}^\ast_m,\mathcal{M}_m)}
}_{\ge 1} \nonumber \\
& \hphantom{=}\,\, \times \underbrace{
\left(\frac{\Delta_m}{\Lambda_m}\right)^{P_m}
\left(\frac{\Lambda_m}{\Delta_m}\right)^{P_{m+1}}
}_{\frac{\Lambda_m}{\Delta_m}\gg 1}\;.
\end{align}
where the likelihood ratio favors the more complex model if it fits the data better, while the Occam factor penalizes the additional parameters.
Here, $\boldsymbol{\theta}^\ast_m$ and $\boldsymbol{\theta}^\ast_{m+1}$ denote the maximum likelihood estimates for models $\mathcal{M}_m$ and $\mathcal{M}_{m+1}$. The quantity $\Delta_m$ is the typical width of the likelihood around the MLE, the region where $p(\mathcal{D}\mid\bm{\theta},\mathcal{M})$ is non-negligible, which scales as $\Delta_m \sim \sigma/\sqrt{N}$ for Gaussian-like likelihoods. The characteristic width is $\Lambda_m$, assumed independent of $N$, so the ratio $\Lambda_m/\Delta_m$ quantifies how much the data concentrates the posterior relative to the prior. Since $\Lambda_m/\Delta_m \gg 1$ for sufficiently large $N$, the Occam factor $(\Lambda_m/\Delta_m)^{P_{m+1}-P_m}$ grows exponentially with the difference in model dimensions, penalizing additional parameters unless they substantially improve the likelihood.

For nested models the likelihood ratio is $\ge 1$, since $\mathcal{M}_{m+1}$ reproduces at least the same fit as $\mathcal{M}_m$ by an appropriate choice of parameters. The Occam factor $(\Lambda_m/\Delta_m)^{P_{m+1}-P_m}$ nonetheless penalizes the added complexity, and because $\Lambda_m/\Delta_m \sim \sqrt{N}$ this penalty grows with the number of data points, making additional parameters increasingly hard to justify without a substantial improvement in fit.

Three remarks close the example. First, the result requires no orthogonality of the basis functions, so the derivation applies to any surrogate model that is linear in its parameters. Second, Eq.~\eqref{eq:evidence-PCE} is exact, whereas the BIC provides only a crude approximation. Third, for multi-output PCE with $\mathcal{M}_m: \mathbb{R}^{D}\to \mathbb{R}^{Q}$ the evidence admits a closed-form expression of similar form when the outputs are uncorrelated, for which we refer to~\cite{ranftl2021bayesian}.

This trade-off embodies \emph{Occam's razor}. Simpler models are preferred unless the data support additional complexity. The posterior odds balance parsimony against explanatory power.

\section{Bayesian experimental design}\label{sec:experimental-design}
We can formulate the core objective of Bayesian experimental design \cite{chaloner1995bayesian,dasgupta199629,Lindley-review} as:
\begin{quotation}
    \textit{Selection of next experimental or simulation conditions that maximize the expected information gain about uncertain model parameters or predictions, thereby enabling more efficient learning from limited data.}
\end{quotation}

\subsection{General concept}
Until now we assumed that the dataset is fixed and given. In many realistic settings, however, data collection is costly, and we may wish to actively select new data points in order to learn as efficiently as possible. This situation arises frequently in mechanics, where obtaining new data may involve running a computationally expensive simulation or performing a labor-intensive experiment. In such cases, we want to use all available information to decide where to sample next --- that is, which experiment or simulation with a particular parameter set will be most informative.

Formally, given a dataset obtained from computer simulations --- see the workflow introduced in Fig.~\ref{fig-workflow} --- we define
\begin{equation}
    \mathcal{D} = \big(\bm{x}^{(n)}, y^{(n)}\big)_{n=1}^N\;,
\end{equation}
we seek the next optimal input parameter set $\bm{x}^{(N+1)}$ at which to query new observations $y^{(N+1)}$. For notational simplicity, we set $y_\ast \equiv y^{(N+1)}$ and $\bm{x}_\ast \equiv \bm{x}^{(N+1)}$, analogous to the notation for new data in GPs (cf. Section~\ref{ssec:GP}).

To make a principled decision about the next query, we introduce the notion of \emph{utility}. A utility function $U(y_\ast, \mathcal{D})$ quantifies the benefit of observing a new output data point $y_\ast$ at design location $\bm{x}_\ast$, given the current dataset $\mathcal{D}$. The choice of utility depends on the modeling objective: it could represent the expected reduction in uncertainty, the improvement in model accuracy, or the information gained about model parameters.

In Bayesian experimental design, we select the \emph{optimal} next query point $\bm{x}^\ast$ --- where, as in previous sections, a subscript asterisk denotes a new point and a superscript asterisk denotes the optimal point --- by maximizing the \emph{expected utility},
\begin{align}\label{eq:def-bayesian-experimental-design}
    \bm{x}^{\ast} &= \arg \max_{\bm{x}_\ast} \mathbb{E}\big[U(y_\ast,\mathcal{D})\big]\;,
\end{align}
where 
\begin{align}
     \mathbb{E}\big[U(y_\ast,\mathcal{D})\big] & = \int U(y_\ast, \mathcal{D}) p(y_\ast \mid \bm{x}_\ast, \mathcal{D}) \: \mathrm{d}y_\ast \nonumber\\ 
     & = \int U(y_\ast, \mathcal{D}) \int  p(y_\ast \mid \bm{x}_\ast, \mathcal{D}, \bm{\theta}) \nonumber\\
     & \hphantom{=}\,\, \times p(\bm{\theta} \mid \mathcal{D}) \; \mathrm{d}\bm{\theta} \: \mathrm{d}y_\ast\;.
\end{align}
Here, $\bm{\theta}$ denotes the parameters of the probabilistic surrogate model. The existing dataset obtained from the computer simulation is used to train this surrogate, which provides a probabilistic approximation of the mapping $\bm{x} \mapsto y$. Consequently, $\bm{\theta}$ represents the surrogate model parameters, and their posterior distribution $p(\bm{\theta} \mid \mathcal{D})$ quantifies the current uncertainty in that approximation.

This expression states that the expected utility is the average usefulness of observing a new data point at $\bm{x}_\ast$, taken over all possible outcomes $y_\ast$ that could occur. The inner integral marginalizes over the current uncertainty in the model parameters $\bm{\theta}$ and the outer integral averages over all potential new observations $y_\ast$, weighted by their predictive likelihood $p(y_\ast \mid \mathcal{D})$. Thus, the expected utility formalizes the benefit of sampling at a proposed design point, accounting for both parameter uncertainty and observation uncertainty.

A popular choice for the utility function is the \emph{information gain}, measured by the Kullback–Leibler (KL) divergence~\cite{mackay1992information,Loredo2004},
\begin{align}
U(y_\ast, \mathcal{D}) & = \int p(\bm{\theta}\mid\mathcal{D},y_\ast) \nonumber\\
& \hphantom{=}\,\, \times \log \Bigg[ \frac{p( \bm{\theta} \mid \mathcal{D},y_\ast)}{p( \bm{\theta} \mid \mathcal{D})} \Bigg] \: \mathrm{d} \bm{\theta}\;,
\end{align} 
which quantifies how much the posterior would change after observing $y_\ast$, relative to either the prior or the posterior at iteration $N$. Intuitively, a larger KL divergence indicates that the potential observation is more \emph{informative} about $\bm{\theta}$. Other utility measures exist, such as Wasserstein distances~\cite{marzouk2025bayesian} or the cross-entropy~\cite{Loredo2004}; see, e.g.,~\cite{chaloner1995bayesian,dasgupta199629,VonToussaint2011} for an overview of utility functions.

Given the KL utility, the corresponding expected utility is
\begin{align}
    \mathbb{E}\big[U(y_\ast,\mathcal{D})\big]
    & = \iint p(\bm{\theta} \mid \mathcal{D},y_\ast) \log \Bigg[ \frac{p(\bm{\theta} \mid \mathcal{D},y_\ast)}{p(\bm{\theta} \mid \mathcal{D})} \Bigg] \nonumber \\
    & \hphantom{=}\,\, \times \int p(y_\ast \mid \mathcal{D}, \bm{\theta}) \nonumber\\ 
    & \hphantom{=}\,\, \times p(\bm{\theta})\: \mathrm{d}\bm{\theta} \:  \mathrm{d}\bm{\theta} \: \mathrm{d}y_\ast\;.
\end{align}
This expression again highlights that we average the resulting information gain over all possible future outcomes $y_\ast$, weighting each outcome by its predictive likelihood.

Because a future observation cannot retroactively change the prior information about the parameters, practical acquisition strategies often approximate $p(\bm{\theta}\mid \mathcal{D},y_\ast) \approx p(\bm{\theta} \mid \mathcal{D})$. 
In practice, only two ingredients are needed: (i) the current posterior $p(\bm{\theta} \mid \mathcal{D})$ and (ii) the predictive model $p(y_\ast \mid \mathcal{D},\bm{\theta})$. Here, the predictive model $p(y_\ast \mid \mathcal{D},\bm{\theta})$ is the surrogate’s probabilistic prediction for a new output at $\bm{x}_\ast$, conditioned on the data and the surrogate parameters. The structure of these integrals naturally leads to nested posterior-sampling schemes~\cite{VonToussaint2011}, but these are often computationally intractable for complex models. This motivates several practical approximations, ultimately leading to what the machine learning community refers to as \textit{Bayesian optimization}.

In summary, when experiments involve an expensive computer simulation, the workflow proceeds as follows: (i) define a probabilistic model with a prior on the unknown parameters and a likelihood linking the design variables to the simulated data, and choose a utility function; (ii) use a surrogate model to approximate the costly simulation and efficiently evaluate the expected utility; and (iii) maximize this utility to select the next simulation point, run the corresponding simulation, update the posterior, and repeat if using a sequential design. 

Among the notable earlier contributions to this field, Shannon-type expected information gain was employed to develop methodologies for optimal experimental design in algebraic nonlinear problems and combustion kinetic models~\cite{Huan2013a} and further applied to assess experiment relevance under model uncertainty and to identify source locations in impedance tomography~\cite{Long2013a,Beck2018a}.
Further contributions proposed measuring the information gain from a hypothetical experiment as the Kullback–Leibler divergence between the prior and posterior probability distributions of the quantity of interest, demonstrating the methodology on a steel wire manufacturing problem~\cite{Pandita2019a}, with follow-up work introducing sequential design strategies based on non-stationary GPs, benchmarked against uncertainty sampling and expected improvement~\cite{Pandita2021a}.
For mechanical experiments, Eberle-Blick and Hyvönen~\cite{EberleBlick2024a} optimize boundary pressure activations to maximize the informational value of the resulting deformations for reconstructing the material parameters, while others optimize the information content of mechanical experiments --- particularly uniaxial tensile tests --- for the calibration of different nonlinear constitutive models~\cite{Ricciardi2024a} or history-dependent constitutive models~\cite{Bhattacharya2026a}. 
The reader is additionally referred to the general overview of Bayesian experimental design concepts provided by~\cite{Ryan2015a}.

\subsection{Bayesian optimization and active learning}\label{ssec:bayesian-optimization}
In contrast to Bayesian experimental design --- which aims to gain information about uncertain model parameters --- the core objective of Bayesian optimization is:
\begin{quotation}
    \textit{Select the next input that is expected to yield the optimal or best objective value, while simultaneously exploring uncertain regions to enhance our overall model understanding.}
\end{quotation}

Thus, Bayesian optimization is fundamentally an optimization method, not a parameter-inference method.

\subsubsection{General concept}
The transition from experimental design to optimization is often motivated by necessity. As previously discussed, the marginalization integrals required for full Bayesian experimental design are frequently computationally intractable. To make the problem manageable, the expected utility is often approximated using a MAP estimate.

In this context, we apply two key simplifications to the utility calculation:
\begin{itemize}
    \item \textit{Parameter approximation}. 
    Instead of integrating over the full parameter posterior $p(\bm{\theta} \mid \mathcal{D})$, we approximate it by a delta function at its mode
  \begin{align}
        p(\bm{\theta} \mid \mathcal{D}) &\approx \delta(\bm{\theta} - \bm{\theta}^\ast)\;, \nonumber \\
        \bm{\theta}^{\ast} = \arg &\max_{\bm{\theta}} p(\bm{\theta} \mid \mathcal{D})\;,
    \end{align}  
    where $\bm{\theta}^\ast$ are the trained surrogate-model parameters.
    \item \textit{Data approximation}. 
    Rather than integrating over all possible future outcomes, we approximate the likelihood for a new data by a delta function placed at the surrogate prediction
    \begin{equation}
        p(y_\ast \mid \mathcal{D},\bm{\theta}^{\ast}) \approx \delta\big(y_\ast - \hat{f}(\bm{x}; \bm{\theta}^{\ast})\big)\;,
    \end{equation}
    where $\hat{f}(\bm{x};\bm{\theta})$ is the surrogate model trained on the dataset $\mathcal{D}$.
\end{itemize}
Together, these approximations replace the full expected utility (Eq.~\eqref{eq:def-bayesian-experimental-design}) by a point estimate
\begin{align}
    \mathbb{E}\big[U(y_\ast,\mathcal{D})\big] \approx U\big(\hat{f}(\bm{x}; \bm{\theta}^{\ast}), \mathcal D\big)\;.
\end{align}
and the next input is chosen as
\begin{align} \label{eq:BO-def}
    \bm{x}^{\ast} &= \arg \max_{\bm{x}_\ast} {U(\hat{f}(\bm{x}_\ast; \bm{\theta}^{\ast}),\mathcal{D})}\;.
\end{align}

In these MAP approximations, we no longer marginalize with respect to the surrogate parameters $\bm{\theta}$ or the possible new observations $y_\ast$. Instead, we transform the original Bayesian experimental design inference into two simpler optimization tasks:
\begin{itemize}
    \item \textit{Inner optimization}. Train the surrogate model by finding the best-fitting parameter vector $\bm{\theta}^\ast$ to learn how the expensive simulation behaves.
    \item \textit{Outer optimization}. Use the trained surrogate to select the next input $\bm{x}^\ast$ by optimizing the utility function $U$. Note that in Bayesian optimization, this utility function is typically referred to as an acquisition function $\alpha$. This step ultimately determines where the next expensive evaluation should be performed.
\end{itemize}
This separation of inference (inner optimization) and design (outer optimization) makes the problem computationally feasible, while still retaining a probabilistic interpretation through the surrogate’s predictive uncertainty. It also establishes a conceptual bridge between rigorous Bayesian experimental design and its practical approximations in engineering problems widely used in machine learning and scientific modeling. In the machine learning community, this approximate procedure is widely known as Bayesian optimization~\cite{Brochu2010, Shahriari2015, Frazier2018} or, more broadly, active learning. In fact, after these two approximations, relatively little \emph{Bayes} remains in the strict sense\footnote{In common practice, a GP prior is typically placed on $\hat{f}$ within this framework, thereby retaining a notion of uncertainty and probabilistic structure. Although not strictly necessary, GPs are particularly convenient for the inner optimization step: unlike generalized linear models, which may require adding new basis functions as new data are incorporated, a GP preserves its functional form and updates only its posterior when new data are observed.}.

To carry out the two-stage procedure in practice, we require a surrogate model that can both fit the available data (inner optimization) and provide predictions for evaluating the acquisition function (outer optimization). For a black-box objective function $f: \mathcal{X} \to \mathbb{R}$, a GP surrogate is often used
\begin{equation}
    \hat{f}_{\rm GP} \sim \mathrm{GP}(\mu(\bm{x}), k(\bm{x},\bm{x}'))\;,
\end{equation}
as introduced in see Section~\ref{ssec:GP}. Conditioning the GP surrogate model on the observed data $\mathcal{D}$ yields the posterior mean prediction $\mu(\bm{x})$ and the predictive uncertainty through the posterior variance $\sigma^2(\bm{x})$, which provide the data-informed prediction of the surrogate and its associated uncertainty at any new input location. Once the GP posterior mean and variance have been computed, they are plugged directly into the acquisition function.

A typically Bayesian optimization workflow consists of five steps:
\begin{enumerate}[label=\roman*).]
    \item Training the surrogate model on the current dataset $\mathcal{D}$.
    \item Computing the acquisition function $\alpha(\bm{x};\mathcal{D})$.
    \item Selecting the next query point $\bm{x}^\ast = \arg\max_{\bm{x} \in \mathcal{X}} \alpha(\bm{x}; \mathcal{D})$.
    \item Evaluating the expensive objective at $\bm{x}^\ast$.
    \item Updating the dataset and repeating. 
\end{enumerate}
The practical application of these steps is demonstrated through an illustrative example in Section~\ref{sec:example4}.

Bayesian optimization enjoys several attractive properties. Importantly, it is known to exhibit global convergence even in non-convex problems~\cite{kawaguchi2015bayesian}. Even more importantly, it does not necessitate a fast method to evaluate the gradient of the objective function to solve the optimization problem (Eq.~\eqref{eq:BO-def}); instead it may (or may not) utilize the gradient of the surrogate.

The literature offers a wide range of acquisition functions, each reflecting a different trade-off between exploration (sampling where uncertainty is high) and exploitation (sampling where improvement is expected). One of the simplest is the upper confidence bound (UCB) criterion
\begin{align}\label{eq:UCB}
    \mathrm{UCB}\big(\hat{f}(\bm{x}, \bm{\theta}^{\ast}), \mathcal D\big) := \mu(\bm{x}) + \beta \sqrt{\sigma^2(\bm{x})}\;,
\end{align}
where $\beta$ is a user-chosen tuning parameter controlling the exploration–exploitation balance: larger values of $\beta$ emphasize exploration by favoring regions with high predictive uncertainty, while smaller values focus on exploitation of areas with high predicted utility; and $\mu(\bm{x})$ and $\sigma^2(\bm{x})$ denote the posterior mean and variance of the GP (cf. Section~\ref{ssec:GP}).

Beyond UCB, many other acquisition functions have been proposed. The probability of improvement (PI) criterion~\cite{jones1998efficient} selects points with the highest probability of exceeding the current best observation, whereas the expected improvement (EI) criterion~\cite{movckus1974bayesian, jones1998efficient} considers the expected magnitude of that improvement. Information-theoretic approaches include the maximum entropy principle~\cite{Loredo2004}, which seeks points that maximally reduce global uncertainty, and more refined methods such as predictive entropy search (PES)~\cite{hernandez2014predictive} and max-value entropy search (MES)~\cite{wang2017max}, which explicitly target information gain about the global optimum. Recently, variance-based criteria such as the global variance approach~\cite{Preuss2021} and robust Bayesian optimization formulations~\cite{hoffer2023robust} have gained traction for addressing noisy or stochastic environments. The choice of the acquisition function depends on the application. In mechanics, where experiments and simulations are typically costly and noisy, exploration-oriented approaches (such as UCB or PES) may be advantageous.

Generally, Bayesian optimization is particularly well suited for applications in mechanics and material design, where experiments or simulations are costly. Conveniently, many modern deep-learning frameworks --- such as \emph{TensorFlow}, \emph{PyTorch}, and \emph{JAX} --- either include or interface directly with Bayesian optimization libraries~\cite{balandat2020botorch}. Moreover, Bayesian optimization can be used to tune the relative weighting of data-loss and physics-loss terms in Eq.~\eqref{eq:def-pinn-lossfunction} or to optimize architectural hyperparameters such as the number of layers or neurons.

Bayesian optimization has been widely adopted across various domains. Typical examples include inverse design of material compositions via alternative acquisition functions~\cite{Tian2025a,Kansara2025a,Rassloff2025a,Chiappetta2026a} or through evolutionary Monte Carlo sampling~\cite{Vangelatos2021a}. It is also used to infer constitutive parameters for material models~\cite{Borowska2022a,MirandaValdez2025a} and tune hyperparameters for NNs~\cite{Snoek2012} and PINNs~\cite{Zhong2026a,Dongxu2026}. More recently, researchers have integrated domain-specific physics into the surrogate model itself; for instance, incorporating physics knowledge as basis functions in a semi-parametric GP has been shown to reduce required evaluations when optimizing numerical parameters in constrained muscle activation simulations~\cite{Huber2024a}.

\subsection{Example: Inverse design for a 3D-printed composite material based on uniaxial extension tests}\label{sec:example4}
We consider a 3D-printed soft material composed of a hydrogel matrix reinforced with dispersed discrete polymeric fibers. The printing process allows control of the fiber dispersion via a scalar design parameter $\bm{\xi} \in \Xi$. The objective is to determine the value of $\bm{\xi} = \kappa$ such that the material attains a prescribed, or optimal, tensile stress value $y_{\mathrm{exp}}^\ast$ at a specific strain.

The available experimental dataset consists of $N$ input--output pairs
\begin{equation}
    \mathcal{D}_{\rm exp} = \big(\kappa_i, y_{\mathrm{exp},i}\big)_{i=1}^{N}\;,
\end{equation}
where $y_{\mathrm{exp},i}$ denotes the measured tensile stress at a certain strain corresponding to the design parameter $\kappa_i$. Due to the costly manufacturing and testing procedure, only $N=2$ initial structures are available.
By assuming that the experimental observations are normally distributed, we can describe them as follows
\begin{equation}
    y_{\mathrm{exp},i} = f(\kappa_i) + \eta\;, \qquad \eta \sim \mathcal{N}(0,\sigma^2)\;,
\end{equation}
where $\sigma$ denotes the measurement variance in this example.

Given the unknown and potentially nonlinear relationship between the design parameter and the tensile stress, a GP surrogate is employed, offering UQ in the small-data regime.
The latent stress–dispersion mapping is modeled as
\begin{equation}
    f(\kappa) \sim \mathrm{GP}\big(0, k(\kappa,\kappa';\boldsymbol{\theta})\big)\;,
\end{equation}
where $k(\kappa,\kappa';\bm{\theta})$ is given in this example by a squared-exponential covariance kernel with hyperparameters
\begin{equation}
    \boldsymbol{\theta} = (\sigma^2_{f}, \iota,\sigma)^{\rm T}\;,
\end{equation}
representing the prior variance magnitude $\sigma^2_{f}$, the correlation length $\iota$, and the variance $\sigma^2$; see Section~\ref{ssec:GP}.
Instead of estimating $\bm{\theta}$ via marginal likelihood maximization, a fully Bayesian approach is adopted in which the hyperparameters are treated as random variables with weakly informative priors. 

The prior scales are chosen based on the empirical scale of the data. The experimental stresses are standardized to zero mean and unit variance, such that the characteristic signal magnitude is of order one. Accordingly, the GP amplitude and noise parameters are assigned weakly informative priors $\sigma_f \sim \mathrm{HalfNormal}(1)$ and $\sigma \sim \mathrm{HalfNormal}(0.3)$, which cover the plausible range of standardized signal and noise magnitudes while assigning vanishing mass to implausibly large values. The length-scale is modeled as $\iota \sim \mathrm{LogNormal}(\mu_\iota, 1)$, where $\mu_\iota$ is chosen such that the prior median corresponds to approximately half of the explored design-parameter range, reflecting smooth stress variations over physically meaningful scales; the resulting prior remains broad enough to let the data determine the precise value.

Given the experimental dataset $\mathcal{D}_{\rm exp}$, the posterior of the hyperparameters is given by
\begin{equation}
    p(\bm{\theta} \mid \mathcal{D}_{\rm exp}) \propto p(\mathcal{D}_{\rm exp} \mid \bm{\theta}) p(\bm{\theta})\;,
\end{equation}
which is inferred using a MAP estimate\footnote{In practice, hyperparameter posteriors are often multimodal with modes existing at different scales. While the ideal Bayesian approach involves using MCMC samplers to produce a collection of posterior samples, capturing all modes accurately can be difficult and computationally expensive. Consequently, a practical choice is often to use the MAP estimate or related methods such as MLE.}.

By approximating the hyperparameter posterior with a delta function at its mode $\bm{\theta}^\ast$, for any design parameter in the design space $\Xi$, the predictive distribution simplifies as follows
\begin{align}
    p(y_\ast \mid \kappa^\ast,\mathcal{D}_{\rm exp}) & = \int 
    p(y_\ast \mid \mathcal{D}_{\rm exp}, \bm{\theta}) \nonumber \\
    & \hphantom{=}\,\,\times p(\boldsymbol{\theta} \mid \mathcal{D}_{\rm exp})
    \, \mathrm{d}\bm{\theta} \nonumber \\
    & \approx  p(y_\ast \mid \mathcal{D}_{\rm exp}, \bm{\theta}^\ast)\;,
\end{align}
effectively eliminating the need for marginalization over $\bm{\theta}$. This approximation yields the predictive mean $\mu_{\rm GP}(\kappa)$ and predictive variance $\sigma_{\rm GP}^2(\kappa)$ directly from the GP conditioned on the fixed hyperparameters. Consequently, the GP provides both a stress prediction and a formal uncertainty estimate for any admissible design parameter.

The inverse design problem consists of identifying the optimal design $\kappa^\ast$ such that
\begin{equation}
    y_{\mathrm{exp}}(\kappa^\ast) \approx y_{\mathrm{exp}}^\ast\;.
\end{equation}

To minimize the number of additional costly experiments, a sequential Bayesian optimization strategy is employed. Bayesian optimization combines the GP surrogate with an acquisition function $\alpha$ that determines where to evaluate the next experiment.

An acquisition function, derived from probabilistic considerations, is implemented using the UCB criterion (Eq.~\eqref{eq:UCB}). To suit the target-matching task, the criterion is adapted by defining exploitation as the minimization of the distance to a target value $y^\ast$
\begin{equation}
    \mathrm{UCB}_t(\kappa)
    =
    - \bigl(\mu_{{\rm GP},t}(\kappa) - y_{\rm exp}^\ast\bigr)^2 + \beta \sqrt{\sigma_{{\rm GP},t}^2(\kappa)}\;,
\end{equation}
where $\beta > 0$ controls the exploration–exploitation trade-off. The first term promotes exploitation by minimizing the squared distance to the prescribed target value $y^\ast$, while the second term encourages exploration in regions of high predictive uncertainty.\\
\begin{figure*}[t!]
\centering
\includegraphics[width=\linewidth]{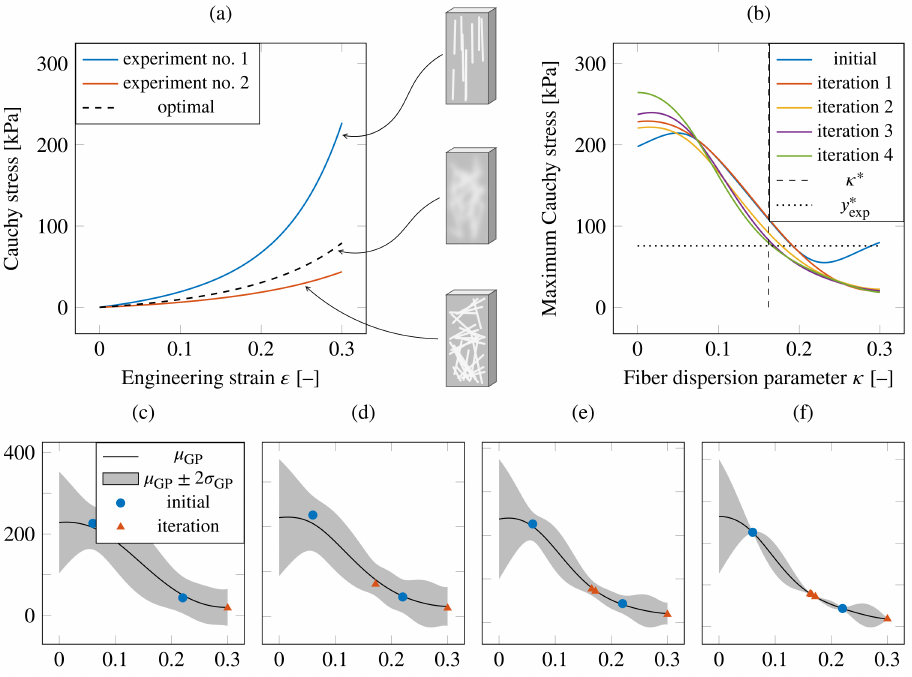}
\caption{
(a) Initial experimental synthetic stress-strain data under uniaxial tension for the virtual 3D-printed composite material. The microstructural representations of selected designs are shown alongside the optimal response predicted using Bayesian optimization. (b) Maximum Cauchy stress at the selected strain level ($\varepsilon = 0.3$) as a function of the design parameter describing fiber dispersion for four Bayesian optimization iterations, illustrating progressive improvement of the Gaussian process (GP) regression models as new data are incorporated; meanwhile, (c–f) present the individual GP regression models for the initial data and iterations, demonstrating the reduction in variance and corresponding predictive uncertainty. (model parameters: $\beta = 0.8$, $\sigma = 0.35$)
}
\label{fig-example_4}
\end{figure*}

\textit{Step 1: Surrogate model.} At iteration $t$, the Bayesian GP surrogate is refitted using the current dataset $\mathcal{D}_{{\rm exp},t}$. The hyperparameters $\bm{\theta}^\ast$ are determined via a MAP estimate, identifying the most probable model configuration. Using these fixed hyperparameters, the predictive mean $\mu_t(\kappa)$ and predictive variance $\sigma_t^2(\kappa)$ are evaluated for any $\kappa \in \Xi$ without the need for numerical marginalization.\\

\textit{Step 2.} The acquisition function is evaluated on a dense uniform discretization of the admissible interval $\Xi$, obtained by subdividing the interval into fine equidistant points. For each candidate value $\kappa$ on this grid, the predictive mean $\mu_{{\rm GP},t}(\kappa)$ and variance $\sigma_{{\rm GP},t}(\kappa)$ are computed and used to evaluate $\mathrm{UCB}_t(\xi)$\footnote{Notably, more advanced global optimization methods can be used to find the maximum of the acquisition function more efficiently, particularly in high-dimensional spaces. Since the GP surrogate is smooth and differentiable, gradient-based optimization can be employed even when gradients of the original experiment or simulation are unavailable.}.\\

\textit{Step 3.} The next candidate design parameter is selected as
$ \kappa_{t+1} = \arg\max_{\kappa \in \Xi} \mathrm{UCB}_t(\kappa)\;.$
Accordingly, the next evaluation point is obtained purely deterministically as the maximizer of the acquisition function.\\

\textit{Step 4.} At the selected design parameter $\kappa_{t+1}$, a new tension experiment is performed to obtain $y_{\mathrm{exp},t+1}$. Alternatively, when experimental testing is impractical, a validated high-fidelity computational model capable of reproducing the tensile test may be used to provide the corresponding observation.\\

\textit{Step 5.} The dataset is updated as
$ \mathcal{D}_{{\rm exp},t+1} = \mathcal{D}_{{\rm exp},t} \cup (\kappa_{t+1}, y_{\mathrm{exp},t+1}),$
after which the GP surrogate is retrained using the augmented dataset, and the Bayesian optimization cycle is repeated from the surrogate update step.

\bigbreak
The algorithm terminates once the experimentally validated mismatch becomes sufficiently small, i.e.,
\begin{equation}
\left|
y_{\mathrm{exp},t+1}
-
y_{\mathrm{exp}}^\ast
\right| / y_{\mathrm{exp}}^\ast
\leq 5\%\;,
\end{equation}
or other predefined tolerance criteria are met.

For the present example, we define the target response as $y_{\mathrm{exp}}^\ast = 19.00\,\mathrm{kPa}$ at $\varepsilon=0.3$. As detailed in Table~\ref{tab-example_4} and illustrated in Figure~\ref{fig-example_4}(a), convergence was achieved after four additional Bayesian optimization iterations, resulting in an optimal parameter $\kappa^\ast = 0.162$ and a corresponding response of $y_{\mathrm{exp}}(\kappa^\ast) = 78.65\,\mathrm{kPa}$. 
Figures~\ref{fig-example_4}(b--d) display the experimental data alongside the evolving GP surrogate. These plots highlight the newly sampled data points and the resulting updates to the GP mean and predictive variance throughout the optimization process.
%
%
\begin{table}[ht]
\centering
\caption{Initial experimental datasets ($N=2$) and iterations of the Bayesian optimization.}
\label{tab-example_4}
\begin{tabular}{cccc}
\toprule
Experiment no. & $\xi$ & $y_{\mathrm{exp}}$ [kPa] & Iterations \\
\midrule
1 & 0.060 & 226.98 &  \\
2 & 0.220 & 43.75 &  \\
\midrule
3 & 0.300 & 18.44 & 1 \\
4 & 0.171 & 71.70 & 2 \\
5 & 0.165 & 76.76 & 3 \\
6 & 0.162 & 78.65 & 4 \\
\bottomrule
\end{tabular}
\end{table}

\section{Bayesian sensitivity analysis}~\label{sec:sensitivity-analysis}
A natural extension of uncertainty propagation is sensitivity analysis, for which the connection can be summarized as:
\begin{quotation}
    \textit{
    A systematic decomposition of output variance into contributions attributable to each input parameter, providing a measure of their relative influence and informing model reduction, calibration, and experimental design.}
\end{quotation}

\subsection{General concept}
Mathematically, (variance-based) sensitivity analysis~\cite{saltelli2008global} and UQ are closely related: They share the same probabilistic formulations, but their primary objectives differ. Consider a deterministic forward model $y = f(\bm{x})$ with input parameter vector $\bm{x}$ treated as a random variable distributed according to a joint probability density $p(\bm{x})$. UQ focuses on describing how uncertainty in the inputs propagates through the model to produce uncertainty in the output, as in Eqs.~\eqref{eq:def-uncertainty-quantification} and~\eqref{eq:def-uncertainty-quantification-deterministic}. In contrast, variance-based sensitivity analysis quantifies how much each individual input $x_i$ contributes to the total output variance. In other words, it studies a specific family of functionals of the (conditional) output distribution --- conditional variances and variances of conditional expectations --- defined as
\begin{align}\label{eq:conditional-variance}
        \mathrm{var}_{x_i}\left[\mathbb{E}_{\bm{x}_{\lnot i}}\left[f(\bm{x}) \mid x_i \right]_{} \right]\;,
\end{align}
where $\bm{x}_{\lnot i}$ denotes the collection of all input variables except $x_i$. Here, we extend the notation for $\mathbb{E}[\cdot]$ and $\var[\cdot]$ introduced in Section~\ref{sec:UQ}: The subscript indicates the marginalized variables, while the vertical bar indicates the conditional variables. More precisely, the conditional expectation is
\begin{align} \label{eq:aux1}
    \mathbb{E}_{\bm{x}_{\lnot i}}\big[f(\bm{x}) \mid x_i\big] & = \int f(\bm{x}) p( \x_{\lnot i} \mid x_i)\:\mathrm{d}\x_{\lnot i} \nonumber \\ 
    & \eqcolon h(x_i)\;.
\end{align}
where $p(\bm{x}_{\lnot i}\mid x_i)$ reads as the conditional density of the remaining variables given $x_i$ and the notation $\mathrm{d}\x_{\lnot i}$ indicates marginalization over all parameters except $x_i$. That means, we keep the variable $x_i$ fixed and integrate out the influence of all remaining parameters. This operation effectively averages the model output over the uncertainty in all inputs other than $x_i$, producing a conditional mean response $h(x_i)$ that depends solely on that parameter. In simple terms, this measures how much variation in the single parameter $x_i$ influences the average model output when the effects of all other parameters are averaged out.

The corresponding variance of the conditional mean response is
\begin{align}\label{eq:aux2}
    \mathrm{var}_{x_i}\left[h(x_i)\right] & = \int \big(h(x_i) - \mathbb{E}_{x_i}[h(x_i)] \big)^{2} \nonumber\\
    &\hphantom{=}\,\, \times p(x_i)\:\mathrm{d}x_i\;,
\end{align}
where $p(x_i)$ is the marginal probability density of input variable $x_i$. The expectation appearing in the expression is
\begin{align}
    \mathbb{E}_{x_i}\big[h(x_i)\big] = \int h(x_i) p(x_i)\: \mathrm{d}x_i\;.
\end{align}

These variances of conditional expectations quantify how much the expected model response changes as a single parameter $x_i$ varies, after averaging out the influence of all remaining inputs. In variance-based sensitivity analysis~\cite{ChapterMelito}, this quantity plays a central role: it represents the contribution of parameter $x_i$ to the overall uncertainty in the model output. To compare this contribution across parameters, it is normalized by the total output variance, which leads directly to the definition of the first-order Sobol’ sensitivity indices.

From these variances of conditional expectations, we define the first-order Sobol’ sensitivity indices~\cite{sobol1990sensitivity} as
\begin{equation}\label{eq:def-first-order-sobol}
    S_i = \frac{\mathrm{var}_{x_i}\left[\mathbb{E}_{\bm{x}_{\lnot i}}
    \big[f(\bm{x}) \mid x_i\big]\right]}
    {\mathrm{var}\left[f(\bm{x})\right]}\;,
\end{equation}
which quantify the fraction of the total output variance that can be attributed to input parameter $x_i$. These quantities --- equivalently, the integral appearing in Eq.~\eqref{eq:conditional-variance} --- can be estimated either by direct Monte Carlo sampling of the simulation model or, more efficiently, by using a surrogate model (cf. Section~\ref{ssec:inverseII}).

A key reason for the popularity of PCEs (cf. Section~\ref{ssec:PCE}) is that, under the assumption of independent input parameters, both the total output variance and the first-order partial variances can be written directly in terms of the PCE coefficients. In particular,
\begin{align}\label{eq:pce_var}
\begin{split}
    \mathrm{var}\left[f(\x)\right] &= \sum_{\bm{\alpha} \in  \mathcal{A}} c^2_{\bm{\alpha}} - c_0^2\;,\\
    \mathrm{var}_{x_{i}}\left[h(x_i)\right] &= \sum_{\bm{\alpha} \in  \mathcal{A}_i} c^2_{\bm{\alpha}} - c_0^2\;,
\end{split}
\end{align}
where $\mathcal{A} = \{\bm{\alpha}\}$ is the set of all multi-indices associated with the polynomial basis functions in the PCE. The coefficient $c_0= \mathbb{E}[f(\vv x)]$ corresponds to the constant basis function, the polynomial of order zero. The subset $\mathcal{A}_i = \{\bm{\alpha} : \alpha_i = 0\}$ contains the multi-indices of basis functions that have degree zero in the variable $x_i$. These coefficients determine how each basis function contributes to the total variance of the model output and how much of that variance is caused by variations in the parameter $x_i$. Importantly, that contribution is a function of not only (i) the dependency of $f$ on $x_i$, but also of (ii) the distribution of $x_i$, in particular also the domain on which it is supported, and (iii) the value of (ii) in relation to (i).

The analytical expressions above are, however, a unique advantage of PCEs. For other surrogate models, such as GPs or PINNs, evaluating Sobol' indices generally requires additional Monte Carlo integration or numerical quadrature, making the computation significantly more expensive. This efficiency is exacerbated for higher-order Sobol' indices, such as second-order interactions,
\begin{equation} \label{eq:sobol-2nd-order-def}
    S_{ij} \propto \mathrm{var}_{x_{i}, x_j}\left[\mathbb{E}_{\bm{x}_{\lnot i,\lnot j}}  \big[ f(\bm{x}) \mid x_i, x_j\big]\right]\;,
\end{equation}
which quantify the joint influence of parameters $x_i$ and $x_j$. Since the number of Sobol' indices grows combinatorially with the number of input parameters, brute-force Monte Carlo estimation of all the indices (Eq.~\eqref{eq:def-first-order-sobol}) quickly becomes infeasible, even more so when non-PCE surrogate models are used.

Nevertheless, the interpretability of these sensitivity indices is inherently limited, since they are based solely on the variance. The variance contains no information about (a)symmetry or the tails of the underlying probability distribution and therefore provides a complete picture only if the distribution is Gaussian~\cite{ChapterWollner}. 
Furthermore, the interpretation of the Sobol' indices is more delicate for correlated input variables. 
Consequently, these variance-based estimates should be interpreted with caution in practice. For a more detailed discussion of sensitivity analysis and Sobol' indices, including total effect indices and generalizations to dependent inputs, we refer the reader to related literature~\cite{ChapterMelito}.

From a Bayesian viewpoint, however, a fully probabilistic version of sensitivity analysis would also marginalize over the uncertainty in the surrogate model parameters. For instance, placing a GP prior on the surrogate model~\cite{oakley2004sensitivity} allows one to propagate parameter uncertainty through the sensitivity indices themselves.

\subsection{Bayesian generalization}
The term \emph{Bayesian sensitivity analysis} is not yet universally defined or established. However, Bayesian theory provides a broader framework. In Eqs.~\eqref{eq:conditional-variance} to~\eqref{eq:sobol-2nd-order-def}, no surrogate model is introduced, except in Eq.~\eqref{eq:pce_var}. The key distinction between Bayesian and vanilla sensitivity analysis lies in the incorporation of priors and the marginalization of hyperparameters. This difference becomes particularly explicit in surrogate-based sensitivity analysis.
 
Let the surrogate parameters be denoted by $\vv\theta$. In Eqs.~\eqref{eq:conditional-variance} to~\eqref{eq:sobol-2nd-order-def}, their role was left implicit. In Eq.~\eqref{eq:pce_var}, for example, we effectively assumed that the PCE coefficients $\vv \theta = (c_\vv \alpha)^{\rm T}$ are known exactly. That is, the variance of $f(\vv x)$ was conditioned on fixed PCE coefficients. Consequently, the Sobol' indices (Eqs.~\eqref{eq:def-first-order-sobol} and~\eqref{eq:sobol-2nd-order-def}) are likewise conditioned on fixed $\vv\theta$. To state this explicitly, Eqs.~\eqref{eq:aux1} and~\eqref{eq:aux2} should be written as
\begin{align} \label{eq:aux3}
    \mathbb{E}_{\bm{x}_{\lnot i}}\big[f(\bm{x}) \mid x_i\big] & \equiv \mathbb{E}_{\bm{x}_{\lnot i}}\big[f(\bm{x}) \mid x_i, \vv \theta\big] \nonumber \\ 
    & = \int f(\bm{x}) p( \x_{\lnot i} \mid x_i, \vv \theta) \:\mathrm{d}\x_{\lnot i} \nonumber \\
    &=: h(x_i)\;.
\end{align}
and
\begin{align}\label{eq:aux4}
        \mathrm{var}_{x_i}\left[h(x_i)\right] & \equiv \mathrm{var}_{x_i}\left[h(x_i) \mid \vv \theta\right] \nonumber \\
        & = \int \big(h(x_i) - \mathbb{E}[h(x_i)] \big)^{2} \nonumber\\
        & \hphantom{=}\,\,\times p(x_i \mid \vv \theta)\:\mathrm{d}x_i\;, 
\end{align}
where all expectations and variances are now explicitly conditioned on fixed surrogate parameters $\vv\theta$.

However, as shown in Sections~\ref{ssec:PCE} and~\ref{sec:model-selection-PCE}, the PCE coefficients are generally not known exactly but must be treated as uncertain random variables.
A fully Bayesian surrogate-based sensitivity analysis therefore assigns a prior to $\vv\theta$ and marginalizes all quantities of interest with respect to it. This marginalization can be performed directly at the level of the Sobol' indices in Eq.~\eqref{eq:def-first-order-sobol}, leading to the Bayesian estimator
\begin{align}\label{eq:def-Bayesian-Sobol-index1}
    \mathbb{E}_{\vv \theta}[S_i]= \int S_ip(\vv \theta) \:\mathrm{d}\vv \theta\;. 
\end{align}

The definition in Eq.~\eqref{eq:def-Bayesian-Sobol-index1} can be impractical due to the fraction in Eq.~\eqref{eq:def-first-order-sobol} appearing inside the integral. A more tractable approach marginalizes the surrogate parameters $\vv\theta$ in the expectations and variances instead. From Eq.~\eqref{eq:def-first-order-sobol}, this leads to the generalized Bayesian Sobol' index
\begin{align} \label{eq:def-Bayesian-Sobol-index2}
    S^{\rm Bayes}_i := \frac{\int \mathrm{var}_{x_i}\left[\mathbb{E}_{\bm{x}_{\lnot i}}
    \big[f(\bm{x}) \mid x_i\big]\right] p(\vv\theta) \:\mathrm{d} \vv\theta}
    {\int \mathrm{var}\left[f(\bm{x})\right] p(\vv\theta) \:\mathrm{d}\vv\theta}\;,
\end{align}
where the integrands are understood to be conditioned on $\vv\theta$, as in Eqs.~\eqref{eq:aux3} and \eqref{eq:aux4}. 
Importantly, the definitions in Eqs.~\eqref{eq:def-Bayesian-Sobol-index1} and \eqref{eq:def-Bayesian-Sobol-index2} are not equivalent and may yield different results in general.
Of these, the later definition is often preferred due to its favorable properties. For instance, in the case of a PCE surrogate with Gaussian likelihood and a flat prior on the coefficients, marginalization amounts to replacing fixed coefficient values in Eq.~\eqref{eq:pce_var} by their posterior expectations, i.e., 
\begin{align} \label{eq:Bayesian-conditoinal-variance}
\begin{split}
        \mathrm{var}\left[f(\x) \mid \vv\theta \right] & = \sum_{\bm{\alpha} \in  \mathcal{A}} c^2_{\bm{\alpha}} - c_0^2\\
        {\xrightarrow{\rm Bayes}}  \quad \mathrm{var}\left[f(\x)\right] & = \sum_{\bm{\alpha} \in  \mathcal{A}} \mathbb{E}_{\vv \theta}[c^2_{\bm{\alpha}}] - \mathbb{E}_{\vv \theta}[c_0^2] \;.
\end{split}
\end{align}
Analogous replacements apply to the variances of the conditional expectations in Eq.~\eqref{eq:pce_var}. On the left-hand side, the conditioning on $\vv\theta$ --- implicit in Eq.~\eqref{eq:pce_var} --- is made explicit; on the right-hand side, the coefficients are marginalized. For this setting, both the posterior expectations and uncertainty estimates of the PCE coefficients --- and thus the Bayesian Sobol' indices in Eq.~\eqref{eq:def-Bayesian-Sobol-index2} --- admit closed-form expressions; see~\cite{ranftl2021bayesian} for details.

The advantage of this Bayesian formulation is that meaningful sensitivity analysis can be performed even when the surrogate is not \emph{sufficiently} accurate. In particular, uncertainty estimates for the Sobol' indices can be derived that explicitly account for surrogate inaccuracy through the posterior of the surrogate parameters, $p(\vv \theta \mid \mathcal{D})$. Rather than conditioning on a single fixed estimate of $\vv \theta$, the Sobol' indices are obtained by marginalizing over this posterior. In this way, uncertainty in the surrogate parameters is systematically propagated into the sensitivity measures. As a result, the sensitivity analysis becomes quantitatively defensible, even when the surrogate’s accuracy is uncertain.

Finally, employing a GP surrogate instead of, for example, a PCE surrogate does not in itself constitute a fully Bayesian sensitivity analysis, contrary to suggestions~\cite{BECKER20111499,melis2017bayesian}. A fully Bayesian treatment would also require marginalization over the GP kernel hyperparameters. In practice, however, these are typically approximated by maximum likelihood point estimates (cf. Section~\ref{ssec:GP}). Moreover, GP-based sensitivity analysis generally relies on numerical integration, since the simple analytical expressions available for PCEs in Eqs.~\eqref{eq:pce_var} and~\eqref{eq:Bayesian-conditoinal-variance} do not apply.

\section{Random fields for constitutive uncertainty}\label{sec:random-fields}
The core objective of random field modeling can be formulated as:

\begin{quotation}
    \textit{Representation of spatially or temporally varying material or model parameters as stochastic priors that encode spatial correlation, uncertainty, and prior knowledge across the domain.}
\end{quotation}

\subsection{General concept and Gaussian random fields} \label{ssec:gaussian-fields}
Random fields~\cite{vanmarcke2010random} provide a useful framework for modeling spatially distributed uncertainties in material properties, particularly for inherently heterogeneous materials such as biological tissue. In this section, we establish that random fields, in this sense and for this purpose, can be understood as a particular and sophisticated form of prior, $p(\bm{x}) \equiv p(\bm{x} \mid \mathbf{X})$, within the general UQ framework introduced in Eq.~\eqref{eq:def-uncertainty-quantification}, i.e., the distribution of the parameters $\vv x$ is now conditioned on spatial coordinates $\mathbf{X}$.

Colloquially, a random field generalizes the concept of a random variable to field quantities, thereby defining a random variable at every point in a spatial domain, with a prescribed correlation between every pair of points within this domain. For example, the GPs introduced in Section~\ref{ssec:GP} can be regarded as a special case of \emph{Gaussian} random fields in parameter space.

In the context of constitutive modeling, a random field assigns a random variable to each point $\mathbf{X} \in \Omega_0$ in the reference domain. In this sense, we merely reinterpret GPs from a different perspective. In surrogate modeling and machine learning (cf. Section~\ref{ssec:GP}), GPs were employed for regression --- that is, to learn an unknown function from data. Here, by contrast, we use GPs for stochastic modeling and to generate samples (realizations) from the modeled probability distribution.

This perspective highlights two key differences relative to Section~\ref{ssec:GP}. First, we now assign a random variable to every point in the \emph{spatial domain}, whereas in surrogate modeling, a random variable was assigned to each point in \emph{parameter space}. Second, our goal here is to sample from the distribution, rather than to infer or learn its parameters in the distribution.

Let $g$ be the variable we aim to model as a random field. A widely used prior model is the Gaussian random field, denoted as
\begin{equation}
    g \sim \mathrm{GP}(\mu(\mathbf{X}), k(\mathbf{X}, \mathbf{X}'))\;,
\end{equation}
or, formally equally,
\begin{equation}
    g \mid \mathbf{X}, \mu, k \sim \mathcal{N}(\mu(\mathbf{X}), k(\mathbf{X}, \mathbf{X}'))\;,
\end{equation}
with mean function $\mu(\mathbf{X})$ and covariance function $k(\mathbf{X}, \mathbf{X}')$, which defines the spatial correlation. Note that $\mu$ and $k$ usually depend on hyperparameters just like in the regression setting in Section~\ref{ssec:GP}, however we omit them in the notation here since the hyperparameters are a matter of \textit{a priori} choice and usually fixed. For $k$, we may again choose, for example, the squared-exponential kernel introduced in Eq.~\eqref{eq:def-SEkernel}. To generate a realization (or sample) from this prior, we evaluate the field $g = \mathcal{G}(\underline{\mathbf{X}})$ on a discretized spatial grid $\underline{\mathbf{X}} = (\mathbf{X}^{(1)}, \dots, \mathbf{X}^{(N_{\mathrm{X}})})^{\rm T}$ of $N_{\mathrm{X}}$ nodes. The resulting vector $g$ is jointly Gaussian distributed with covariance matrix
\begin{equation}
    g \sim \mathcal{N}(0, K)\;, \quad [K]_{ij} = k(\mathbf{X}^{(i)}, \mathbf{X}^{(j)})\;.
\end{equation}
Identifying $g \coloneqq \vv x$, this construction defines a prior for our physical parameters $\vv x$ at locations $\mathbf{X}$, $\vv x \mid \mathbf{X} \sim \mathcal{N}(0, K)$, for uncertain input parameters in a general UQ problem, as discussed in Section~\ref{ssec:propagation}. 
In order to evaluate the UQ formulations in Eqs.~\eqref{eq:def-uncertainty-quantification} and~\eqref{eq:def-uncertainty-quantification-deterministic}, we therefore must be able to sample random-field realizations from $p(\vv x \mid \mathbf{X})$. Since $\vv x$ now is a random field, we often, but not necessarily, find that $y$ is also a random field. For example, the Cauchy stress would be a random field, while its spatial average would be a simple scalar random variable.

In the following, we outline how such realizations can be generated for $\vv x$ for Gaussian and non-Gaussian random fields. We may use these realizations then to generate samples of $y$ in order to obtain Monte Carlo estimates for the UQ equations (Eq.~\eqref{eq:def-uncertainty-quantification-deterministic}), which are then given by $y(\vv x)$ or $\hat{y}(\vv x)$, depending on whether we can evaluate the desired number of samples with the original simulation $y$ or need a surrogate $\hat{y}$ instead. 

We highlight a few representative examples in biomechanics. Gaussian random fields have been adopted, for instance by Bosnjak et al.~\cite{Bosnjak2025a}, who perturbed patient-specific aortic geometries through displacement maps derived from Gaussian fields to quantify boundary uncertainty. Hauseux et al.~\cite{Hauseux2018a} modeled spatial variability in anisotropic brain tissue properties using correlated Gaussian fluctuation. Tran et al.~\cite{tran2019uncertainty} represented spatial variations in the arterial stiffness of coronary artery bypass grafts via Gaussian field approximations. 
In engineering fracture mechanics, Gaussian random fields have been applied to model spatial variability in phase-field simulations, including: critical energy release rate distributions~\cite{He2025a}, random pore configurations in porous materials~\cite{Su2023a}, and spatially varying failure strength and fracture toughness in quasi-brittle materials~\cite{Hai2022a}. Beyond phase-field simulations, Gaussian Markov random fields have been applied to reconstruct spatially varying fields such as porosity in large-scale finite element simulations on three-dimensional domains~\cite{Nitzler2026b}. Additionally, researchers have extracted Gaussian random field hyperparameters from scanning electron microscope images of fiber distributions in reinforced composite plates to perform finite element-based microscopic fracture analysis, thereby estimating local apparent strength while considering random microstructure morphologies~\cite{Stefanou2022a,Sakata2026a}

\begin{figure*}
    \centering
    \includegraphics[scale=1.0]{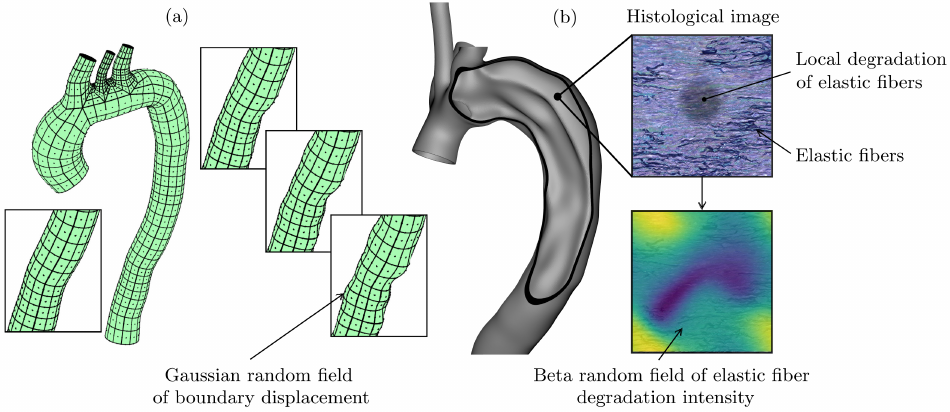}
    \caption{Exemplary applications of (non-Gaussian) random fields in biomechanics: (a) stochastic perturbation of a patient-specific aortic surface, in which the nodes of a structured hexahedral mesh are displaced using maps derived from Gaussian random fields to model local boundary uncertainty; (b) modeling local pathological degradation of elastic fibers in aortic dissection as observed in histology, where the spatially varying degradation is represented by a beta random field.}
    \label{fig-random_field}
\end{figure*}

Beyond Gaussian fields, non-Gaussian random fields have been introduced to capture more complex or bounded spatial variability, such as beta-distributed fields modeling elastic fiber degradation in the pathological aortic wall~\cite{ranftl2022stochastic}, spatially dependent fiber-orientation distributions required for anisotropic constitutive models of the aortic wall in patient-specific simulations~\cite{Staber2018a}, or heterogeneous isotropic material parameters in abdominal aortic aneurysm models~\cite{Biehler2015a}. Two such representative examples are illustrated in Fig.~\ref{fig-random_field}.

\subsection{Sampling random field realizations}\label{ssec:sampling-random-fields}
There are multitude of ways to draw samples $g^{(s)}$. These realizations serve as spatially correlated uncertain model inputs, such as spatially inhomogenous parameters in a constitutive model. Here, we introduce three commonly used strategies: the Cholesky decomposition, spectral methods, and the Karhunen–Loève expansion.

\subsubsection{Cholesky decomposition}\label{sssec:cholesky-decompostion}
To sample a finite-dimensional realization from a GP prior, we make use of the fact that the GP marginal at a set of input points $\underline{\mathbf{X}}$ follows a multivariate normal distribution. By definition of a GP, this can be written as 
\begin{gather}
    \bm{g} \sim \mathcal{N}(\boldsymbol{\mu}, K)\;,  \quad \boldsymbol{\mu}^{(i)} = \mu(\mathbf{X}^{(i)})\;, \nonumber\\
    \quad [K]_{ij} = k(\mathbf{X}^{(i)}, \mathbf{X}^{(j)})\;.
\end{gather}
where $\mu$ is the mean vector and $K$ is the covariance matrix describing spatial correlations between all points.

To sample from this distribution, we use the Cholesky decomposition of the covariance matrix. Let $L \in \mathbb{R}^{N \times N}$ denote the lower-triangular matrix obtained from
\begin{equation}
K = LL^{\rm T}\;,
\end{equation}
where $K$ is positive definite by construction.
The sampling procedure then consists of two simple steps
\begin{enumerate}[label=\roman*).]
    \item Draw a standard i.i.d. normal random vector $\vv\zeta^{(s)}$ from $\vv \zeta \sim \mathcal{N}(\mathbf{0}, I_N)$, where $I_N$ is the $N \times N$ identity matrix.
    \item Transform this sample using $\bm{g}^{(s)} = \boldsymbol{\mu} + L \vv \zeta^{(s)}$.
\end{enumerate}
This transformation produces a sample $\bm{g}^{(s)}$ from $\mathcal{N}(\boldsymbol{\mu}, K)$ that has the desired mean and covariance. Indeed, we can easily check that
\begin{align}
    \mathbb{E}\big[\bm{g}\big] = \boldsymbol{\mu} + L\, \mathbb{E}\big[\vv \zeta \big] = \boldsymbol{\mu}\;,
\end{align}
and
\begin{align}
    \mathrm{cov}[\bm{g}] & =  \mathbb{E}\big[(L\vv\zeta)(L\vv\zeta)^{\rm T} \big] = L  \mathbb{E}\big[\vv\zeta\vv\zeta^{\rm T} \big] L^{\rm T} \nonumber \\
    & = L I_N L^{\rm T} = LL^{\rm T} = K\;.
\end{align}
Since $\bm{g}$ has mean $\boldsymbol{\mu}$ and covariance $K$, and linear transformations of Gaussian variables remain Gaussian, $\bm{g}^{(s)}$ is a valid sample from the GP marginal $\mathcal{N}(\boldsymbol{\mu},K)$.

Note that we would arrive at the same result if we used the square root of the inverse covariance matrix, $K^{-1/2}$, instead of $L$. In other words, we could also use a singular value decomposition instead of a Cholesky decomposition of the covariance for the purposes of generating samples from the distribution defined by that covariance. However, this approach has an important limitation. 

The computational cost of the Cholesky factorization, and of general matrix inversions, scales as $\mathcal{O}(N_x^3)$, where $N_x$ is the number of collocation points, i.e., the number of spatial nodes in the finite element or finite volume discretization where the random field is evaluated. The same limitation appears in GP regression in the machine learning setting (Section \ref{ssec:GP}), where the scaling depends on the number of data points rather than the number of spatial nodes.
As a consequence, sampling random-field realizations over large or finely discretized domains becomes computationally expensive. This bottleneck is especially pronounced in biomechanics, where finely discretized spatial domains or three-dimensional patient-specific computational models often involve thousands or millions of degrees of freedom. In the next section, we introduce an alternative sampling method that achieves a more favorable computational scaling.

\subsubsection{Spectral methods (Shinozuka's method)}\label{sssec:spectral-methods}
Spectral methods, and in particular Shinozuka’s method~\cite{Shinozuka1991a,Shinozuka1996a}, offer an efficient alternative for simulating \emph{stationary} Gaussian random fields without explicitly forming or factorizing the covariance matrix.

Shinozuka’s method is based on representing a stationary GP in the frequency domain, where the covariance structure can be expressed through its power spectral density. A random field $g(\mathbf{X})$ is called \emph{stationary} when its covariance depends only on the relative distance $\vv r := \mathbf{X}-\mathbf{X}'$ between two points rather than their absolute positions
\begin{equation}
    k(\mathbf{X},\mathbf{X}') = k(\mathbf{X} - \mathbf{X}') = k(\vv r)\;. 
\end{equation}

For zero-mean stationary GP $g(\mathbf{X})$ with covariance function 
\begin{equation}
    k(\mathbf{r}) = \mathbb{E}\big[g(\mathbf{X}) g(\mathbf{X} + \mathbf{r})\big]\;,
\end{equation}
the corresponding power spectral density $S(\boldsymbol{\omega})$ is given by the Fourier transform of $k$, i.e., 
\begin{equation}
    S(\boldsymbol{\omega})  = \int_{\mathbb{R}^D} k(\mathbf{r}) e^{-i \boldsymbol{\omega} \cdot \mathbf{r}} \: \mathrm{d}\mathbf{r}\;.
\end{equation}
The power spectral density quantifies how the total variance of the random field is distributed across spatial frequencies, analogous to how an energy spectrum describes the contribution of different modes to a physical signal.

\newcommand{\nuorp}{p}
\newcommand{\nuorpN}{P}
To construct a sample realization of $g(\mathbf{X})$ over a spatial domain, Shinozuka’s method discretizes the frequency space into a finite set of wavevectors $( \boldsymbol{\omega}_\nuorp )_{\nuorp=1}^{\nuorpN}$. Each wavevector $\boldsymbol{\omega}_\nuorp$ defines a plane wave with a specific direction and spatial frequency: its magnitude $||\boldsymbol{\omega}_\nuorp|| = 2\pi/\lambda_\nuorp$ corresponds to the wavelength $\lambda_\nuorp$, and its orientation determines the direction of spatial oscillation.
For instance, in one dimension with a uniform grid, we could discretize as $\omega_\nuorp = \nuorp\Delta\omega$ with $\Delta\omega = \omega_{\rm max}/\nuorpN$.

For each frequency component, a random phase 
\begin{equation}\label{eq:shinozuka-aux-uniform}
    \vartheta_\nuorp \sim \mathrm{Uniform}[0, 2\pi)
\end{equation}
is independently drawn to introduce randomness while preserving the desired second-order statistics. Each component is then assigned an amplitude with volume weights $\Delta \boldsymbol{\omega}_\nuorp$, and the field is reconstructed as a weighted sum of cosine waves
\begin{equation} \label{eq:shinozuka}
    g(\mathbf{X}) \approx \sum_{\nuorp=1}^{\nuorpN} \sqrt{2 S(\boldsymbol{\omega}_\nuorp) \Delta \boldsymbol\omega_\nuorp} \cos(\boldsymbol{\omega}_\nuorp \cdot \mathbf{X} + \vartheta_\nuorp)\;,
\end{equation}
i.e., a discrete (Riemann-)approximation to the inverse Fourier transform.
This harmonic superposition, where the discretizations of the spatial and frequency domains need not necessarily coincide, produces a spatially correlated random field whose covariance matches the target $k(\mathbf{r})$ in the limit of fine frequency discretization, which can be shown, for example, via the central limit theorem.

Shinozuka’s spectral method offers several appealing advantages over direct covariance-based approaches. It is computationally efficient, since it bypasses the need to assemble and factorize large covariance matrices, leading to significant savings in both time and memory. The approach also scales better to higher dimensions and finely discretized domains, especially when implemented using fast Fourier transform techniques on regular grids. Furthermore, the spectral representation provides an intuitive interpretation of spatial variability in terms of underlying frequency content, enabling direct control over the correlation structure through the power spectral density. 

Owing to these properties, Shinozuka’s method has become a practical choice for generating large-scale realizations of stationary random fields, where it facilitates efficient modeling of spatial heterogeneity in material properties, tissue structure, or loading conditions. Alas, Shinozuka's method assumes stationarity. For non-stationary fields, usually more sophisticated, and often less scalable, methods are necessary, one of which we will discuss next.

\subsubsection{Karhunen-Loève expansion}\label{sssec:Karhunen-Loeven-expansion}
The Karhunen–Loève expansion~\cite{ghanem1991Stochastic,xiu2010numerical,maitre2010spectral} --- a representation for random fields closely related to Shinozuka’s method --- offers a mathematically elegant and physically interpretable decomposition of a random field in terms of deterministic spatial modes and random coefficients.
It provides an eigen-spectral representation of any second-order random field (i.e., one with finite variance) whose covariance function is known\footnote{Specifying the covariance function is a major practical limitation in, for example, biomechanics. Empirical estimation of spatial correlations is often restricted by limited sample sizes and the high degree of inter-subject variability inherent in biological tissues.}.

Let $g(\mathbf{X})$ be a zero-mean, square-integrable random field defined over a (spatial) domain $\Omega_0 \subset \mathbb{R}^D$, with covariance function
\begin{equation} \label{eq:karhunen-loeve-covariance-aux}
    k(\mathbf{X}, \mathbf{X}') = \mathbb{E}\big[ g(\mathbf{X}) g(\mathbf{X}')\big]\;. 
\end{equation}
Under suitable regularity conditions on $k$, the field can be expressed \emph{exactly} as an infinite series
\begin{equation}
    g(\mathbf{X}) = \sum_{p=1}^\infty \sqrt{\lambda_p} \, \zeta_p \, \phi_p(\mathbf{X})\;,
\end{equation}
where $\zeta_p$ are i.i.d. standard normal random variables, $\zeta_p \sim \mathcal{N}(0,1)$, and $(\lambda_p, \phi_p(\mathbf{X}))$ are the eigenvalue–eigenfunction pairs of the generalized eigenvalue problem given by the integral equation
\begin{equation} \label{eq:KL-eigenproblem}
    \int_{\Omega_0} k(\mathbf{X}, \mathbf{X}') \phi_p(\mathbf{X}') \:\mathrm{d}\mathbf{X}' = \lambda_p \phi_p(\mathbf{X})\;.
\end{equation}
Each term in the Karhunen-Loève expansion represents a spatial mode $\phi_p(\mathbf{X})$ modulated by a random coefficient $\sqrt{\lambda_p}\xi_p$. The eigenvalue $\lambda_p$ quantify the mean contribution, or mean \emph{energy}, of each mode to the total field variance, meaning that modes associated with small eigenvalues contribute little to the overall variability. We notice that through Eq.~\eqref{eq:KL-eigenproblem}, the expansion in general depends on the domain $\Omega_0$.

In practice, the series is truncated after a finite number of terms $P$, yielding the approximation
\begin{equation} \label{eq:KL-expansion}
    g(\mathbf{X}) \approx \sum_{p=1}^P \sqrt{\lambda_p} \xi_p \phi_p(\mathbf{X})\;,
\end{equation}
where $P$ is chosen so that the cumulative energy of the retained modes captures a desired proportion of the total variance. Owing to the orthonormality of the eigenfunctions, the variance of the truncated expansion is given by the sum of the retained eigenvalues. Consequently, the relative variance represented by the first $P$ modes reads
\begin{equation}
    \var[g] \propto \frac{\sum_{p=1}^P\lambda_p}{\sum_{p=1}^\infty\lambda_p} \approx 1\;.
\end{equation}
This truncation provides a compact, low-dimensional representation of the stochastic field, useful for UQ. 

Shinozuka’s method can be interpreted as a special case of the Karhunen–Loève expansion formulated in the Fourier domain and truncated to a finite number of Fourier modes. This enables efficient computation but restricts the method to stationary covariance functions and infinite or periodic domains. In contrast, the Karhunen–Loève expansion requires the solution of the eigenvalue problem in Eq.~\eqref{eq:KL-eigenproblem}. Although computationally more demanding, it permits the modeling of more general and spatially complex random fields.

There is also a structural similarity between the Karhunen–Loève expansion and PCE (cf. Section~\ref{ssec:PCE}). However, their purposes differ fundamentally. In Section~\ref{ssec:GP}, PCE was used to learn an approximation $\hat{y}(\vv x)$ from data, similar in spirit to GPs in machine learning. Here, in contrast, the expansion is employed to \emph{generate} samples from $p(\vv x\mid \mathbf{X})$. Furthermore, although both approaches constitute orthogonal decompositions (typically in a Hilbert space), they differ in their objectives, underlying assumptions, and construction.

Furthermore, the Karhunen-Loève expansion is conceptually related to the representer theorem (cf. Eq.~\eqref{eq:representer-theorem}) and Mercer’s theorem, shown below, which also connect positive-definite kernels to orthogonal basis functions. In fact, for GPs, the Karhunen-Loève expansion can be interpreted as decomposing the GP prior into orthogonal spatial modes, where the eigenfunctions correspond to the principal directions of variability implied by the covariance kernel.

\bigbreak
\textit{Mercer's theorem}. In simple terms, the Mercer's theorem, which is typically formulated in the setting of Hilbert spaces in a manner similar to the representer theorem, states that a kernel function $k$ can be represented via its non-negative eigenvalues $(\lambda_p)_{p=1}^\infty$ and corresponding orthogonal eigenfunctions $(\phi_p)_{p=1}^\infty$ as follows
\begin{equation}
    k(\mathbf{X}, \mathbf{X}') = \sum_{p=1}^\infty \lambda_p \phi_p(\mathbf{X})\phi_p(\mathbf{X}')\;.
\end{equation}
In this expression, each eigenfunction $\phi_p(\mathbf{X})$ represents a deterministic spatial mode, while its associated eigenvalue $\lambda_p$ quantifies the magnitude, or \emph{energy}, of that mode in the kernel’s representation. With this knowledge, we can also understand why Eq.~\eqref{eq:KL-expansion} yields samples whose covariance follows Eq.~\eqref{eq:karhunen-loeve-covariance-aux} by abstracting the proof for the samples obtained from Cholesky decomposition in Section~\ref{sssec:cholesky-decompostion}. 
In the context of GPs, the kernel function $k(\mathbf{X}, \mathbf{X}')$ defines the covariance structure of the prior distribution over functions. Mercer’s theorem implies that this covariance kernel can be decomposed spectrally as above, i.e., it admits an eigenfunction decomposition, which in turn means that any sample from the GP prior $g \sim \mathrm{GP}(0, k(\mathbf{x}, \mathbf{x}'))$ can be represented (in distribution) as a above series expansion.

Thus, a GP prior can be viewed as a random series expansion over the orthonormal basis, a perspective that is closely related to the popular \emph{random features} approach \cite{rahimi2007random}. In this representation, each eigenfunction $\phi_p$ contributes a spatial mode of variability, and the associated eigenvalue $\lambda_p$ determines how strongly that mode influences the overall process. The larger the eigenvalue, the greater the contribution of that eigenfunction to the field’s variability.
This viewpoint provides a function-space interpretation of GPs. Instead of treating a GP as a distribution over random variables, it can be understood as a distribution over functions spanned by the basis $\phi_p$. The smoothness, statistics, complexity, and expressiveness of the GP are therefore determined directly by the spectral properties of the kernel $k(\mathbf{X},\mathbf{X}')$.

\subsection{Non-Gaussian random fields}\label{ssec:random-fields-nG}
Until now, we have only explicitly discussed Gaussian random fields in Section~\ref{ssec:gaussian-fields} and, separately, sampling methods in Section~\ref{ssec:sampling-random-fields}. Among these methods, only the Cholesky decomposition is strictly limited to Gaussian fields, while Shinozuka’s method yields only approximately Gaussian samples in the limit of a large number of modes, due to the central limit theorem.
In contrast, the Karhunen–Loève expansion (Section~\ref{sssec:Karhunen-Loeven-expansion}) does not rely on any assumption of Gaussianity. It merely ensures that the second-order statistics (Eq.~\eqref{eq:karhunen-loeve-covariance-aux}) are satisfied for a given random field, irrespective of its higher-order statistics. Only if all higher-order cumulants vanish is a Gaussian field recovered.
However, none of these methods is well suited for the explicit modeling of non-Gaussian random fields or their higher-order statistics, which we briefly introduce next.

The realization of non-Gaussian random fields is indeed an active area of research~\cite{Liu2019a}. In many classical engineering and biomechanical applications, random spatial fields do not follow a purely Gaussian distribution. For instance, variables such as material properties, biological tissue microstructure, or loading conditions may exhibit asymmetric or heavy-tailed spatial distributions that cannot be accurately represented by a Gaussian model. A common strategy to construct such fields is to apply a probability transformation to an underlying Gaussian random field.

Let $g$ denote a Gaussian random field and $g_{\rm non}$ the corresponding transformed non-Gaussian random field, with ${\mathcal{C}}_g$ and ${\mathcal{C}}_{g_{\rm non}}^{-1}$ denoting the cumulative distribution function of $g$ and the inverse cumulative distribution function (quantile function) of $g_{\rm non}$, respectively. The transformation between the two fields can then be written as~\cite{Grigoriu1995a,Grigoriu1998a,Kim2015a}
\begin{equation}
    g_{\rm non}(\Omega_0) = {\mathcal{C}}^{-1}_{g_{\rm non}}\Big[ {\mathcal{C}}_{g} \big[g(\Omega_0) \big]
    \Big]\;.
\end{equation}
In this mapping, each realization of the Gaussian field is first transformed to a uniform variable through its cumulative distribution function ${\mathcal{C}}_g$, and then mapped to the target non-Gaussian distribution using the inverse cumulative distribution function ${\mathcal{C}}_{g_{\rm non}}^{-1}$. This approach ensures that the marginal distribution of the transformed field follows the desired non-Gaussian form while inheriting the spatial correlation structure of the underlying Gaussian field in transformed form.

However, evaluating this transformation in practice can be challenging. Computing ${\mathcal{C}}_g$ may be computationally expensive, especially for large-scale correlated Gaussian fields, and closed-form expressions for the inverse ${\mathcal{C}}_{g_{\rm non}}^{-1}$ are often unavailable. Consequently, elaborate numerical procedures are typically required to generate approximate samples from non-Gaussian random fields~\cite{Vio2001b,Demetriu2005a,Bocchini2008a,Shields2011a,Kim2015a}. Despite these difficulties, such transformations remain a widely used and flexible approach for modeling complex spatial variability in engineering and biomechanical systems.

\subsubsection{Pointwise transformation methods}\label{sssec:pointwise-transformation-methods}
Non-Gaussian transformations can also be applied pointwise when specific constraints on the parameter range of a random field must be enforced. Such transformations are particularly useful when the modeled quantity must remain within physical or probabilistic bounds, as is often required in material modeling or probabilistic simulations~\cite{ranftl2022stochastic}.

To illustrate the concept, consider first a univariate example. Let $\mathrm{g}_1$ and $\mathrm{g}_2$ be two independent Gaussian random variables. The sum of their squares, $\mathrm{g}_1^2+\mathrm{g}_2^2$, follows a Gamma distribution, more precisely $\chi^2$-distribution. If we define $\Gamma_1 = \mathrm{g}_1^2+\mathrm{g}_2^2$ and $\Gamma_2 = \mathrm{g}_3^2+\mathrm{g}_4^2$ as  two independent Gamma-distributed variables, then their ratio $\Gamma_1/(\Gamma_1 + \Gamma_2)$ follows a Beta distribution, in this case, a uniform distribution, as shown in~\cite{ranftl2022stochastic}. This simple example demonstrates how non-Gaussian random variables with bounded support can be generated through algebraic combinations of Gaussian random variables~\cite{Hasofer1998a,Vio2001b,Vio2002a,Demetriu2005a}.

We now extend this concept to spatially correlated random fields. Let $\{g_{\nu}(\mathbf{X})\}$, $\nu = 1,\dots,2R$, $R \in \mathbb{N}$, be a collection of independent Gaussian random fields, and let $g_{\nu}(\underline{\mathbf{X}})$ be their corresponding discretization on a set of nodes $\underline{\mathbf{X}}$ as in Section~\ref{ssec:gaussian-fields}.  Each field $g_{\nu}$ is assumed to have identical statistical properties but remains independent of the others. For simplicity, we omit the superscripts $s$ denoting individual realizations, following the notation used in Section~\ref{ssec:sampling-random-fields}.

A Gamma-type random field is then obtained by pointwise transformation as
\begin{equation}\label{eq:Gammafield_definition}
     \Gamma_{R}(\mathbf{X}) = \frac{1}{2} \sum_{\nu=1}^{2R} g_{\nu}^2(\mathbf{X})\;,
\end{equation}
where the squaring and summation are applied element-wise, ensuring that $\Gamma_{R}$ and $g_{\nu}$ share the same spatial support and discretization. Consequently, a sample of a Gamma field, $\Gamma_{R}^{(s)}(\underline{\mathbf{X}})$, can be generated from at least two independent Gaussian random field samples, $g_{1}^{(s)}({\underline{\mathbf{X}}})$ and $g_{2}^{(s)}({\underline{\mathbf{X}}})$, by applying a pointwise transformation at each $\mathbf{X}^{(n)} \in \underline{\mathbf{X}}$ according to Eq.~\eqref{eq:Gammafield_definition}. Note that the Gamma distribution includes the exponential and $\chi$-distributions as special cases.

With two independent Gamma fields, $\Gamma_{R}(\mathbf{X})$ and $\Gamma_{R^\prime}(\mathbf{X})$, characterized by the same covariance structure, a Beta-type random field ${\beta}_{R,{R^\prime}}(\mathbf{X})$ can be constructed as
\begin{equation}\label{eq:Beta_random_field_definition}
    {\beta}_{R,{R^\prime}}(\mathbf{X}) = \frac{\Gamma_R(\mathbf{X})}{\Gamma_R(\mathbf{X}) + \Gamma_{R^\prime}(\mathbf{X})}\;,
\end{equation}
where the division and addition are likewise applied element-wise. Samples of ${\beta}_{R,{R^\prime}}^{(s)}(\underline{\mathbf{X}})$ can then be generated on this set of nodes, as before, from two independent Gamma random field samples, $\Gamma_{R}^{(s)}(\underline{\mathbf{X}})$ and $\Gamma_{R'}^{(s)}(\underline{\mathbf{X}})$, via a pointwise transformation according to Eq.~\eqref{eq:Beta_random_field_definition}.
The univariate marginal PDF of this field --- i.e., at a fixed spatial location $\mathbf{X}$ (cf. Eq.~\eqref{eq:Beta_random_field_definition}) --- follows a Beta distribution, 
\begin{gather} \label{eq:beta-marginal}
    \p{\beta}{\mathbf{X}} = \frac{1}{{\cal{B}}(R,R^\prime)} \beta^{R-1} (1-\beta)^{{R^\prime}-1}\;, \nonumber \\
    \quad \forall\,\mathbf{X} \in \underline{\mathbf{X}}\;, \quad 0 \leq \beta \leq 1\;.
\end{gather}
We have thus constructed a random field realization whose values are either strictly positive (Eq.~\ref{eq:Gammafield_definition}) or strictly bounded between 0 and 1 (Eq.~\ref{eq:Beta_random_field_definition}) at every point in the domain. These constraints are crucial in mechanical problems, where negative values of Young's modulus or unbounded order parameters such as damage variables are physically meaningless. Moreover, even a negative value at a single point in the domain can have detrimental implications for the numerical solution of the boundary value problem.

While generating Gaussian random fields on simple domains is typically straightforward, the computational effort can increase substantially for complex geometries or irregular meshes. Moreover, closed-form expressions for the correlation structure of transformed fields have been proposed in~\cite{Hasofer1998a,Vio2001b,Shields2011a}, yet practical applications may still face numerical challenges, especially when enforcing a prescribed correlation structure~\cite{Benowitz2015a}.

In summary, pointwise transformation methods provide a practical means of constructing non-Gaussian random fields with bounded support, enabling the modeling of physically constrained parameters. In practice, these methods are often combined with Gaussian field generation methods such as the Shinozuka spectral approach or the Karhunen–Loève expansion, which can serve as the approximately Gaussian bases for subsequent transformations\footnote{We note a word of caution that transformations mapping random variables from infinite to finite support can introduce numerical artifacts, causing the transformed fields to deviate from their expected statistical properties. Accordingly, practitioners should always verify the statistics of transformed fields numerically.}.

Finally, the random field generation methods presented here are not exhaustive. Readers interested in more sophisticated methods are referred to related excellent literature~\cite{Kramer2007a}. These include methods for anisotropic fields~\cite{Fuglstad2015b, Fuglstad2015a} or large computational domains~\cite{DeCarvalhoPaludo2019a, Panunzio2018a}, approaches based on fast Markovian approximations~\cite{Rue2001}, formulations that represent the random field as the solution of a PDE with stochastic forcing on the same spatial discretization~\cite{Lindgren2011a}, as well as Lanczos-type methods \cite{Chow2014a} and contour-integral-based methods~\cite{Aune2013a}.

\section{Discussion}
In this review article, we introduced UQ for mechanical problems through the lens of Bayesian probability theory. Specifically, we presented it as a unifying framework for the two central challenges in mechanics: forward problems and inverse problems, backed by a significant body of literature and simple yet illustrative examples. We place a specific emphasis on biomechanical applications, where inherent variability and uncertainty necessitates such a framework, and provide illustrative examples to clarify these concepts.

The Bayesian perspective offers several conceptual and practical advantages for mechanics, specifically within the field of biomechanics. First, Bayesian inference dissolves the artificial boundaries between UQ and propagation, parameter estimation, and surrogate modeling, enabling the seamless integration of sensitivity analysis and scientific machine learning, where data-driven models complement physics-based approaches. Second, by representing uncertainties as probability distributions rather than point estimates, Bayesian inference supports engineering design and decision-making. In biomechanics, this facilitates clinical decision-making under uncertainty, acting as a key enabler for clinical translation. Third, it allows for flexible and \emph{consistent, rigorous} integration of data and models --- priors, likelihoods, and surrogates act as modular components that can be tailored to specific mechanical problems. Finally, by making modeling choices explicit and quantifiable, Bayesian methods reduce hidden assumptions and clarify the distinction between \emph{epistemic} and \emph{aleatoric} uncertainty, and the mixing thereof, an ongoing debate often clouded by ambiguity~\cite{DerKiureghian2009a,ChapterWollner}. This improves interpretability, transparency, reproducibility, and predictive reliability. Furthermore, Bayesian frameworks naturally interface with computational tools like sampling, optimization, and experimental design, all of which are relevant in (bio)mechanics.

As there are still many open problems, several promising research directions emerge for advancing the Bayesian perspective in mechanics. The selection of priors, likelihoods, and surrogate models remains problem dependent and is often complicated by limited or inaccessible data. For many parameters, suitable priors are simply unavailable and substantial research is necessary to specify meaningful \emph{and} useful input distributions from experimental or clinical information. Progress in scalable Bayesian computation through efficient sampling, variational methods, and multifidelity strategies will be essential to handle high dimensional, nonlinear, and large scale biomechanical models. Closer integration with scientific machine learning, including NOs, PINNs, and hybrid physics–data models, offers new opportunities but still requires rigorous Bayesian formulations and explicit treatment of surrogate uncertainty, which is still in its infancy. Especially surrogate modeling approaches that are explicitly tailored to and informed by mechanics remain rare. Bayesian experimental design provides an underexplored means to maximize information gain and optimize data collection. Addressing random fields and multiscale uncertainty will be crucial for modeling spatially heterogeneous and hierarchical materials and for linking macroscale variability to microscale structure. Finally, advancing clinical translation by embedding Bayesian UQ into decision support systems can enhance the credibility and impact of computational biomechanics in personalized medicine and medical device development.

%

\section*{Declaration of competing interest}
The authors declare that they have no known competing financial interests or personal relationships that could have appeared to influence the work reported in this paper.

\section*{Author contribution}
S.R. conceived and developed the theoretical framework, and led the overall integration and writing of the manuscript. 
M.R. shaped the (bio)mechanics perspective, connected the theoretical framework to and conceptualized application examples, drafted parts of the manuscript, and refined the formal presentation for didactic clarity and managed the manuscript preparation. 
G.A.H., and E.K. contributed field-specific expertise, discussions, and manuscript revision. 
All authors reviewed and approved the final manuscript.

\section*{Acknowledgements}
S.R. was financially supported by the Austrian Science Fund (FWF) under grand no.~10.55776/J4774.
M.R. and E.K. acknowledge support from the European Research Council (ERC) under grant no.~101141626 (DISCOVER), funded by the European Union. Views and opinions expressed are, however, those of the authors only and do not necessarily reflect those of the European Union or the European Research Council Executive Agency. Neither the European Union nor the granting authority can be held responsible for them.
The authors wish to acknowledge Maximilian P. Wollner (Institute of Biomechanics, Graz University of Technology, Graz, Austria) for his many valuable discussions, constructive feedback, and critical revisions of earlier drafts of this manuscript.

%






\bibliography{literature}

\end{document}